%% file: harmonic.tex
\documentclass[a4paper]{article}
\usepackage{amsmath, amsfonts, amsthm,amssymb}
\usepackage{mathrsfs}
\usepackage[margin=2cm]{geometry}
\usepackage{xcolor}
\usepackage{colortbl}
\usepackage{graphicx}
\usepackage{caption}
\usepackage{subfig}
\usepackage{multirow,makecell,multicol}
\usepackage{hhline}
\usepackage{stackengine}
\usepackage{lscape}
\usepackage[T1]{fontenc}

\newcommand\xrowht[2][0]{\addstackgap[.5\dimexpr#2\relax]{\vphantom{#1}}}
\allowdisplaybreaks
\input{tcilatex}
\begin{document}

\title{Stress and power as a response to harmonic excitation of a \\
fractional anti-Zener and Zener type viscoelastic body}
\author{Sla\dj an Jeli\'{c}\thanks{
Department of Physics, Faculty of Sciences, University of Novi Sad, Trg D.
Obradovi\'{c}a 4, 21000 Novi Sad, Serbia, sladjan.jelic@df.uns.ac.rs} \quad
Du\v{s}an Zorica\thanks{
Department of Physics, Faculty of Sciences, University of Novi Sad, Trg D.
Obradovi\'{c}a 4, 21000 Novi Sad, Serbia, dusan.zorica@df.uns.ac.rs} \
\thanks{%
Mathematical Institute, Serbian Academy of Arts and Sciences, Kneza Mihaila
36, 11000 Belgrade, Serbia, dusan\textunderscore zorica@mi.sanu.ac.rs}}
\maketitle

\begin{abstract}
\noindent The stress as a response to strain prescribed as a harmonic
excitation is examined in both transient and steady state regime for the
viscoelastic body modeled by thermodynamically consistent fractional
anti-Zener and Zener models by the use of the Laplace transform method.
Assuming strain as a sine function, the time evolution of power per unit
volume, previously derived as a sum of time derivative of a conserved term,
which represents the rate of change of stored energy, and a dissipative
term, which represents dissipated power, is investigated when expressed
through the relaxation modulus and creep compliance. Further, two forms of energy and two forms of dissipated power per unit volume are examined
in order to see whether they coincide.

\end{abstract}

\section{Introduction}

The fractional anti-Zener and Zener models are derived in \cite{SD-1} by
considering the corresponding rheological schemes, having the classical
spring and dash-pot replaced by the fractional ones, modeling the fractional
spring by expressing stress $\sigma $ in terms of fractional integral of
strain $\varepsilon $, as well as by modeling the fractional dash-pot by
expressing stress in terms of Riemann-Liouville fractional derivative of
strain, so that%
\begin{equation*}
\sigma \left( t\right) =E\,_{0}\mathrm{I}_{t}^{\xi }\varepsilon \left(
t\right) \quad \text{and}\quad \sigma \left( t\right) =\eta \,_{0}\mathrm{D}%
_{t}^{\zeta }\varepsilon \left( t\right) ,\quad \xi ,\zeta \in \left(
0,1\right) ,
\end{equation*}%
with $E$ and $\eta $ denoting generalized Young modulus and coefficient of
viscosity having the fractional integral of order $\xi >0$ and
Riemann-Liouville fractional derivative of order $\zeta \in \left(
0,1\right) $ respectively defined by 
\begin{equation*}
{}_{0}\mathrm{I}_{t}^{\xi }f\left( t\right) =\frac{t^{\xi -1}}{\Gamma \left(
\xi \right) }\ast f\left( t\right) =\frac{1}{\Gamma (\xi )}\int_{0}^{t}\frac{%
f(t^{\prime })}{(t-t^{\prime })^{1-\xi }}\mathrm{d}t^{\prime }\quad \text{and%
}\quad {}_{0}\mathrm{D}_{t}^{\zeta }f\left( t\right) =\frac{\mathrm{d}}{%
\mathrm{d}t}\,{}_{0}\mathrm{I}_{t}^{1-\zeta }f\left( t\right) =\frac{\mathrm{%
d}}{\mathrm{d}t}\left( \frac{t^{-\zeta }}{\Gamma \left( 1-\zeta \right) }%
\ast f\left( t\right) \right) ,
\end{equation*}%
see \cite{TAFDE}. 
Fractional dash-pot, also known as the Scott-Blair element, is extensively
used in rheological schemes in order to formulate constitutive models of
viscoelastic body, see \cite{Mai-10,hilf,SchiesselMetzlerBlumenNonnemacher}.
In particular, the fractional dash-pot is used in classical rheological
scheme of the Burgers model in order to derive its fractional counterpart,
see \cite{OZ-1}, while the Bessel model is derived in \cite%
{ColombaroGiustiMainardi,Giusti,GiustiMainardi} considering the rheological
scheme with infinite number of springs and dash-pots.

Further, the models obtained according to the rheological schemes are tested
for thermodynamical consistency, implying the fractional anti-Zener and
Zener models listed in Appendix \ref{FAZ-ZM}, along with the restrictions on
model parameters. The energy balance properties of viscoelastic body are
examined in \cite{SD-2} by transforming the power per unit volume into the
sum of time derivative of a quantity that can be interpreted as the energy
per unit volume, originating from material's elastic properties, and a
quantity that can be interpreted as the dissipated power per unit volume,
originating from material's viscous properties, by the use of constitutive
equation written in the form%
\begin{equation}
\sigma \left( t\right) =\frac{\mathrm{d}}{\mathrm{d}t}\left( \sigma
_{sr}\left( t\right) \ast \varepsilon \left( t\right) \right) ,\quad \text{or%
}\quad \varepsilon \left( t\right) =\dot{\varepsilon}_{cr}\left( t\right)
\ast \sigma \left( t\right) ,\quad \text{with}\quad \dot{\varepsilon}%
_{cr}\left( t\right) =\frac{\mathrm{d}}{\mathrm{d}t}\varepsilon _{cr}\left(
t\right) ,  \label{sigma-epsilon}
\end{equation}%
assuming $\varepsilon _{cr}^{\left( g\right) }=\varepsilon _{cr}\left(
0\right) =0$. In the equivalent forms (\ref{sigma-epsilon}) of constitutive
model, the relaxation modulus, a function representing the time evolution of
stress in a stress relaxation experiment, i.e., the stress being a response
to the strain assumed as the Heaviside function, is denoted by $\sigma _{sr},
$ while $\varepsilon _{cr}$ denotes the creep compliance, a function
representing the time evolution of strain in a creep compliance experiment,
i.e., the strain being a response to the stress assumed as the Heaviside
function. Both energy per unit volume and dissipated power per unit volume
proved to have two forms, the first one expressed in terms of relaxation
modulus and arbitrary strain, and the second one expressed in terms of creep
compliance and arbitrary stress. Moreover, the explicit forms of relaxation
modulus and creep compliance are calculated in \cite{SD-2} and used in order
to formulate narrowed thermodynamical restrictions guaranteeing that
relaxation modulus is a completely monotonic function and creep compliance
is a Bernstein function.

The aim is to investigate time evolution of stress as a response to strain
assumed as a harmonic excitation of viscoelastic body modeled by
thermodynamically consistent fractional anti-Zener and Zener models, as well
as the transition from transient to steady state regime. Further, the
equivalence between two forms of energy and dissipated power per unit
volume, as well as the equivalence between power per unit volume written in
terms of relaxation modulus and creep compliance, is analyzed assuming
either the harmonic strain, or stress as a response to such a strain.

The restrictions on model parameters guaranteeing dissipativity in the
steady state regime are based upon the request of storage and loss modulus
non-negativity for any frequency, see \cite{b-t}. These type of requests are
used in \cite{SD-1} for the fractional anti-Zener and Zener models, while
for the fractional Burgers models the same approach is adopted in \cite{OZ-1}
and \cite{BazhlekovaTsocheva}. In \cite{AKOZ}, this method is used in order
to formulate restrictions on parameters in fractional models of viscoelastic
body having the orders not exceeding the first order, as well as to classify
these models. The dissipation inequality in time domain is considered in 
\cite{a-2002,AJP}, while \cite{ZO} examine the dissipativity of the
fractional wave equations a priori. Energy dissipation in Bessel-type
material, the thermodynamical consistency concerning the wave equation, as
well as the account on the Zener wave equation are examined in \cite%
{ColombaroGiustiMentrelli,EndeLionLammering,HolmHolm,NH}. The account on
thermodynamics of materials with memory is found in \cite%
{AmendolaFabrizioGolden}.

The property of relaxation modulus to be completely monotonic and
simultaneously the property of creep compliance to be a Bernstein function
is considered in \cite{Mai-10} in general, while in \cite%
{BazhlekovaBazhlekov} these properties are considered for distributed-order
fractional Zener model. Moreover, the request for the mentioned properties
of relaxation modulus and creep compliance implied in \cite{OZ-2} that the
thermodynamical restrictions for fractional Burgers models formulated in 
\cite{OZ-1} must be narrowed. In \cite%
{GlockleNonnenmacher1,GlockleNonnenmacher2} the response of viscoelastic
body to prescribed stress or strain is obtained in terms of the Fox
function, while in \cite{DemirciTonuk,grah} creep and stress relaxation
tests are experimentally conducted in order to model biological material.
Creep and stress relaxation tests on materials including the inertial
effects are analytically described in \cite{APZ-4,APZ-3} in the case of
distributed-order model, as well as in \cite{SD} in the case of
thermodynamically consistent fractional Burgers models.

Zener, modified Zener, and modified Maxwell models are used in \cite%
{R-S1,R-S-2001,R-S,R-S2,R-S-2008} in order to study damped oscillations and
wave propagation. Fractional Zener and Burgers wave equations are analyzed
in \cite{OparnicaBroucke,KOZ10} and \cite{OZO}, while in \cite%
{OparnicaBroucke1,KOZ19} the distributed-order wave equation is
investigated. Multidimensional fractional Zener wave equation is considered
in \cite{AitIchou,CunhaFilho,OparnicaSuli}. Fractional order constitutive
models of viscoelastic materials, wave propagation, dispersion, and
attenuation processes are reviewed in \cite%
{APSZ-1,APSZ-2,Holm-book,Mai-10,R-S-2010}, while the acoustic waves are
surveyed in \cite{Cai2018}.

\section{Transient response and steady state regime\label{TRandSSR}}

\subsection{Transient response to harmonic excitation}

Assuming that a viscoelastic body, modeled by the fractional anti-Zener and
Zener models listed in Appendix \ref{FAZ-ZM}, is subject to an excitation in
a form of strain prescribed as a harmonic function%
\begin{equation}
\varepsilon \left( t\right) =\varepsilon _{0}\cos \left( \omega t\right) ,
\label{Harmeks}
\end{equation}%
with constant amplitude $\varepsilon _{0}$ and angular frequency $\omega ,$
the stress induced in a body is sought as a response in the transient
regime. Therefore, using the Laplace transform 
\begin{equation*}
\tilde{f}(s)=\mathcal{L}[f(t)](s)=\int_{0}^{\infty }f(t)\,\mathrm{e}^{-st}%
\mathrm{d}t,\quad \text{for}\quad \mathrm{Re}\,s>0,
\end{equation*}%
the constitutive models of fractional anti-Zener and Zener type from
Appendix \ref{FAZ-ZM} are transformed into%
\begin{equation}
\tilde{\sigma}\left( s\right) =s^{\xi }\frac{\phi _{\varepsilon }\left(
s\right) }{\phi _{\sigma }\left( s\right) }\tilde{\varepsilon}\left(
s\right) ,  \label{konst-jednacina}
\end{equation}%
where $\xi $ is model dependent constitutive parameter, while $\phi
_{\varepsilon }$ and $\phi _{\sigma }$ are model dependent constitutive
functions in Laplace domain, listed in Table \ref{skupina}, so that the
stress in Laplace domain takes the form%
\begin{equation}
\tilde{\sigma}\left( s\right) =\varepsilon _{0}s^{\xi }\frac{\phi
_{\varepsilon }\left( s\right) }{\phi _{\sigma }\left( s\right) }\frac{s}{%
s^{2}+\omega ^{2}},  \label{sigma-harm-ld}
\end{equation}%
by the Laplace transform of a cosine function in harmonic excitation, given
by (\ref{Harmeks}), and the constitutive equation in Laplace domain (\ref%
{konst-jednacina}).

The stress in time domain as a response to strain as a harmonic excitation,
after performing the inverse Laplace transform in (\ref{sigma-harm-ld}), is
obtained as%
\begin{equation}
\sigma (t)=\sigma ^{\left( \mathrm{V}\right) }\left( t\right) +\sigma
^{\left( \mathrm{H}\right) }\left( t\right) +\left\{ \!\!\!%
\begin{tabular}{ll}
$0$, & if $\phi _{\sigma }$ has no zeros, \smallskip \\ 
$\sigma ^{\left( \mathrm{RP}\right) }\left( t\right) $, & if $\phi _{\sigma
} $ has a negative real zero, \smallskip \\ 
$\sigma ^{\left( \mathrm{CCP}\right) }\left( t\right) $, & if $\phi _{\sigma
}$ has a pair of complex conjugated zeros,%
\end{tabular}%
\right.  \label{resp-to-HE}
\end{equation}%
with functions appearing in (\ref{resp-to-HE}) defined by%
\begin{align}
\sigma ^{\left( \mathrm{V}\right) }\left( t\right) & =\frac{\varepsilon _{0}%
}{\pi }\int_{0}^{\infty }\frac{K\left( \rho \right) }{\left\vert \phi
_{\sigma }\left( \rho \mathrm{e}^{\mathrm{i}\pi }\right) \right\vert ^{2}}%
\frac{\rho ^{1+\xi }}{\rho ^{2}+\omega ^{2}}\mathrm{e}^{-\rho t}\mathrm{d}%
\rho ,\quad \text{or}  \label{sigma-V1} \\
\sigma ^{\left( \mathrm{V}\right) }\left( t\right) & =\frac{\varepsilon _{0}%
}{\pi }\int_{0}^{\infty }\frac{\left\vert \phi _{\varepsilon }\left( \rho 
\mathrm{e}^{\mathrm{i}\pi }\right) \right\vert }{\left\vert \phi _{\sigma
}\left( \rho \mathrm{e}^{\mathrm{i}\pi }\right) \right\vert }\sin \left(
\arg \phi _{\varepsilon }\left( \rho \mathrm{e}^{\mathrm{i}\pi }\right)
-\arg \phi _{\sigma }\left( \rho \mathrm{e}^{\mathrm{i}\pi }\right) +\xi \pi
\right) \frac{\rho ^{1+\xi }}{\rho ^{2}+\omega ^{2}}\mathrm{e}^{-\rho t}%
\mathrm{d}\rho ,  \label{sigma-V2} \\
\sigma ^{\left( \mathrm{H}\right) }\left( t\right) & =\varepsilon
_{0}\left\vert \hat{E}\left( \omega \right) \right\vert \cos \left( \omega
t+\arg \phi _{\varepsilon }\left( \mathrm{i}\omega \right) -\arg \phi
_{\sigma }\left( \mathrm{i}\omega \right) +\frac{\xi \pi }{2}\right) ,\ 
\text{with}\ \left\vert \hat{E}\left( \omega \right) \right\vert =\omega
^{\xi }\frac{\left\vert \phi _{\varepsilon }\left( \mathrm{i}\omega \right)
\right\vert }{\left\vert \phi _{\sigma }\left( \mathrm{i}\omega \right)
\right\vert },  \label{sigma-H} \\
\sigma ^{\left( \mathrm{RP}\right) }\left( t\right) & =-\varepsilon _{0}%
\frac{\left\vert \phi _{\varepsilon }\left( s_{\scriptscriptstyle\mathrm{R}{%
\mathrm{P}}}\right) \right\vert }{\left\vert \phi _{\sigma }^{\prime }\left(
s_{\scriptscriptstyle\mathrm{R}{\mathrm{P}}}\right) \right\vert }\mathrm{%
\cos }\left( \arg \phi _{\varepsilon }\left( s_{\scriptscriptstyle\mathrm{R}{%
\mathrm{P}}}\right) -\arg \phi _{\sigma }^{\prime }\left( s_{%
\scriptscriptstyle\mathrm{R}{\mathrm{P}}}\right) +\xi \pi \right) \frac{\rho
_{\scriptscriptstyle\mathrm{R}{\mathrm{P}}}^{1+\xi }}{\rho _{%
\scriptscriptstyle\mathrm{R}{\mathrm{P}}}^{2}+\omega ^{2}}\mathrm{e}^{-\rho
_{\scriptscriptstyle\mathrm{R}{\mathrm{P}}}t},  \label{sigma-rp} \\
\sigma ^{\left( \mathrm{CCP}\right) }\left( t\right) & =2\varepsilon _{0}%
\frac{\left\vert \phi _{\varepsilon }\left( s_{\scriptscriptstyle{\mathrm{CCP%
}}}\right) \right\vert }{\left\vert \phi _{\sigma }^{\prime }\left( s_{%
\scriptscriptstyle{\mathrm{CCP}}}\right) \right\vert }\frac{\rho _{%
\scriptscriptstyle{\mathrm{CCP}}}^{1+\xi }}{\left\vert s_{\scriptscriptstyle{%
\mathrm{CCP}}}^{2}+\omega ^{2}\right\vert }\mathrm{e}^{-\left\vert \func{Re}%
s_{\scriptscriptstyle{\mathrm{CCP}}}\right\vert t}  \notag \\
& \qquad \times \cos \left( \func{Im}s_{\scriptscriptstyle{\mathrm{CCP}}%
}t+\arg \phi _{\varepsilon }\left( s_{\scriptscriptstyle{\mathrm{CCP}}%
}\right) -\arg \phi _{\sigma }^{\prime }\left( s_{\scriptscriptstyle{\mathrm{%
CCP}}}\right) +\left( 1+\xi \right) \varphi _{\scriptscriptstyle{\mathrm{CCP}%
}}-\phi \left( \omega \right) \right) ,  \label{sigma-ccp}
\end{align}%
where function $K$ is given by%
\begin{align}
K\left( \rho \right) & =\frac{1}{2\mathrm{i}}\left( \mathrm{e}^{\mathrm{i}%
\xi \pi }\phi _{\varepsilon }\left( \rho \mathrm{e}^{\mathrm{i}\pi }\right) 
\bar{\phi}_{\sigma }\left( \rho \mathrm{e}^{\mathrm{i}\pi }\right) -\mathrm{e%
}^{-\mathrm{i}\xi \pi }\bar{\phi}_{\varepsilon }\left( \rho \mathrm{e}^{%
\mathrm{i}\pi }\right) \phi _{\sigma }\left( \rho \mathrm{e}^{\mathrm{i}\pi
}\right) \right)  \notag \\
& =\left\vert \phi _{\varepsilon }\left( \rho \mathrm{e}^{\mathrm{i}\pi
}\right) \right\vert \left\vert \phi _{\sigma }\left( \rho \mathrm{e}^{%
\mathrm{i}\pi }\right) \right\vert \sin \left( \arg \phi _{\varepsilon
}\left( \rho \mathrm{e}^{\mathrm{i}\pi }\right) -\arg \phi _{\sigma }\left(
\rho \mathrm{e}^{\mathrm{i}\pi }\right) +\xi \pi \right)  \notag \\
& =\cos \left( \xi \pi \right) \left( \mathrm{\func{Im}}\phi _{\varepsilon
}\left( \rho \mathrm{e}^{\mathrm{i}\pi }\right) \func{Re}\phi _{\sigma
}\left( \rho \mathrm{e}^{\mathrm{i}\pi }\right) -\func{Re}\phi _{\varepsilon
}\left( \rho \mathrm{e}^{\mathrm{i}\pi }\right) \func{Im}\phi _{\sigma
}\left( \rho \mathrm{e}^{\mathrm{i}\pi }\right) \right)  \notag \\
& \quad +\sin \left( \xi \pi \right) \left( \mathrm{\func{Re}}\phi
_{\varepsilon }\left( \rho \mathrm{e}^{\mathrm{i}\pi }\right) \func{Re}\phi
_{\sigma }\left( \rho \mathrm{e}^{\mathrm{i}\pi }\right) +\func{Im}\phi
_{\varepsilon }\left( \rho \mathrm{e}^{\mathrm{i}\pi }\right) \func{Im}\phi
_{\sigma }\left( \rho \mathrm{e}^{\mathrm{i}\pi }\right) \right) ,  \label{K}
\end{align}%
while $\phi _{\sigma }^{\prime }\left( s\right) =\frac{\mathrm{d}}{\mathrm{d}%
s}\phi _{\sigma }\left( s\right) $ and where $s_{\scriptscriptstyle\mathrm{R}%
{\mathrm{P}}}=\rho _{\scriptscriptstyle\mathrm{R}{\mathrm{P}}}\,\mathrm{e}^{%
\mathrm{i}\pi }$ is a negative real zero of function $\phi _{\sigma },$
while $s_{\scriptscriptstyle{\mathrm{CCP}}}=\rho _{\scriptscriptstyle{%
\mathrm{CCP}}}\,\mathrm{e}^{\mathrm{i}\varphi _{\scriptscriptstyle{\mathrm{%
CCP}}}}$ and its complex conjugate $\bar{s}_{\scriptscriptstyle{\mathrm{CCP}}%
}$ are complex zeros of function $\phi _{\sigma }$ having negative real
part, with%
\begin{equation}
\cot \phi \left( \omega \right) =\cot \left( 2\varphi _{\scriptscriptstyle{%
\mathrm{CCP}}}\right) +\frac{\omega ^{2}}{\rho _{\scriptscriptstyle{\mathrm{%
CCP}}}^{2}\sin \left( 2\varphi _{\scriptscriptstyle{\mathrm{CCP}}}\right) }.
\label{faza}
\end{equation}

The function $\sigma ^{\left( \mathrm{V}\right) }$ is completely monotonic,
i.e., it satisfies%
\begin{equation*}
\sigma ^{\left( \mathrm{V}\right) }\left( t\right) \geqslant 0\quad \text{and%
}\quad \left( -1\right) ^{k}\frac{\mathrm{d}^{k}}{\mathrm{d}t^{k}}\dot{\sigma%
}^{\left( \mathrm{V}\right) }\left( t\right) \leqslant 0,\quad \text{for}%
\quad t>0,\quad k\in 
\mathbb{N}
_{0},
\end{equation*}%
if $K\left( \rho \right) \geqslant 0$, see (\ref{K}), i.e., if $\sin (\arg
\phi _{\varepsilon }(\rho \mathrm{e}^{\mathrm{i}\pi })-\arg \phi _{\sigma
}(\rho \mathrm{e}^{\mathrm{i}\pi })+\xi \pi )\geqslant 0,$ while if $K\left(
\rho \right) <0$, i.e., if $\sin (\arg \phi _{\varepsilon }(\rho \mathrm{e}^{%
\mathrm{i}\pi })-\arg \phi _{\sigma }(\rho \mathrm{e}^{\mathrm{i}\pi })+\xi
\pi )<0,$ then function $\sigma ^{\left( \mathrm{V}\right) }$ can be at most
non-monotonic, however always tending to zero. Function $\sigma ^{\left( 
\mathrm{H}\right) }$ is an oscillatory function with angular frequency $%
\omega $ equal to the frequency of the prescribed harmonic deformation, see (%
\ref{Harmeks}), and it is phase-shifted, depending on the model parameters,
see (\ref{sigma-H}). Function $\sigma ^{\left( \mathrm{RP}\right) }$ is
either a positive exponentially decreasing function tending to zero, or a
negative exponentially increasing function also tending to zero, with time
constant $\rho _{\scriptscriptstyle\mathrm{R}{\mathrm{P}}}$, obtained as a
negative real zero of function $\phi _{\sigma }$, see (\ref{sigma-rp}),
while function $\sigma ^{\left( \mathrm{CCP}\right) }$ is an oscillatory
function of an exponentially decreasing amplitude, having angular frequency
defined by the imaginary part of a complex zero of function $\phi _{\sigma }$%
, i.e., by $\func{Im}s_{\scriptscriptstyle{\mathrm{CCP}}}$, and damping
parameter defined by the real part of a complex zero of function $\phi
_{\sigma }$, i.e., by $\left\vert \func{Re}s_{\scriptscriptstyle{\mathrm{CCP}%
}}\right\vert $ , see (\ref{sigma-ccp}).

\begin{landscape}
\input{tabela-svih-modela.tex} 
\end{landscape}

The stress in the form (\ref{resp-to-HE}), containing functions given by (%
\ref{sigma-V1}) - (\ref{sigma-ccp}), is obtained by the use of definition of
the inverse Laplace transform%
\begin{equation}
\sigma \left( t\right) =\mathcal{L}^{-1}\left[ \tilde{\sigma}\left( s\right) %
\right] \left( t\right) =\frac{1}{2\pi \mathrm{i}}\int_{p_{0}-\mathrm{i}%
\infty }^{p_{0}+\mathrm{i}\infty }\tilde{\sigma}\left( s\right) \mathrm{e}%
^{st}\mathrm{d}s,  \label{Lap-transf-sigma}
\end{equation}%
and the Cauchy residue theorem, having the integration performed either
along the contour $\Gamma ^{(\mathrm{I})}$, depicted in Figure \ref{nemaTG},
giving%
\begin{align}
\frac{1}{2\pi \mathrm{i}}\oint_{\Gamma ^{(\mathrm{I})}}\tilde{\sigma}\left(
s\right) \mathrm{e}^{st}\mathrm{d}s& =\func{Res}\left( \tilde{\sigma}\left(
s\right) \mathrm{e}^{st},\mathrm{i}\omega \right) +\func{Res}\left( \tilde{%
\sigma}\left( s\right) \mathrm{e}^{st},-\mathrm{i}\omega \right)  \notag \\
& \quad +\left\{ \!\!\!%
\begin{tabular}{ll}
$0$, & \negthinspace \negthinspace \negthinspace\ if $\phi _{\sigma }$ has
no zeros, \smallskip \\ 
$\func{Res}\left( \tilde{\sigma}\left( s\right) \mathrm{e}^{st},s_{%
\scriptscriptstyle{\mathrm{CCP}}}\right) +\func{Res}\left( \tilde{\sigma}%
\left( s\right) \mathrm{e}^{st},\bar{s}_{\scriptscriptstyle{\mathrm{CCP}}%
}\right) $, & \negthinspace \negthinspace \negthinspace\ if $\phi _{\sigma }$
has $s_{\scriptscriptstyle{\mathrm{CCP}}}$ and $\bar{s}_{\scriptscriptstyle{%
\mathrm{CCP}}}$ as zeros,%
\end{tabular}%
\right.  \label{KRT-HE}
\end{align}%
or along the contour $\Gamma ^{(\mathrm{II})}$, depicted in Figure \ref%
{negativnaTG}, giving%
\begin{equation}
\frac{1}{2\pi \mathrm{i}}\oint_{\Gamma ^{(\mathrm{II})}}\tilde{\sigma}\left(
s\right) \mathrm{e}^{st}\mathrm{d}s=\func{Res}\left( \tilde{\sigma}\left(
s\right) \mathrm{e}^{st},\mathrm{i}\omega \right) +\func{Res}\left( \tilde{%
\sigma}\left( s\right) \mathrm{e}^{st},-\mathrm{i}\omega \right) ,
\label{CRT-HE}
\end{equation}%
if $\phi _{\sigma }$ has a negative real zero $s_{\scriptscriptstyle{\mathrm{%
RP}}}=-\rho _{\scriptscriptstyle{\mathrm{RP}}},$ since besides the
possibility that function $\tilde{\sigma}$, see (\ref{sigma-harm-ld}), has a
negative real pole\ $s_{\scriptscriptstyle\mathrm{R}{\mathrm{P}}}$ or a pair
of complex conjugated poles $s_{\scriptscriptstyle{\mathrm{CCP}}}$ and $\bar{%
s}_{\scriptscriptstyle{\mathrm{CCP}}}$, representing the zeros of function $%
\phi _{\sigma }$, the function $\tilde{\sigma}$ additionally has two poles,
namely $\pm \mathrm{i}\omega $, that originate from cosine function in
Laplace domain.

\noindent 
\begin{minipage}{\columnwidth}
\begin{minipage}[c]{0.4\columnwidth}
\centering
\includegraphics[width=0.7\columnwidth]{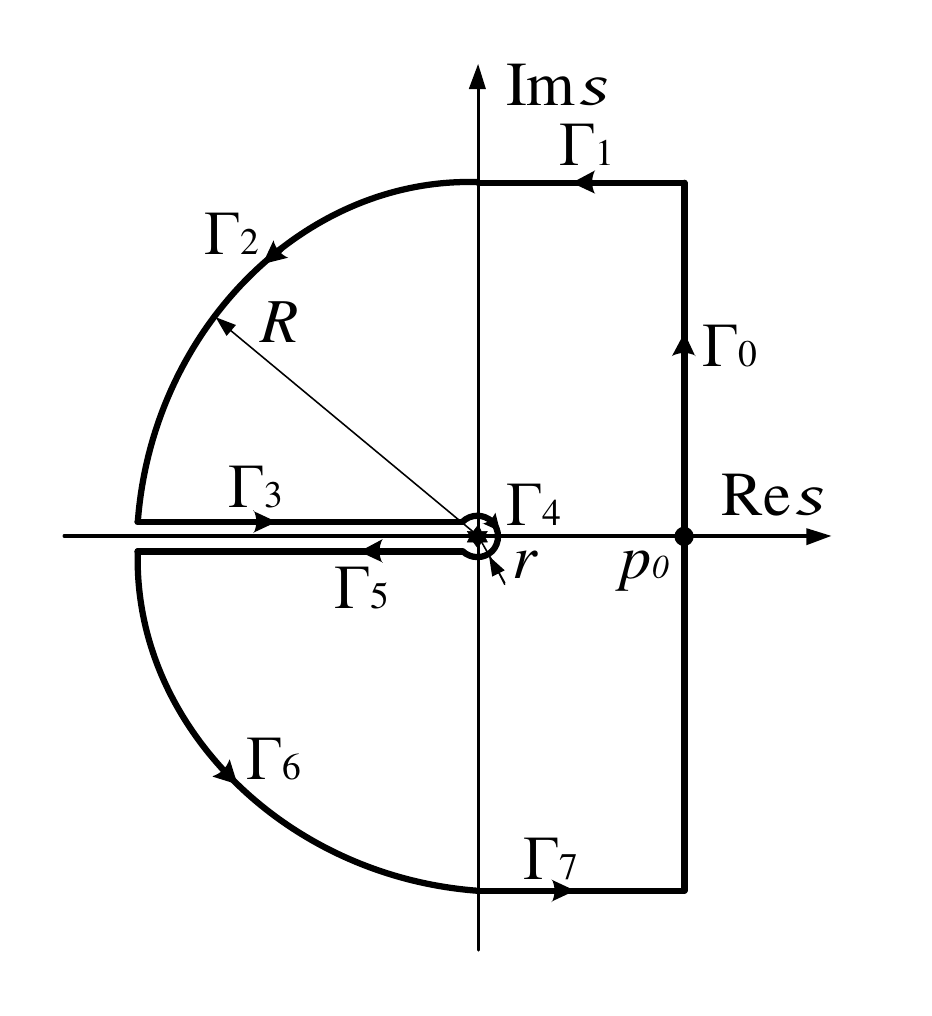}
\captionof{figure}{Integration contour $\Gamma^{(\mathrm{I})}$.}
\label{nemaTG}
\end{minipage}
\hfil
\begin{minipage}[c]{0.55\columnwidth}
\centering
\begin{tabular}{rll}
$\Gamma _{0}:$ & Bromwich path, &  \\ 
$\Gamma _{1}:$ & $s=p+\mathrm{i}R,$ & $p\in \left[ 0,p_{0}\right],\, p_0\geq 0$ arbitrary, \\ 
$\Gamma _{2}:$ & $s=R\mathrm{e}^{\mathrm{i}\varphi },$ & $\varphi \in \left[ 
\frac{\pi }{2},\pi \right] ,$ \\ 
$\Gamma _{3}:$ & $s=\rho \mathrm{e}^{\mathrm{i}\pi },$ & $\rho \in \left[ r,R%
\right] ,$ \\ 
$\Gamma _{4}:$ & $s=r\mathrm{e}^{\mathrm{i}\varphi },$ & $\varphi \in \left[ -\pi
,\pi \right] ,$ \\ 
$\Gamma _{5}:$ & $s=\rho \mathrm{e}^{-\mathrm{i}\pi },$ & $\rho \in \left[ r,R%
\right] ,$ \\
$\Gamma _{6}:$  & $s=R\mathrm{e}^{\mathrm{i}\varphi },$ & $\varphi \in \left[ 
-\pi, -\frac{\pi }{2} \right] ,$ \\
$\Gamma _{7}:$ & $s=p-\mathrm{i}R,$ & $p\in \left[ 0,p_{0}\right],\, p_0\geq 0$ arbitrary.  
\end{tabular}
\captionof{table}{Parametrization of integration contour $\Gamma^{(\mathrm{I})}$.}
\label{nemaTG-param}
\end{minipage}
\end{minipage}\smallskip

\noindent 
\begin{minipage}{\columnwidth}
\begin{minipage}[c]{0.4\columnwidth}
\centering
\includegraphics[width=0.7\columnwidth]{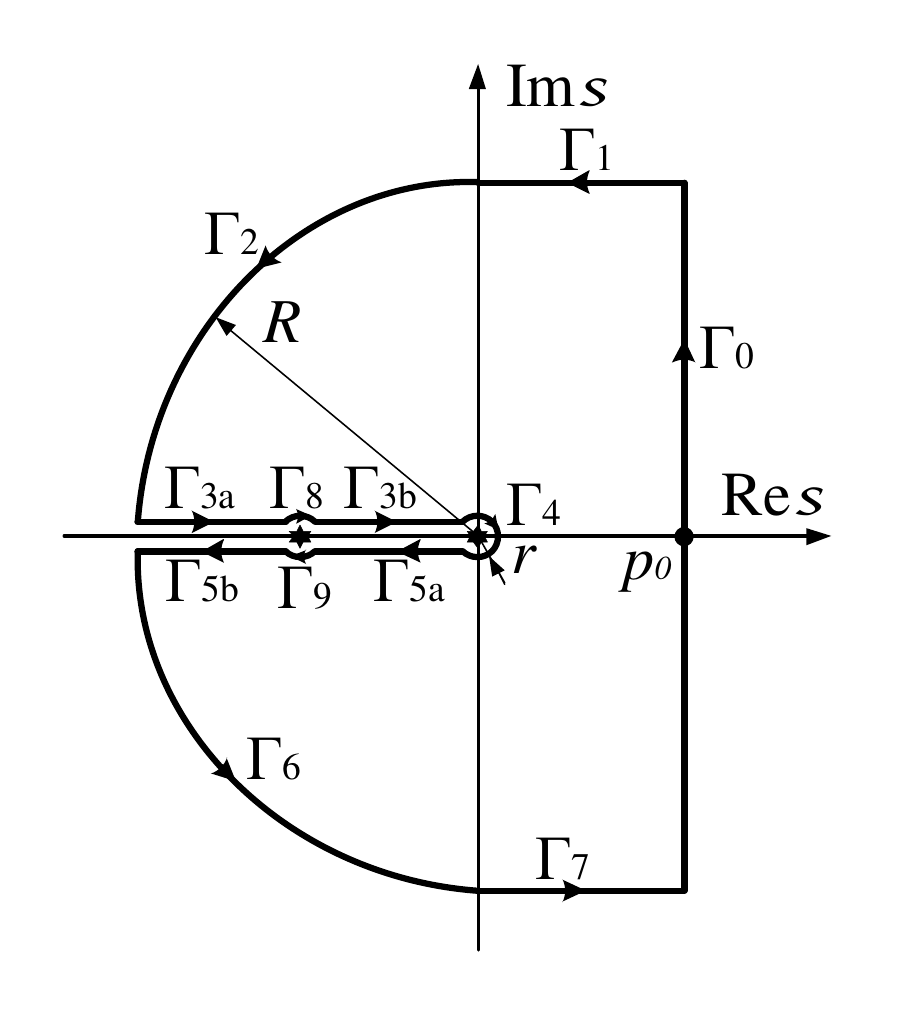}
\captionof{figure}{Integration contour $\Gamma^{(\mathrm{II})}$.}
\label{negativnaTG}
\end{minipage}
\hfil
\begin{minipage}[c]{0.55\columnwidth}
\centering
\begin{tabular}{rll}
$\Gamma _{0}:$ & Bromwich path, &  \\ 
$\Gamma _{1}:$ & $s=p+\mathrm{i}R,$ & $p\in \left[ 0,p_{0}\right],\, p_0\geq 0$ arbitrary, \\ 
$\Gamma _{2}:$ & $s=R\mathrm{e}^{\mathrm{i}\varphi },$ & $\varphi \in \left[ 
\frac{\pi }{2},\pi \right] ,$ \\ 
$\Gamma _{3a}\cup\Gamma _{3b}:$ & $s=\rho \mathrm{e}^{\mathrm{i}\pi },$ & $\rho \in \left[ r,R%
\right] ,$ \\ 
$\Gamma _{4}:$ & $s=r\mathrm{e}^{\mathrm{i}\varphi },$ & $\varphi \in \left[ -\pi
,\pi \right] ,$ \\ 
$\Gamma _{5a}\cup\Gamma _{5b}:$ & $s=\rho \mathrm{e}^{-\mathrm{i}\pi },$ & $\rho \in \left[ r,R%
\right] ,$ \\
$\Gamma _{6}:$  & $s=R\mathrm{e}^{\mathrm{i}\varphi },$ & $\varphi \in \left[ 
-\pi, -\frac{\pi }{2} \right] ,$ \\
$\Gamma _{7}:$ & $s=p-\mathrm{i}R,$ & $p\in \left[ 0,p_{0}\right],\, p_0\geq 0$ arbitrary,\\
$\Gamma _{8}:$  & $s=-\rho^*+r\mathrm{e}^{\mathrm{i}\varphi },$ & $\varphi \in \left[0,
\pi\right] ,$ \\
$\Gamma _{9}:$  & $s=-\rho^*+r\mathrm{e}^{\mathrm{i}\varphi },$ & $\varphi \in \left[ -\pi,
0 \right]$.  
\end{tabular}
\captionof{table}{Parametrization of integration contour $\Gamma^{(\mathrm{II})}$.}
\label{negativnaTG-param}
\end{minipage}
\end{minipage}\smallskip

The Cauchy residue theorems (\ref{KRT-HE}) and (\ref{CRT-HE}), with the
integrals having non-zero contributions in the limit when $r\rightarrow 0$
and $R\rightarrow \infty $, take the respective forms%
\begin{align}
\frac{1}{2\pi \mathrm{i}}\int_{\Gamma _{0}}\tilde{\sigma}\left( s\right) 
\mathrm{e}^{st}\mathrm{d}s+\frac{1}{2\pi \mathrm{i}}& \int_{\Gamma _{3}}%
\tilde{\sigma}\left( s\right) \mathrm{e}^{st}\mathrm{d}s+\frac{1}{2\pi 
\mathrm{i}}\int_{\Gamma _{5}}\tilde{\sigma}\left( s\right) \mathrm{e}^{st}%
\mathrm{d}s  \notag \\
& =\sigma ^{\left( \mathrm{H}\right) }\left( t\right) +\left\{ \!\!\!%
\begin{tabular}{ll}
$0$, & \negthinspace \negthinspace \negthinspace\ if $\phi _{\sigma }$ has
no zeros, \smallskip \\ 
$\sigma ^{\left( {\mathrm{CCP}}\right) }\left( t\right) $, & \negthinspace
\negthinspace \negthinspace\ if $\phi _{\sigma }$ has $s_{\scriptscriptstyle{%
\mathrm{CCP}}}$ and $\bar{s}_{\scriptscriptstyle{\mathrm{CCP}}}$ as zeros,%
\end{tabular}%
\right.  \label{integrali-i-reziduumi}
\end{align}%
and%
\begin{align}
\frac{1}{2\pi \mathrm{i}}\int_{\Gamma _{0}}\tilde{\sigma}\left( s\right) 
\mathrm{e}^{st}\mathrm{d}s& +\frac{1}{2\pi \mathrm{i}}\int_{\Gamma _{3a}\cup
\Gamma _{3b}}\tilde{\sigma}\left( s\right) \mathrm{e}^{st}\mathrm{d}s+\frac{1%
}{2\pi \mathrm{i}}\int_{\Gamma _{5a}\cup \Gamma _{5b}}\tilde{\sigma}\left(
s\right) \mathrm{e}^{st}\mathrm{d}s  \notag \\
& +\frac{1}{2\pi \mathrm{i}}\int_{\Gamma _{8}}\tilde{\sigma}\left( s\right) 
\mathrm{e}^{st}\mathrm{d}s+\frac{1}{2\pi \mathrm{i}}\int_{\Gamma _{9}}\tilde{%
\sigma}\left( s\right) \mathrm{e}^{st}\mathrm{d}s=\sigma ^{\left( \mathrm{H}%
\right) }\left( t\right) ,  \label{int-i-rez}
\end{align}%
where functions $\sigma ^{\left( \mathrm{H}\right) }$ and $\sigma ^{\left( {%
\mathrm{CCP}}\right) }$ originate form the residues in expressions (\ref%
{integrali-i-reziduumi}) and (\ref{int-i-rez}), so that%
\begin{align}
\sigma ^{\left( \mathrm{H}\right) }\left( t\right) &=\func{Res}\left( \tilde{%
\sigma}\left( s\right) \mathrm{e}^{st},\omega \mathrm{e}^{\mathrm{i}\frac{%
\pi }{2}}\right) +\func{Res}\left( \tilde{\sigma}\left( s\right) \mathrm{e}%
^{st},\omega \mathrm{e}^{-\mathrm{i}\frac{\pi }{2}}\right) \quad \text{and}
\label{sigma-ha} \\
\sigma ^{\left( {\mathrm{CCP}}\right) }\left( t\right) &=\func{Res}\left( 
\tilde{\sigma}\left( s\right) \mathrm{e}^{st},s_{\scriptscriptstyle{\mathrm{%
CCP}}}\right) +\func{Res}\left( \tilde{\sigma}\left( s\right) \mathrm{e}%
^{st},\bar{s}_{\scriptscriptstyle{\mathrm{CCP}}}\right) .
\label{sigma-ce-ce-pe}
\end{align}
One transforms the Cauchy residue theorems (\ref{integrali-i-reziduumi}) and
(\ref{int-i-rez}) into%
\begin{equation*}
\sigma \left( t\right) -\sigma ^{\left( \mathrm{V}\right) }\left( t\right)
=\sigma ^{\left( \mathrm{H}\right) }\left( t\right) +\left\{ \!\!\!%
\begin{tabular}{ll}
$0$, & \negthinspace \negthinspace \negthinspace\ if $\phi _{\sigma }$ has
no zeros, \smallskip \\ 
$\sigma ^{\left( {\mathrm{CCP}}\right) }\left( t\right) $, & \negthinspace
\negthinspace \negthinspace\ if $\phi _{\sigma }$ has $s_{\scriptscriptstyle{%
\mathrm{CCP}}}$ and $\bar{s}_{\scriptscriptstyle{\mathrm{CCP}}}$ as zeros,%
\end{tabular}%
\right.
\end{equation*}%
and, if $\phi _{\sigma }$ has a negative real zero, into%
\begin{equation*}
\sigma \left( t\right) -\sigma ^{\left( \mathrm{V}\right) }\left( t\right)
-\sigma ^{\left( \mathrm{RP}\right) }\left( t\right) =\sigma ^{\left( 
\mathrm{H}\right) }\left( t\right) ,
\end{equation*}%
in the limit when $r\rightarrow 0$ and $R\rightarrow \infty ,$ with 
\begin{equation*}
\sigma \left( t\right) =\frac{1}{2\pi \mathrm{i}}\lim_{R\rightarrow \infty
}\int_{\Gamma _{0}}\tilde{\sigma}\left( s\right) \mathrm{e}^{st}\mathrm{d}s,
\end{equation*}%
due to the inverse Laplace transform (\ref{Lap-transf-sigma}), as well as
with the functions $\sigma ^{\left( \mathrm{V}\right) }$ and $\sigma
^{\left( \mathrm{RP}\right) },$ defined by%
\begin{align}
\sigma ^{\left( \mathrm{V}\right) }\left( t\right) &=-\lim_{\substack{ %
r\rightarrow 0  \\ R\rightarrow \infty }}\left( \frac{1}{2\pi \mathrm{i}}%
\int_{\Gamma _{3},\Gamma _{3a}\cup \Gamma _{3b}}\tilde{\sigma}\left(
s\right) \mathrm{e}^{st}\mathrm{d}s+\frac{1}{2\pi \mathrm{i}}\int_{\Gamma
_{5},\Gamma _{5a}\cup \Gamma _{5b}}\tilde{\sigma}\left( s\right) \mathrm{e}%
^{st}\mathrm{d}s\right) \quad \text{and}  \label{sigma-ve} \\
\sigma ^{\left( \mathrm{RP}\right) }\left( t\right) &=-\lim_{r\rightarrow
0}\left( \frac{1}{2\pi \mathrm{i}}\int_{\Gamma _{8}}\tilde{\sigma}\left(
s\right) \mathrm{e}^{st}\mathrm{d}s+\frac{1}{2\pi \mathrm{i}}\int_{\Gamma
_{9}}\tilde{\sigma}\left( s\right) \mathrm{e}^{st}\mathrm{d}s\right) .
\label{sigma-er-pe}
\end{align}

Function $\sigma ^{\left( \mathrm{H}\right) }$, defined by (\ref{sigma-ha}),
is calculated as 
\begin{align*}
\sigma ^{\left( \mathrm{H}\right) }\left( t\right) & =\func{Res}\left( 
\tilde{\sigma}\left( s\right) \mathrm{e}^{st},\omega \mathrm{e}^{\mathrm{i}%
\frac{\pi }{2}}\right) +\func{Res}\left( \tilde{\sigma}\left( s\right) 
\mathrm{e}^{st},\omega \mathrm{e}^{-\mathrm{i}\frac{\pi }{2}}\right)  \\
& =\varepsilon _{0}\left( \omega \mathrm{e}^{\mathrm{i}\frac{\pi }{2}%
}\right) ^{\xi }\frac{\phi _{\varepsilon }\left( \omega \mathrm{e}^{\mathrm{i%
}\frac{\pi }{2}}\right) }{\phi _{\sigma }\left( \omega \mathrm{e}^{\mathrm{i}%
\frac{\pi }{2}}\right) }\left. \frac{s}{\frac{\mathrm{d}}{\mathrm{d}s}\left(
s^{2}+\omega ^{2}\right) }\right\vert _{s=\omega \mathrm{e}^{\mathrm{i}\frac{%
\pi }{2}}}\mathrm{e}^{\omega \mathrm{e}^{\mathrm{i}\frac{\pi }{2}}t} \\
& \qquad +\varepsilon _{0}\left( \omega \mathrm{e}^{-\mathrm{i}\frac{\pi }{2}%
}\right) ^{\xi }\frac{\phi _{\varepsilon }\left( \omega \mathrm{e}^{-\mathrm{%
i}\frac{\pi }{2}}\right) }{\phi _{\sigma }\left( \omega \mathrm{e}^{-\mathrm{%
i}\frac{\pi }{2}}\right) }\left. \frac{s}{\frac{\mathrm{d}}{\mathrm{d}s}%
\left( s^{2}+\omega ^{2}\right) }\right\vert _{s=\omega \mathrm{e}^{-\mathrm{%
i}\frac{\pi }{2}}}\mathrm{e}^{\omega \mathrm{e}^{-\mathrm{i}\frac{\pi }{2}}t}
\\
& =\frac{1}{2}\varepsilon _{0}\omega ^{\xi }\frac{\phi _{\varepsilon }\left(
\omega \mathrm{e}^{\mathrm{i}\frac{\pi }{2}}\right) }{\phi _{\sigma }\left(
\omega \mathrm{e}^{\mathrm{i}\frac{\pi }{2}}\right) }\mathrm{e}^{\mathrm{i}%
\frac{\xi \pi }{2}}\mathrm{e}^{\mathrm{i}\omega t}+\frac{1}{2}\varepsilon
_{0}\omega ^{\xi }\frac{\phi _{\varepsilon }\left( \omega \mathrm{e}^{-%
\mathrm{i}\frac{\pi }{2}}\right) }{\phi _{\sigma }\left( \omega \mathrm{e}^{-%
\mathrm{i}\frac{\pi }{2}}\right) }\mathrm{e}^{-\mathrm{i}\frac{\xi \pi }{2}}%
\mathrm{e}^{-\mathrm{i}\omega t} \\
& =\frac{1}{2}\varepsilon _{0}\omega ^{\xi }\frac{\left\vert \phi
_{\varepsilon }\left( \omega \mathrm{e}^{\mathrm{i}\frac{\pi }{2}}\right)
\right\vert }{\left\vert \phi _{\sigma }\left( \omega \mathrm{e}^{\mathrm{i}%
\frac{\pi }{2}}\right) \right\vert }\left( \frac{\mathrm{e}^{\mathrm{i}\arg
\phi _{\varepsilon }\left( \omega \mathrm{e}^{\mathrm{i}\frac{\pi }{2}%
}\right) }}{\mathrm{e}^{\mathrm{i}\arg \phi _{\sigma }\left( \omega \mathrm{e%
}^{\mathrm{i}\frac{\pi }{2}}\right) }}\mathrm{e}^{\mathrm{i}\frac{\xi \pi }{2%
}}\mathrm{e}^{\mathrm{i}\omega t}+\mathrm{e}^{-\mathrm{i}\frac{\xi \pi }{2}}%
\frac{\mathrm{e}^{\mathrm{i}\arg \phi _{\varepsilon }\left( \omega \mathrm{e}%
^{-\mathrm{i}\frac{\pi }{2}}\right) }}{\mathrm{e}^{\mathrm{i}\arg \phi
_{\sigma }\left( \omega \mathrm{e}^{\mathrm{-}\frac{\pi }{2}}\right) }}%
\mathrm{e}^{-\mathrm{i}\omega t}\right)  \\
& =\varepsilon _{0}\omega ^{\xi }\frac{\left\vert \phi _{\varepsilon }\left(
\omega \mathrm{e}^{\mathrm{i}\frac{\pi }{2}}\right) \right\vert }{\left\vert
\phi _{\sigma }\left( \omega \mathrm{e}^{\mathrm{i}\frac{\pi }{2}}\right)
\right\vert }\cos \left( \omega t+\arg \phi _{\varepsilon }\left( \omega 
\mathrm{e}^{\mathrm{i}\frac{\pi }{2}}\right) -\arg \phi _{\sigma }\left(
\omega \mathrm{e}^{\mathrm{i}\frac{\pi }{2}}\right) +\frac{\xi \pi }{2}%
\right) ,
\end{align*}%
see also (\ref{sigma-H}), since $\pm \mathrm{i}\omega $ are poles of
function $\tilde{\sigma}$, given by (\ref{sigma-harm-ld}), while function $%
\sigma ^{\left( \mathrm{CCP}\right) }$, defined by (\ref{sigma-ce-ce-pe}),
is calculated as%
\begin{align}
\sigma ^{\left( \mathrm{CCP}\right) }\left( t\right) & =\func{Res}\left( 
\tilde{\sigma}\left( s\right) \mathrm{e}^{st},s_{\scriptscriptstyle{\mathrm{%
CCP}}}\right) +\func{Res}\left( \tilde{\sigma}\left( s\right) \mathrm{e}%
^{st},\bar{s}_{\scriptscriptstyle{\mathrm{CCP}}}\right)   \notag \\
& =\varepsilon _{0}s_{\scriptscriptstyle{\mathrm{CCP}}}^{\xi }\frac{\phi
_{\varepsilon }\left( s_{\scriptscriptstyle{\mathrm{CCP}}}\right) }{\phi
_{\sigma }^{\prime }\left( s_{\scriptscriptstyle{\mathrm{CCP}}}\right) }%
\frac{s_{\scriptscriptstyle{\mathrm{CCP}}}}{s_{\scriptscriptstyle{\mathrm{CCP%
}}}^{2}+\omega ^{2}}\mathrm{e}^{s_{\scriptscriptstyle{\mathrm{CCP}}%
}t}+\varepsilon _{0}\bar{s}_{\scriptscriptstyle{\mathrm{CCP}}}^{\xi }\frac{%
\phi _{\varepsilon }\left( \bar{s}_{\scriptscriptstyle{\mathrm{CCP}}}\right) 
}{\phi _{\sigma }^{\prime }\left( \bar{s}_{\scriptscriptstyle{\mathrm{CCP}}%
}\right) }\frac{\bar{s}_{\scriptscriptstyle{\mathrm{CCP}}}}{\bar{s}_{%
\scriptscriptstyle{\mathrm{CCP}}}^{2}+\omega ^{2}}\mathrm{e}^{\bar{s}_{%
\scriptscriptstyle{\mathrm{CCP}}}t}  \notag \\
& =\varepsilon _{0}\frac{\left\vert \phi _{\varepsilon }\left( s_{%
\scriptscriptstyle{\mathrm{CCP}}}\right) \right\vert }{\left\vert \phi
_{\sigma }^{\prime }\left( s_{\scriptscriptstyle{\mathrm{CCP}}}\right)
\right\vert }\frac{\rho _{\scriptscriptstyle{\mathrm{CCP}}}^{1+\xi }}{%
\left\vert s_{\scriptscriptstyle{\mathrm{CCP}}}^{2}+\omega ^{2}\right\vert
^{2}}\mathrm{e}^{\func{Re}s_{\scriptscriptstyle{\mathrm{CCP}}}t}  \notag \\
& \qquad \times \left( \frac{\mathrm{e}^{\mathrm{i}\arg \phi _{\varepsilon
}\left( s_{\scriptscriptstyle{\mathrm{CCP}}}\right) }}{\mathrm{e}^{\mathrm{i}%
\arg \phi _{\sigma }^{\prime }\left( s_{\scriptscriptstyle{\mathrm{CCP}}%
}\right) }}\left( \rho _{\scriptscriptstyle{\mathrm{CCP}}}^{2}\mathrm{e}^{-2%
\mathrm{i}\varphi _{\scriptscriptstyle{\mathrm{CCP}}}}+\omega ^{2}\right) 
\mathrm{e}^{\mathrm{i}\left( 1+\xi \right) \varphi _{\scriptscriptstyle{%
\mathrm{CCP}}}}\mathrm{e}^{\mathrm{i}\func{Im}s_{\scriptscriptstyle{\mathrm{%
CCP}}}t}\right.   \notag \\
& \qquad \qquad \left. +\frac{\mathrm{e}^{-\mathrm{i}\arg \phi _{\varepsilon
}\left( s_{\scriptscriptstyle{\mathrm{CCP}}}\right) }}{\mathrm{e}^{-\mathrm{i%
}\arg \phi _{\sigma }^{\prime }\left( s_{\scriptscriptstyle{\mathrm{CCP}}%
}\right) }}\left( \rho _{\scriptscriptstyle{\mathrm{CCP}}}^{2}\mathrm{e}^{2%
\mathrm{i}\varphi _{\scriptscriptstyle{\mathrm{CCP}}}}+\omega ^{2}\right) 
\mathrm{e}^{-\mathrm{i}\left( 1+\xi \right) \varphi _{\scriptscriptstyle{%
\mathrm{CCP}}}}\mathrm{e}^{-\mathrm{i}\func{Im}s_{\scriptscriptstyle{\mathrm{%
CCP}}}t}\right)   \notag \\
& =2\varepsilon _{0}\frac{\left\vert \phi _{\varepsilon }\left( s_{%
\scriptscriptstyle{\mathrm{CCP}}}\right) \right\vert }{\left\vert \phi
_{\sigma }^{\prime }\left( s_{\scriptscriptstyle{\mathrm{CCP}}}\right)
\right\vert }\frac{\rho _{\scriptscriptstyle{\mathrm{CCP}}}^{1+\xi }}{%
\left\vert s_{\scriptscriptstyle{\mathrm{CCP}}}^{2}+\omega ^{2}\right\vert
^{2}}\mathrm{e}^{\func{Re}s_{\scriptscriptstyle{\mathrm{CCP}}}t}  \notag \\
& \qquad \times \Big(\rho _{\scriptscriptstyle{\mathrm{CCP}}}^{2}\cos \left( 
\func{Im}s_{\scriptscriptstyle{\mathrm{CCP}}}t+\arg \phi _{\varepsilon
}\left( s_{\scriptscriptstyle{\mathrm{CCP}}}\right) -\arg \phi _{\sigma
}^{\prime }\left( s_{\scriptscriptstyle{\mathrm{CCP}}}\right) +\left( 1+\xi
\right) \varphi _{\scriptscriptstyle{\mathrm{CCP}}}-2\varphi _{%
\scriptscriptstyle{\mathrm{CCP}}}\right)   \notag \\
& \qquad \qquad +\omega ^{2}\cos \left( \func{Im}s_{\scriptscriptstyle{%
\mathrm{CCP}}}t+\arg \phi _{\varepsilon }\left( s_{\scriptscriptstyle{%
\mathrm{CCP}}}\right) -\arg \phi _{\sigma }^{\prime }\left( s_{%
\scriptscriptstyle{\mathrm{CCP}}}\right) +\left( 1+\xi \right) \varphi _{%
\scriptscriptstyle{\mathrm{CCP}}}\right) \Big)  \notag \\
& =2\varepsilon _{0}\frac{\left\vert \phi _{\varepsilon }\left( s_{%
\scriptscriptstyle{\mathrm{CCP}}}\right) \right\vert }{\left\vert \phi
_{\sigma }^{\prime }\left( s_{\scriptscriptstyle{\mathrm{CCP}}}\right)
\right\vert }\frac{\rho _{\scriptscriptstyle{\mathrm{CCP}}}^{1+\xi }}{%
\left\vert s_{\scriptscriptstyle{\mathrm{CCP}}}^{2}+\omega ^{2}\right\vert
^{2}}\mathrm{e}^{\func{Re}s_{\scriptscriptstyle{\mathrm{CCP}}}t}  \notag \\
& \qquad \times \Big(\left( \rho _{\scriptscriptstyle{\mathrm{CCP}}}^{2}\cos
\left( 2\varphi _{\scriptscriptstyle{\mathrm{CCP}}}\right) +\omega
^{2}\right) \cos \left( \func{Im}s_{\scriptscriptstyle{\mathrm{CCP}}}t+\arg
\phi _{\varepsilon }\left( s_{\scriptscriptstyle{\mathrm{CCP}}}\right) -\arg
\phi _{\sigma }^{\prime }\left( s_{\scriptscriptstyle{\mathrm{CCP}}}\right)
+\left( 1+\xi \right) \varphi _{\scriptscriptstyle{\mathrm{CCP}}}\right)  
\notag \\
& \qquad \qquad +\rho _{\scriptscriptstyle{\mathrm{CCP}}}^{2}\sin \left(
2\varphi _{\scriptscriptstyle{\mathrm{CCP}}}\right) \sin \left( \func{Im}s_{%
\scriptscriptstyle{\mathrm{CCP}}}t+\arg \phi _{\varepsilon }\left( s_{%
\scriptscriptstyle{\mathrm{CCP}}}\right) -\arg \phi _{\sigma }^{\prime
}\left( s_{\scriptscriptstyle{\mathrm{CCP}}}\right) +\left( 1+\xi \right)
\varphi _{\scriptscriptstyle{\mathrm{CCP}}}\right) \Big),
\label{sigma-ccp-1}
\end{align}%
since $s_{\scriptscriptstyle{\mathrm{CCP}}}=\rho _{\scriptscriptstyle{%
\mathrm{CCP}}}\mathrm{e}^{\mathrm{i}\varphi _{\scriptscriptstyle{\mathrm{CCP}%
}}}$ and $\bar{s}_{\scriptscriptstyle{\mathrm{CCP}}}=\rho _{%
\scriptscriptstyle{\mathrm{CCP}}}\mathrm{e}^{-\mathrm{i}\varphi _{%
\scriptscriptstyle{\mathrm{CCP}}}}\ $are poles of function $\tilde{\sigma}$,
given by (\ref{sigma-harm-ld}), originating from the zeros of function $\phi
_{\sigma },$ so that%
\begin{align*}
\sigma ^{\left( \mathrm{CCP}\right) }\left( t\right) & =2\varepsilon _{0}%
\frac{\left\vert \phi _{\varepsilon }\left( s_{\scriptscriptstyle{\mathrm{CCP%
}}}\right) \right\vert }{\left\vert \phi _{\sigma }^{\prime }\left( s_{%
\scriptscriptstyle{\mathrm{CCP}}}\right) \right\vert }\frac{\rho _{%
\scriptscriptstyle{\mathrm{CCP}}}^{1+\xi }}{\left\vert s_{\scriptscriptstyle{%
\mathrm{CCP}}}^{2}+\omega ^{2}\right\vert }\mathrm{e}^{-\left\vert \func{Re}%
s_{\scriptscriptstyle{\mathrm{CCP}}}\right\vert t} \\
& \qquad \times \cos \left( \func{Im}s_{\scriptscriptstyle{\mathrm{CCP}}%
}t+\arg \phi _{\varepsilon }\left( s_{\scriptscriptstyle{\mathrm{CCP}}%
}\right) -\arg \phi _{\sigma }^{\prime }\left( s_{\scriptscriptstyle{\mathrm{%
CCP}}}\right) +\left( 1+\xi \right) \varphi _{\scriptscriptstyle{\mathrm{CCP}%
}}-\phi \left( \omega \right) \right) ,
\end{align*}%
see (\ref{sigma-ccp}), is obtained using the substitution%
\begin{align*}
A\left( \omega \right) \cos \phi \left( \omega \right) & =\rho _{%
\scriptscriptstyle{\mathrm{CCP}}}^{2}\cos \left( 2\varphi _{%
\scriptscriptstyle{\mathrm{CCP}}}\right) +\omega ^{2}, \\
A\left( \omega \right) \sin \phi \left( \omega \right) & =\rho _{%
\scriptscriptstyle{\mathrm{CCP}}}^{2}\sin \left( 2\varphi _{%
\scriptscriptstyle{\mathrm{CCP}}}\right) ,
\end{align*}%
i.e.,%
\begin{gather*}
A\left( \omega \right) =\left\vert s_{\scriptscriptstyle{\mathrm{CCP}}%
}^{2}+\omega ^{2}\right\vert =\sqrt{\rho _{\scriptscriptstyle{\mathrm{CCP}}%
}^{4}+2\rho _{\scriptscriptstyle{\mathrm{CCP}}}^{2}\omega ^{2}\cos \left(
2\varphi _{\scriptscriptstyle{\mathrm{CCP}}}\right) +\omega ^{4}}, \\
\cot \phi \left( \omega \right) =\cot \left( 2\varphi _{\scriptscriptstyle{%
\mathrm{CCP}}}\right) +\frac{\omega ^{2}}{\rho _{\scriptscriptstyle{\mathrm{%
CCP}}}^{2}\sin \left( 2\varphi _{\scriptscriptstyle{\mathrm{CCP}}}\right) }
\end{gather*}%
in the expression (\ref{sigma-ccp-1}).

The function $\sigma ^{\left( \mathrm{V}\right) },$ defined by (\ref%
{sigma-ve}) and representing a contribution from integrals along contours $%
\Gamma _{3}$ and $\Gamma _{5}$ in (\ref{integrali-i-reziduumi}), as well as
from integrals along contours $\Gamma _{3a}\cup \Gamma _{3b}$ and $\Gamma
_{5a}\cup \Gamma _{5b}\ $in (\ref{int-i-rez}), respectively belonging to
contours $\Gamma ^{(\mathrm{I})}$ and $\Gamma ^{(\mathrm{II})},$ depicted in
Figures \ref{nemaTG} and \ref{negativnaTG}, in the limit when $r\rightarrow
0 $ and $R\rightarrow \infty ,$ is calculated as%
\begin{align*}
\sigma ^{\left( \mathrm{V}\right) }\left( t\right) & =-\lim_{\substack{ %
r\rightarrow 0  \\ R\rightarrow \infty }}\left( \frac{1}{2\pi \mathrm{i}}%
\int_{\Gamma _{3},\Gamma _{3a}\cup \Gamma _{3b}}\tilde{\sigma}\left(
s\right) \mathrm{e}^{st}\mathrm{d}s+\frac{1}{2\pi \mathrm{i}}\int_{\Gamma
_{5},\Gamma _{5a}\cup \Gamma _{5b}}\tilde{\sigma}\left( s\right) \mathrm{e}%
^{st}\mathrm{d}s\right) \\
& =-\frac{\varepsilon _{0}}{2\pi \mathrm{i}}\left( \int_{\infty
}^{0}\!\!\rho ^{\xi }\mathrm{e}^{\mathrm{i}\xi \pi }\frac{\phi _{\varepsilon
}\left( \rho \mathrm{e}^{\mathrm{i}\pi }\right) }{\phi _{\sigma }\left( \rho 
\mathrm{e}^{\mathrm{i}\pi }\right) }\frac{\rho \mathrm{e}^{\mathrm{i}\pi }}{%
\rho ^{2}\mathrm{e}^{2\mathrm{i}\pi }+\omega ^{2}}\mathrm{e}^{\rho t\mathrm{e%
}^{\mathrm{i}\pi }}\mathrm{e}^{\mathrm{i}\pi }\mathrm{d}\rho
+\int_{0}^{\infty }\!\!\rho ^{\xi }\mathrm{e}^{-\mathrm{i}\xi \pi }\frac{%
\phi _{\varepsilon }\left( \rho \mathrm{e}^{-\mathrm{i}\pi }\right) }{\phi
_{\sigma }\left( \rho \mathrm{e}^{-\mathrm{i}\pi }\right) }\frac{\rho 
\mathrm{e}^{-\mathrm{i}\pi }}{\rho ^{2}\mathrm{e}^{-2\mathrm{i}\pi }+\omega
^{2}}\mathrm{e}^{\rho t\mathrm{e}^{-\mathrm{i}\pi }}\mathrm{e}^{-\mathrm{i}%
\pi }\mathrm{d}\rho \right) \\
& =\frac{\varepsilon _{0}}{2\pi \mathrm{i}}\int_{0}^{\infty }\left( \mathrm{e%
}^{\mathrm{i}\xi \pi }\frac{\phi _{\varepsilon }\left( \rho \mathrm{e}^{%
\mathrm{i}\pi }\right) }{\phi _{\sigma }\left( \rho \mathrm{e}^{\mathrm{i}%
\pi }\right) }-\mathrm{e}^{-\mathrm{i}\xi \pi }\frac{\phi _{\varepsilon
}\left( \rho \mathrm{e}^{-\mathrm{i}\pi }\right) }{\phi _{\sigma }\left(
\rho \mathrm{e}^{-\mathrm{i}\pi }\right) }\right) \frac{\rho ^{1+\xi }}{\rho
^{2}+\omega ^{2}}\mathrm{e}^{-\rho t}\mathrm{d}\rho \\
& =\frac{\varepsilon _{0}}{2\pi \mathrm{i}}\int_{0}^{\infty }\frac{%
\left\vert \phi _{\varepsilon }\left( \rho \mathrm{e}^{\mathrm{i}\pi
}\right) \right\vert }{\left\vert \phi _{\sigma }\left( \rho \mathrm{e}^{%
\mathrm{i}\pi }\right) \right\vert }\left( \mathrm{e}^{\mathrm{i}\xi \pi }%
\frac{\mathrm{e}^{\mathrm{i}\arg \phi _{\varepsilon }\left( \rho \mathrm{e}^{%
\mathrm{i}\pi }\right) }}{\mathrm{e}^{\mathrm{i}\arg \phi _{\sigma }\left(
\rho \mathrm{e}^{\mathrm{i}\pi }\right) }}-\mathrm{e}^{-\mathrm{i}\xi \pi }%
\frac{\mathrm{e}^{\mathrm{i}\arg \phi _{\varepsilon }\left( \rho \mathrm{e}%
^{-\mathrm{i}\pi }\right) }}{\mathrm{e}^{\mathrm{i}\arg \phi _{\sigma
}\left( \rho \mathrm{e}^{-\mathrm{i}\pi }\right) }}\right) \frac{\rho
^{1+\xi }}{\rho ^{2}+\omega ^{2}}\mathrm{e}^{-\rho t}\mathrm{d}\rho \\
& =\frac{\varepsilon _{0}}{\pi }\int_{0}^{\infty }\frac{\left\vert \phi
_{\varepsilon }\left( \rho \mathrm{e}^{\mathrm{i}\pi }\right) \right\vert }{%
\left\vert \phi _{\sigma }\left( \rho \mathrm{e}^{\mathrm{i}\pi }\right)
\right\vert }\sin \left( \arg \phi _{\varepsilon }\left( \rho \mathrm{e}^{%
\mathrm{i}\pi }\right) -\arg \phi _{\sigma }\left( \rho \mathrm{e}^{\mathrm{i%
}\pi }\right) +\xi \pi \right) \frac{\rho ^{1+\xi }}{\rho ^{2}+\omega ^{2}}%
\mathrm{e}^{-\rho t}\mathrm{d}\rho \quad \text{i.e.,} \\
& =\frac{\varepsilon _{0}}{\pi }\int_{0}^{\infty }\frac{K\left( \rho \right) 
}{\left\vert \phi _{\sigma }\left( \rho \mathrm{e}^{\mathrm{i}\pi }\right)
\right\vert ^{2}}\frac{\rho ^{1+\xi }}{\rho ^{2}+\omega ^{2}}\mathrm{e}%
^{-\rho t}\mathrm{d}\rho ,
\end{align*}%
see the defining relation (\ref{K}) for the function $K\ $and also (\ref%
{sigma-V1}) and (\ref{sigma-V2}), while the contribution from integrals
along contours $\Gamma _{8}$ and $\Gamma _{9},$ encircling zeros $s_{%
\scriptscriptstyle{\mathrm{RP}}}=\rho _{\scriptscriptstyle{\mathrm{RP}}}%
\mathrm{e}^{\mathrm{i}\pi }$ and $\bar{s}_{\scriptscriptstyle{\mathrm{RP}}%
}=\rho _{\scriptscriptstyle{\mathrm{RP}}}\mathrm{e}^{-\mathrm{i}\pi }$ of
function $\phi _{\sigma }$ and belonging to contour $\Gamma ^{(\mathrm{II})}$
depicted in Figures \ref{negativnaTG}, is calculated as%
\begin{align*}
& \int_{\Gamma _{8}}\tilde{\sigma}\left( s\right) \mathrm{e}^{st}\mathrm{d}%
s+\int_{\Gamma _{9}}\tilde{\sigma}\left( s\right) \mathrm{e}^{st}\mathrm{d}s
\\
& \qquad \qquad =\varepsilon _{0}\int_{\pi }^{0}\left( s_{\scriptscriptstyle%
\mathrm{R}{\mathrm{P}}}+r\mathrm{e}^{\mathrm{i}\varphi }\right) ^{\xi }\frac{%
\phi _{\varepsilon }\left( s_{\scriptscriptstyle\mathrm{R}{\mathrm{P}}}+r%
\mathrm{e}^{\mathrm{i}\varphi }\right) }{\phi _{\sigma }\left( s_{%
\scriptscriptstyle\mathrm{R}{\mathrm{P}}}+r\mathrm{e}^{\mathrm{i}\varphi
}\right) }\frac{s_{\scriptscriptstyle\mathrm{R}{\mathrm{P}}}+r\mathrm{e}^{%
\mathrm{i}\varphi }}{\left( s_{\scriptscriptstyle\mathrm{R}{\mathrm{P}}}+r%
\mathrm{e}^{\mathrm{i}\varphi }\right) ^{2}+\omega ^{2}}\mathrm{e}^{\left(
s_{\scriptscriptstyle\mathrm{R}{\mathrm{P}}}+r\mathrm{e}^{\mathrm{i}\varphi
}\right) t}\mathrm{i}r\mathrm{e}^{\mathrm{i}\varphi }\mathrm{d}\varphi \\
& \qquad \qquad \qquad +\varepsilon _{0}\int_{0}^{-\pi }\left( s_{%
\scriptscriptstyle\mathrm{R}{\mathrm{P}}}+r\mathrm{e}^{\mathrm{i}\varphi
}\right) ^{\xi }\frac{\phi _{\varepsilon }\left( \bar{s}_{\scriptscriptstyle{%
\mathrm{RP}}}+r\mathrm{e}^{\mathrm{i}\varphi }\right) }{\phi _{\sigma
}\left( \bar{s}_{\scriptscriptstyle{\mathrm{RP}}}+r\mathrm{e}^{\mathrm{i}%
\varphi }\right) }\frac{\bar{s}_{\scriptscriptstyle{\mathrm{RP}}}+r\mathrm{e}%
^{\mathrm{i}\varphi }}{\left( \bar{s}_{\scriptscriptstyle{\mathrm{RP}}}+r%
\mathrm{e}^{\mathrm{i}\varphi }\right) ^{2}+\omega ^{2}}\mathrm{e}^{\left( 
\bar{s}_{\scriptscriptstyle\mathrm{R}{\mathrm{P}}}+r\mathrm{e}^{\mathrm{i}%
\varphi }\right) t}\mathrm{i}r\mathrm{e}^{\mathrm{i}\varphi }\mathrm{d}%
\varphi ,
\end{align*}%
yielding, after taking the limit $r\rightarrow 0$ in the previous
expression, the function $\sigma ^{\left( \mathrm{RP}\right) },$ defined by (%
\ref{sigma-er-pe}), in the form%
\begin{align*}
\sigma ^{\left( \mathrm{RP}\right) }\left( t\right) & =-\lim_{r\rightarrow
0}\left( \frac{1}{2\pi \mathrm{i}}\int_{\Gamma _{8}}\tilde{\sigma}\left(
s\right) \mathrm{e}^{st}\mathrm{d}s+\frac{1}{2\pi \mathrm{i}}\int_{\Gamma
_{9}}\tilde{\sigma}\left( s\right) \mathrm{e}^{st}\mathrm{d}s\right) \\
& =-\frac{\varepsilon _{0}}{2\pi \mathrm{i}} \\
& \qquad \lim_{r\rightarrow 0}\Bigg(\int_{\pi }^{0}\frac{\phi _{\varepsilon
}\left( s_{\scriptscriptstyle\mathrm{R}{\mathrm{P}}}+r\mathrm{e}^{\mathrm{i}%
\varphi }\right) }{\phi _{\sigma }\left( s_{\scriptscriptstyle\mathrm{R}{%
\mathrm{P}}}\right) +\left. \phi _{\sigma }^{\prime }\left( s\right) \left(
s-s_{\scriptscriptstyle\mathrm{R}{\mathrm{P}}}\right) \right\vert _{s=s_{%
\scriptscriptstyle\mathrm{R}{\mathrm{P}}}+r\mathrm{e}^{\mathrm{i}\varphi
}}+\ldots }\frac{\left( s_{\scriptscriptstyle\mathrm{R}{\mathrm{P}}}+r%
\mathrm{e}^{\mathrm{i}\varphi }\right) ^{1+\xi }}{\left( s_{%
\scriptscriptstyle\mathrm{R}{\mathrm{P}}}+r\mathrm{e}^{\mathrm{i}\varphi
}\right) ^{2}+\omega ^{2}}\mathrm{e}^{\left( s_{\scriptscriptstyle\mathrm{R}{%
\mathrm{P}}}+r\mathrm{e}^{\mathrm{i}\varphi }\right) t}\mathrm{i}r\mathrm{e}%
^{\mathrm{i}\varphi }\mathrm{d}\varphi \\
& \qquad \qquad +\int_{0}^{-\pi }\frac{\phi _{\varepsilon }\left( \bar{s}_{%
\scriptscriptstyle\mathrm{R}{\mathrm{P}}}+r\mathrm{e}^{\mathrm{i}\varphi
}\right) }{\phi _{\sigma }\left( \bar{s}_{\scriptscriptstyle\mathrm{R}{%
\mathrm{P}}}\right) +\left. \phi _{\sigma }^{\prime }\left( s\right) \left(
s-\bar{s}_{\scriptscriptstyle\mathrm{R}{\mathrm{P}}}\right) \right\vert _{s=%
\bar{s}_{\scriptscriptstyle\mathrm{R}{\mathrm{P}}}+r\mathrm{e}^{\mathrm{i}%
\varphi }}+\ldots }\frac{\left( \bar{s}_{\scriptscriptstyle\mathrm{R}{%
\mathrm{P}}}+r\mathrm{e}^{\mathrm{i}\varphi }\right) ^{1+\xi }}{\left( \bar{s%
}_{\scriptscriptstyle\mathrm{R}{\mathrm{P}}}+r\mathrm{e}^{\mathrm{i}\varphi
}\right) ^{2}+\omega ^{2}}\mathrm{e}^{\left( \bar{s}_{\scriptscriptstyle%
\mathrm{R}{\mathrm{P}}}+r\mathrm{e}^{\mathrm{i}\varphi }\right) t}\mathrm{i}r%
\mathrm{e}^{\mathrm{i}\varphi }\mathrm{d}\varphi \Bigg) \\
& =-\frac{\varepsilon _{0}}{2\pi }\int_{\pi }^{0}\frac{\phi _{\varepsilon
}\left( s_{\scriptscriptstyle\mathrm{R}{\mathrm{P}}}\right) }{\phi _{\sigma
}^{\prime }\left( s_{\scriptscriptstyle\mathrm{R}{\mathrm{P}}}\right) }\frac{%
s_{\scriptscriptstyle\mathrm{R}{\mathrm{P}}}^{1+\xi }}{s_{\scriptscriptstyle%
\mathrm{R}{\mathrm{P}}}^{2}+\omega ^{2}}\mathrm{e}^{s_{\scriptscriptstyle%
\mathrm{R}{\mathrm{P}}}t}\mathrm{d}\varphi -\frac{\varepsilon _{0}}{2\pi }%
\int_{0}^{-\pi }\frac{\phi _{\varepsilon }\left( \bar{s}_{\scriptscriptstyle%
\mathrm{R}{\mathrm{P}}}\right) }{\phi _{\sigma }^{\prime }\left( \bar{s}_{%
\scriptscriptstyle\mathrm{R}{\mathrm{P}}}\right) }\frac{\bar{s}_{%
\scriptscriptstyle\mathrm{R}{\mathrm{P}}}^{1+\xi }}{\bar{s}_{%
\scriptscriptstyle\mathrm{R}{\mathrm{P}}}^{2}+\omega ^{2}}\mathrm{e}^{\bar{s}%
_{\scriptscriptstyle\mathrm{R}{\mathrm{P}}}t}\mathrm{d}\varphi \\
& =-\varepsilon _{0}\frac{\left\vert \phi _{\varepsilon }\left( s_{%
\scriptscriptstyle\mathrm{R}{\mathrm{P}}}\right) \right\vert }{\left\vert
\phi _{\sigma }^{\prime }\left( s_{\scriptscriptstyle\mathrm{R}{\mathrm{P}}%
}\right) \right\vert }\mathrm{\cos }\left( \arg \phi _{\varepsilon }\left(
s_{\scriptscriptstyle\mathrm{R}{\mathrm{P}}}\right) -\arg \phi _{\sigma
}^{\prime }\left( s_{\scriptscriptstyle\mathrm{R}{\mathrm{P}}}\right) +\xi
\pi \right) \frac{\rho _{\scriptscriptstyle\mathrm{R}{\mathrm{P}}}^{1+\xi }}{%
\rho _{\scriptscriptstyle\mathrm{R}{\mathrm{P}}}^{2}+\omega ^{2}}\mathrm{e}%
^{-\rho _{\scriptscriptstyle\mathrm{R}{\mathrm{P}}}t},
\end{align*}%
see also (\ref{sigma-rp}).

It is left to prove that the integrals in Cauchy residue theorems (\ref%
{KRT-HE}) and (\ref{CRT-HE}) along the contours $\Gamma _{1}$ ($\Gamma _{7}$%
), $\Gamma _{2}$ ($\Gamma _{6}$), and $\Gamma _{4}$, being parts of contours 
$\Gamma ^{(\mathrm{I})}$ and $\Gamma ^{(\mathrm{II})}$ depicted in Figures %
\ref{nemaTG} and \ref{negativnaTG}, tend to zero in the limit when $%
r\rightarrow 0$ and $R\rightarrow \infty .$ Constitutive functions $\phi
_{\varepsilon }$ and $\phi _{\sigma }$, listed in Table \ref{skupina}, that
are power-type functions, are required in \cite{SD-2} to satisfy conditions 
\begin{gather}
\frac{\left\vert \phi _{\varepsilon }\left( p+\mathrm{i}R\right) \right\vert 
}{\left\vert \phi _{\sigma }\left( p+\mathrm{i}R\right) \right\vert }\sim
R^{\zeta _{R}},\quad \text{when}\quad R\rightarrow \infty \quad \text{for}%
\quad p\in \left[ 0,p_{0}\right] ,  \label{kondisn-1} \\
\frac{\left\vert \phi _{\varepsilon }\left( r\mathrm{e}^{\mathrm{i}\varphi
}\right) \right\vert }{\left\vert \phi _{\sigma }\left( r\mathrm{e}^{\mathrm{%
i}\varphi }\right) \right\vert }\sim \frac{1}{r^{\zeta _{r}}},\quad \text{%
when}\quad r\rightarrow 0\quad \text{for}\quad \varphi \in \left[ -\pi ,\pi %
\right] ,  \label{kondisn-2}
\end{gather}%
with the orders $\zeta _{r}$ and $\zeta _{R}$ obeying 
\begin{equation}
-\xi <\zeta _{R}<1-\xi \quad \text{and}\quad -\left( 1-\xi \right) <\zeta
_{r}<\xi  \label{kondisn-3}
\end{equation}%
in order to calculate the relaxation modulus and creep compliance in their
integral representations.

The integral along contour $\Gamma _{1},$ taking the form%
\begin{equation*}
I_{\Gamma _{1}}=\varepsilon _{0}\int_{p_{0}}^{0}\frac{\phi _{\varepsilon
}\left( p+\mathrm{i}R\right) }{\phi _{\sigma }\left( p+\mathrm{i}R\right) }%
\frac{\left( p+\mathrm{i}R\right) ^{1+\xi }}{\left( p+\mathrm{i}R\right)
^{2}+\omega ^{2}}\mathrm{e}^{\left( p+\mathrm{i}R\right) t}\mathrm{d}p,
\end{equation*}%
according to the parametrization given in Tables \ref{nemaTG-param} and \ref%
{negativnaTG-param}, has a zero contribution, since its absolute value is%
\begin{align*}
\left\vert I_{\Gamma _{1}}\right\vert & \leqslant \varepsilon
_{0}\int_{0}^{p_{0}}\frac{\left\vert \phi _{\varepsilon }\left( p+\mathrm{i}%
R\right) \right\vert }{\left\vert \phi _{\sigma }\left( p+\mathrm{i}R\right)
\right\vert }\frac{\left\vert 1-\mathrm{i}\frac{p}{R}\right\vert ^{1-\xi }}{%
R^{1-\xi }\left\vert \left( 1-\mathrm{i}\frac{p}{R}\right) ^{2}-\left( 
\mathrm{i}\frac{\omega }{R}\right) ^{2}\right\vert }\mathrm{e}^{pt}\mathrm{d}%
p \\
& \leqslant \varepsilon _{0}\int_{0}^{p_{0}}\frac{1}{R^{1-\left( \xi +\zeta
_{R}\right) }}\mathrm{e}^{pt}\mathrm{d}p\rightarrow 0,\quad \text{for}\quad
R\rightarrow \infty ,
\end{align*}%
according to condition (\ref{kondisn-1}), with $\zeta _{R}<1-\xi $ due to
the condition (\ref{kondisn-3}). Using the similar argumentation, one can
prove that the integral along the contour $\Gamma _{7}$ also\ has zero
contribution. The zero contribution of the integral along the contour $%
\Gamma _{2},$ obtained as%
\begin{equation*}
I_{\Gamma _{2}}=\varepsilon _{0}\int_{\frac{\pi }{2}}^{\pi }\frac{\phi
_{\varepsilon }\left( R\mathrm{e}^{\mathrm{i}\varphi }\right) }{\phi
_{\sigma }\left( R\mathrm{e}^{\mathrm{i}\varphi }\right) }\frac{R^{1+\xi }%
\mathrm{e}^{\mathrm{i}\left( 1+\xi \right) \varphi }}{R^{2}\mathrm{e}^{2%
\mathrm{i}\varphi }+\omega ^{2}}\mathrm{e}^{Rt\mathrm{e}^{\mathrm{i}\varphi
}}\mathrm{i}R\mathrm{e}^{\mathrm{i}\varphi }\mathrm{d}\varphi
\end{equation*}%
according to the parametrization given in Tables \ref{nemaTG-param} and \ref%
{negativnaTG-param}, is guaranteed by the fact that $\phi _{\varepsilon }$
and $\phi _{\sigma }$ are power-type functions, since%
\begin{align*}
\left\vert I_{\Gamma _{2}}\right\vert & \leqslant \varepsilon _{0}\int_{%
\frac{\pi }{2}}^{\pi }\frac{\left\vert \phi _{\varepsilon }\left( R\mathrm{e}%
^{\mathrm{i}\varphi }\right) \right\vert }{\left\vert \phi _{\sigma }\left( R%
\mathrm{e}^{\mathrm{i}\varphi }\right) \right\vert }\frac{R^{2+\xi }}{%
\left\vert R^{2}\mathrm{e}^{2\mathrm{i}\varphi }+\omega ^{2}\right\vert }%
\mathrm{e}^{Rt\cos \varphi }\mathrm{d}\varphi \\
& \leqslant \varepsilon _{0}\int_{\frac{\pi }{2}}^{\pi }\frac{\left\vert
\phi _{\varepsilon }\left( R\mathrm{e}^{\mathrm{i}\varphi }\right)
\right\vert }{\left\vert \phi _{\sigma }\left( R\mathrm{e}^{\mathrm{i}%
\varphi }\right) \right\vert }\frac{R^{\xi }}{\left\vert 1+\frac{\omega ^{2}%
}{R^{2}\mathrm{e}^{2\mathrm{i}\varphi }}\right\vert }\mathrm{e}^{Rt\cos
\varphi }\mathrm{d}\varphi \\
& \leqslant \varepsilon _{0}\int_{\frac{\pi }{2}}^{\pi }R^{\xi +\zeta _{R}}%
\mathrm{e}^{Rt\cos \varphi }\mathrm{d}\varphi \rightarrow 0,\quad \text{for}%
\quad R\rightarrow \infty .
\end{align*}%
The integral along contour $\Gamma _{6}$ also has zero contribution, that
can be proved by using the similar argumentation. According to the condition
on the ratio of functions $\phi _{\varepsilon }$ and $\phi _{\sigma }$,
given by (\ref{kondisn-2}), and the condition $\zeta _{r}<\xi ,$ given by (%
\ref{kondisn-3}), the integral along the contour $\Gamma _{4},$ reading%
\begin{align*}
I_{\Gamma _{4}}& =\varepsilon _{0}\int_{\pi }^{-\pi }\frac{\phi
_{\varepsilon }\left( r\mathrm{e}^{\mathrm{i}\varphi }\right) }{\phi
_{\sigma }\left( r\mathrm{e}^{\mathrm{i}\varphi }\right) }\frac{r^{1+\xi }%
\mathrm{e}^{\mathrm{i}\left( 1+\xi \right) \varphi }}{r\mathrm{e}^{\mathrm{i}%
\varphi }+\omega ^{2}}\mathrm{e}^{rt\mathrm{e}^{\mathrm{i}\varphi }}\mathrm{i%
}r\mathrm{e}^{\mathrm{i}\varphi }\mathrm{d}\varphi \\
& =\varepsilon _{0}\int_{\pi }^{-\pi }\frac{\phi _{\varepsilon }\left( r%
\mathrm{e}^{\mathrm{i}\varphi }\right) }{\phi _{\sigma }\left( r\mathrm{e}^{%
\mathrm{i}\varphi }\right) }\frac{r^{2+\xi }\mathrm{e}^{\mathrm{i}\left(
2+\xi \right) \varphi }}{r\mathrm{e}^{\mathrm{i}\varphi }+\omega ^{2}}%
\mathrm{e}^{rt\mathrm{e}^{\mathrm{i}\varphi }}\mathrm{id}\varphi ,
\end{align*}%
according to the parametrization given in Tables \ref{nemaTG-param} and \ref%
{negativnaTG-param}, has also zero contribution, since%
\begin{align*}
\left\vert I_{\Gamma _{4}}\right\vert & \leqslant \varepsilon _{0}\int_{-\pi
}^{\pi }\frac{\left\vert \phi _{\varepsilon }\left( r\mathrm{e}^{\mathrm{i}%
\varphi }\right) \right\vert }{\left\vert \phi _{\sigma }\left( r\mathrm{e}^{%
\mathrm{i}\varphi }\right) \right\vert }\frac{r^{2+\xi }}{\left\vert r%
\mathrm{e}^{\mathrm{i}\varphi }+\omega ^{2}\right\vert }\mathrm{e}^{rt%
\mathrm{\cos }\varphi }\mathrm{d}\varphi \\
& \leqslant \frac{\varepsilon _{0}}{\omega ^{2}}\int_{-\pi }^{\pi }r^{2+\xi
-\zeta _{r}}\mathrm{d}\varphi \rightarrow 0,\quad \text{for}\quad
r\rightarrow 0.
\end{align*}%
\bigskip

\subsection{Steady state response to harmonic excitation}

If a viscoelastic body is subject to a strain in the form of a harmonic
function of angular frequency $\omega $ and amplitude $\mathcal{\varepsilon }%
_{0},$ then, due to the energy dissipation properties of the material and
after sufficiently long time, the stress will also be a harmonic function of
the same angular frequency as the strain, however shifted by the phase angle 
$\varphi _{\sigma }$ and having amplitude $\sigma _{0}$, so that one has%
\begin{equation}
\underline{\varepsilon }(t)=\varepsilon _{0}\,\mathrm{e}^{\mathrm{i}\omega
t}\quad \text{and}\quad \underline{\sigma }(t)=\sigma _{0}\left( \omega
\right) \,\mathrm{e}^{\mathrm{i}(\omega t+\varphi _{\sigma }\left( \omega
\right) )},  \label{fazori}
\end{equation}%
with the quantities 
\begin{equation}
\varepsilon \left( t\right) =\func{Re}\underline{\varepsilon }%
(t)=\varepsilon _{0}\,\cos \left( \omega t\right) \quad \text{and}\quad
\sigma \left( t\right) =\func{Re}\underline{\sigma }(t)=\sigma _{0}\left(
\omega \right) \,\cos \left( \omega t+\varphi _{\sigma }\left( \omega
\right) \right) ,  \label{fazori-1}
\end{equation}%
having physical meaning. Assuming that the constitutive equation describing
mechanical properties of the material is linear, the complex modulus is
defined by 
\begin{gather}
\hat{E}\left( \omega \right) =\frac{\underline{\sigma }(t)}{\underline{%
\varepsilon }(t)}=\frac{\sigma _{0}\left( \omega \right) }{\varepsilon _{0}}%
\mathrm{e}^{\mathrm{i}\varphi _{\sigma }\left( \omega \right) }=\left\vert 
\hat{E}\left( \omega \right) \right\vert \mathrm{e}^{\mathrm{i}\varphi
_{\sigma }\left( \omega \right) }=\hat{E}^{\prime }\left( \omega \right) +%
\mathrm{i}\hat{E}^{\prime \prime }\left( \omega \right) ,\quad \text{with}
\label{komplex-modulus} \\
\hat{E}^{\prime }\left( \omega \right) =\frac{\sigma _{0}\left( \omega
\right) }{\varepsilon _{0}}\cos \varphi _{\sigma }\left( \omega \right)
\quad \text{and}\quad \hat{E}^{\prime \prime }\left( \omega \right) =\frac{%
\sigma _{0}\left( \omega \right) }{\varepsilon _{0}}\mathrm{\sin }\varphi
_{\sigma }\left( \omega \right) ,  \notag
\end{gather}%
and the energy dissipation requirement%
\begin{equation*}
\hat{E}^{\prime }\left( \omega \right) \geqslant 0\quad \text{and}\quad \hat{%
E}^{\prime \prime }\left( \omega \right) \geqslant 0,\quad \forall \omega
\geqslant 0,
\end{equation*}%
posed on the storage and loss modulus $\hat{E}^{\prime }$ and $\hat{E}%
^{\prime \prime }\ $in \cite{b-t} and used in \cite{SD-1} in order to
formulate thermodynamically consistent constitutive models originating from
the rheological schemes corresponding to Zener and anti-Zener models can be
equivalently stated as%
\begin{equation*}
\cos \varphi _{\sigma }\left( \omega \right) \geqslant 0\quad \text{and}%
\quad \mathrm{\sin }\varphi _{\sigma }\left( \omega \right) \geqslant
0,\quad \forall \omega \geqslant 0.
\end{equation*}

On the other hand, the complex modulus is obtained from the constitutive
model in the Laplace domain by setting $s=\mathrm{i}\omega ,$ so that the
equation (\ref{konst-jednacina}) yields 
\begin{equation*}
\hat{E}\left( \omega \right) =\left. \frac{\tilde{\sigma}\left( s\right) }{%
\tilde{\varepsilon}\left( s\right) }\right\vert _{s=\mathrm{i}\omega
}=\left( \mathrm{i}\omega \right) ^{\xi }\frac{\phi _{\varepsilon }\left( 
\mathrm{i}\omega \right) }{\phi _{\sigma }\left( \mathrm{i}\omega \right) }%
=\omega ^{\xi }\frac{\left\vert \phi _{\varepsilon }\left( \mathrm{i}\omega
\right) \right\vert }{\left\vert \phi _{\sigma }\left( \mathrm{i}\omega
\right) \right\vert }\mathrm{e}^{\mathrm{i}\left( \arg \phi _{\varepsilon
}\left( \mathrm{i}\omega \right) -\arg \phi _{\sigma }\left( \mathrm{i}%
\omega \right) +\frac{\xi \pi }{2}\right) },
\end{equation*}%
that combined with $\hat{E}\left( \omega \right) =\frac{\underline{\sigma }%
(t)}{\underline{\varepsilon }(t)},$ see (\ref{komplex-modulus}), according
to the assumption on the harmonicity of strain and stress (\ref{fazori})
gives%
\begin{align}
\underline{\sigma }(t)& =\varepsilon _{0}\,\omega ^{\xi }\frac{\left\vert
\phi _{\varepsilon }\left( \mathrm{i}\omega \right) \right\vert }{\left\vert
\phi _{\sigma }\left( \mathrm{i}\omega \right) \right\vert }\mathrm{e}^{%
\mathrm{i}\left( \omega t+\arg \phi _{\varepsilon }\left( \mathrm{i}\omega
\right) -\arg \phi _{\sigma }\left( \mathrm{i}\omega \right) +\frac{\xi \pi 
}{2}\right) },\quad \text{i.e.,}  \notag \\
\sigma \left( t\right) & =\func{Re}\underline{\sigma }(t)=\varepsilon
_{0}\,\omega ^{\xi }\frac{\left\vert \phi _{\varepsilon }\left( \mathrm{i}%
\omega \right) \right\vert }{\left\vert \phi _{\sigma }\left( \mathrm{i}%
\omega \right) \right\vert }\cos \left( \omega t+\arg \phi _{\varepsilon
}\left( \mathrm{i}\omega \right) -\arg \phi _{\sigma }\left( \mathrm{i}%
\omega \right) +\frac{\xi \pi }{2}\right) ,  \label{sigma-isto-kao-sigma-h}
\end{align}%
so that the stress amplitude and phase angle, appearing in (\ref{fazori})
and (\ref{fazori-1}), are%
\begin{align*}
\sigma _{0}\left( \omega \right) & =\varepsilon _{0}\,\omega ^{\xi }\frac{%
\left\vert \phi _{\varepsilon }\left( \mathrm{i}\omega \right) \right\vert }{%
\left\vert \phi _{\sigma }\left( \mathrm{i}\omega \right) \right\vert }%
=\varepsilon _{0}\,\left\vert \hat{E}\left( \omega \right) \right\vert , \\
\varphi _{\sigma }\left( \omega \right) & =\arg \phi _{\varepsilon }\left( 
\mathrm{i}\omega \right) -\arg \phi _{\sigma }\left( \mathrm{i}\omega
\right) +\frac{\xi \pi }{2}.
\end{align*}%
The stress in the steady state regime, given by (\ref{sigma-isto-kao-sigma-h}%
), has exactly the same form as the function $\sigma ^{\left( \mathrm{H}%
\right) },$ given by (\ref{sigma-H}), representing the term in the transient
response of stress originating from the harmonic strain excitation, that
does not decay to zero after sufficiently long time.

\subsection{Numerical examples}

Time evolution of stress as a response to a strain assumed as a harmonic
function, along with its transition from transient to steady state regime is
examined for the thermodynamically consistent fractional anti-Zener and
Zener model I$^{+}$ID.ID, given by 
\begin{equation}
\left( a_{1}\,{}_{0}\mathrm{I}_{t}^{\alpha +\beta +\nu }+a_{2}\,{}_{0}%
\mathrm{I}_{t}^{\nu }+a_{3}\,{}_{0}\mathrm{D}_{t}^{\alpha +\beta -\nu
}\right) \sigma \left( t\right) =\left( b_{1}\,{}_{0}\mathrm{I}_{t}^{\alpha
}+b_{2}\,{}_{0}\mathrm{D}_{t}^{\beta }\right) \varepsilon \left( t\right) ,
\label{model}
\end{equation}%
with the narrowed thermodynamical restrictions on the model parameters%
\begin{gather}
0\leqslant \alpha +\beta -\nu \leqslant 1,\quad 1\leqslant \alpha +\beta
+\nu \leqslant 2,\quad \alpha \leqslant \nu \leqslant 1-\beta ,  \notag \\
\frac{a_{1}}{a_{2}}\frac{\left\vert \cos \frac{\left( \alpha +2\beta +\nu
\right) \pi }{2}\right\vert }{\cos \frac{\left( \nu -\alpha \right) \pi }{2}}%
\leqslant \frac{a_{1}}{a_{2}}\frac{\sin \frac{\left( \alpha +2\beta +\nu
\right) \pi }{2}}{\sin \frac{\left( \nu -\alpha \right) \pi }{2}}\frac{%
\left\vert \cos \frac{\left( \alpha +2\beta +\nu \right) \pi }{2}\right\vert 
}{\cos \frac{\left( \nu -\alpha \right) \pi }{2}}\leqslant \frac{b_{1}}{b_{2}%
},  \label{suzena-1} \\
\frac{b_{1}}{b_{2}}\leqslant \frac{a_{2}}{a_{3}}\frac{\sin \frac{\left(
\beta +\nu \right) \pi }{2}}{\sin \frac{\left( 2\alpha +\beta -\nu \right)
\pi }{2}}\frac{\cos \frac{\left( \beta +\nu \right) \pi }{2}}{\cos \frac{%
\left( 2\alpha +\beta -\nu \right) \pi }{2}}\leqslant \frac{a_{2}}{a_{3}}%
\frac{\sin \frac{\left( \beta +\nu \right) \pi }{2}}{\sin \frac{\left(
2\alpha +\beta -\nu \right) \pi }{2}},  \label{suzena-2}
\end{gather}%
see also Appendix \ref{FAZ-ZM}.

\begin{figure}[h]
\begin{center}
\begin{minipage}{0.46\columnwidth}
				\subfloat[Comparison of transient and steady state responses, depicted by solid line and dots respectively.]{
				\includegraphics[width=\columnwidth]{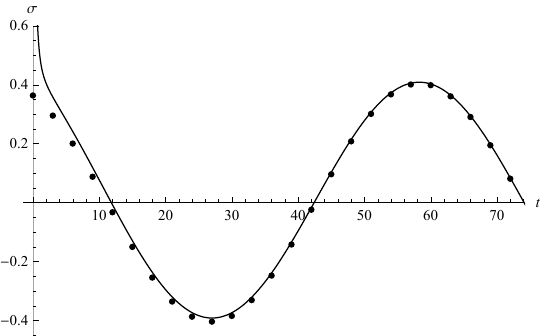}
				\label{sigma-harm-NP}}
		\end{minipage}
\hfill 
\begin{minipage}{0.46\columnwidth}
				\subfloat[Comparison of transient response and short time asymptotics, depicted by solid and dashed lines respectively.]{
				\includegraphics[width=\columnwidth]{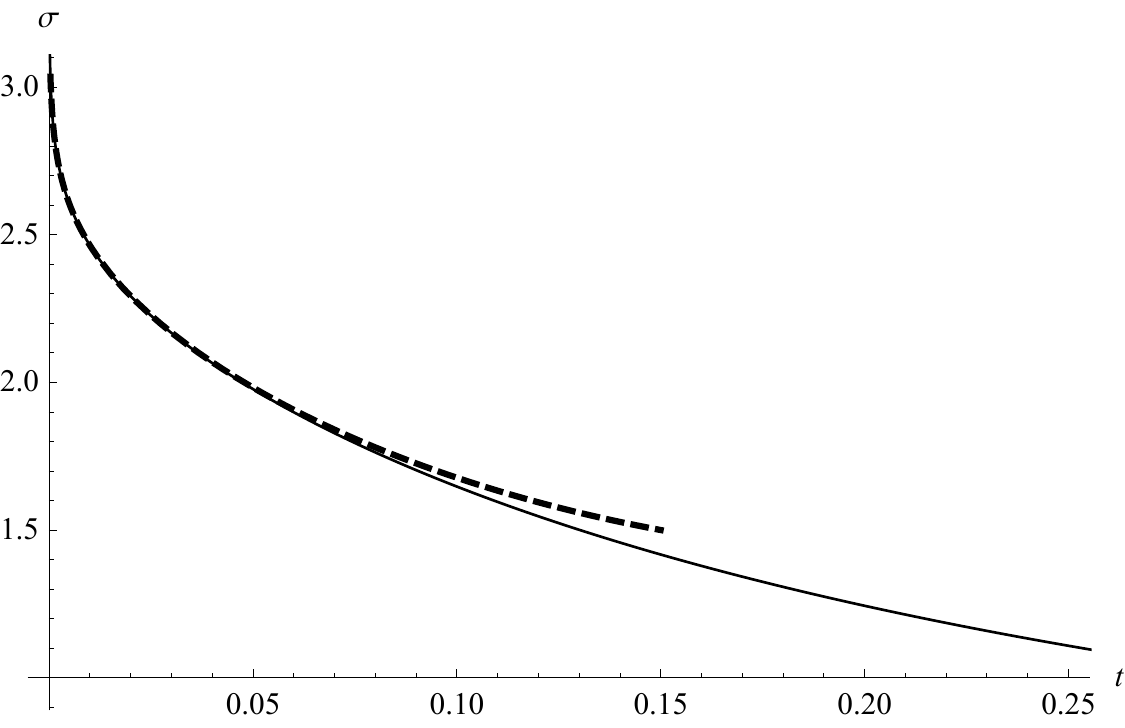}
				\label{sigma-harm-NP-srsha}}
		\end{minipage}
\end{center}
\caption{Time evolution of stress as a response to strain acting as harmonic
forcing, $\protect\varepsilon \left( t\right) =\protect\varepsilon _{0}\cos
\left( \protect\omega t\right) $, in the case when $\protect\phi _{\protect%
\sigma }$ has no zeros, obtained for model parameters as in Table \protect
\ref{parametri} and angular frequency $\protect\omega =0.1$.}
\label{harmonici-NP}
\end{figure}

\begin{figure}[h]
\begin{center}
\begin{minipage}{0.46\columnwidth}
				\subfloat[Comparison of transient and steady state responses, depicted by solid line and dots respectively.]{
				\includegraphics[width=\columnwidth]{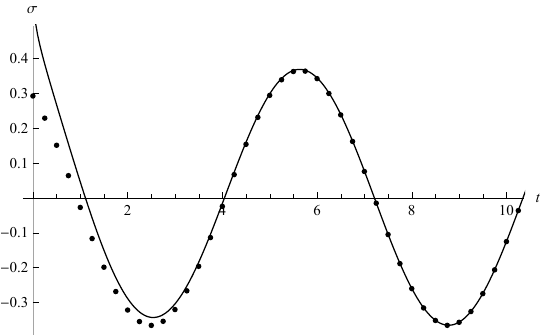}
				\label{sigma-harm-RP}}
		\end{minipage}
\hfill 
\begin{minipage}{0.46\columnwidth}
				\subfloat[Comparison of transient response and short time asymptotics, depicted by solid and dashed lines respectively.]{
				\includegraphics[width=\columnwidth]{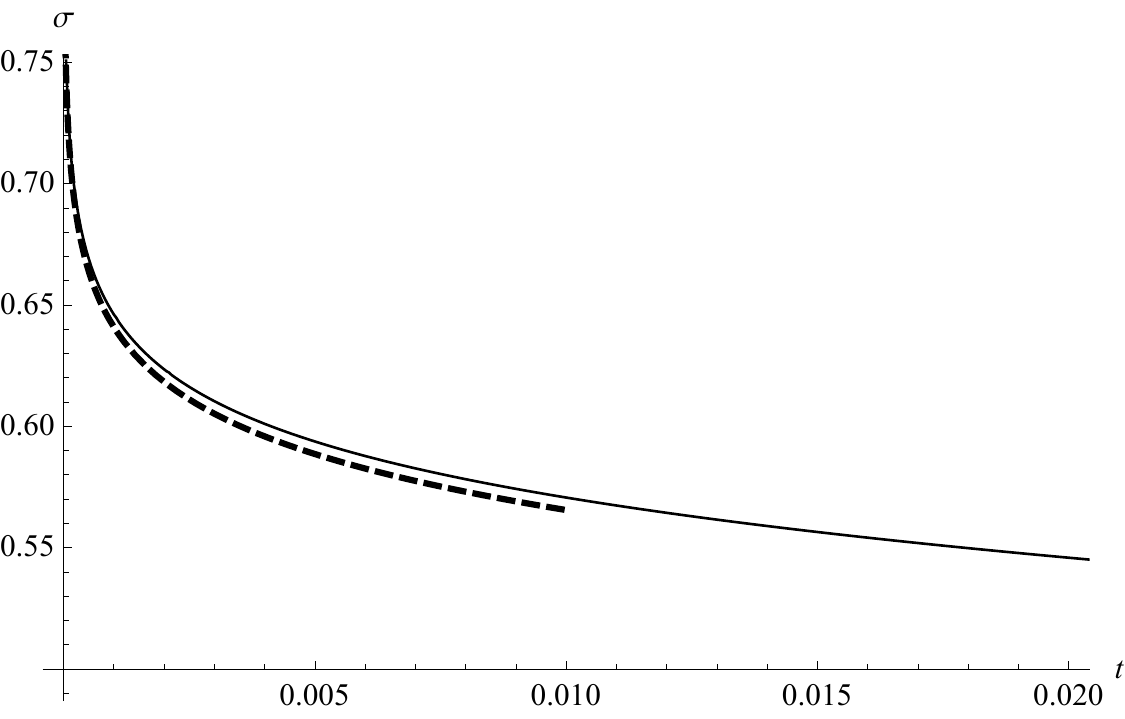}
				\label{sigma-harm-RP-srsha}}
		\end{minipage}
\end{center}
\caption{Time evolution of stress as a response to strain acting as harmonic
forcing, $\protect\varepsilon \left( t\right) =\protect\varepsilon _{0}\cos
\left( \protect\omega t\right) $, in the case when $\protect\phi _{\protect%
\sigma }$ has a negative real zero, obtained for model parameters as in
Table \protect\ref{parametri} and angular frequency $\protect\omega =1$.}
\label{harmonici-RP}
\end{figure}

\begin{figure}[h]
\begin{center}
\begin{minipage}{0.46\columnwidth}
				\subfloat[Comparison of transient and steady state responses, depicted by solid line and dots respectively.]{
				\includegraphics[width=\columnwidth]{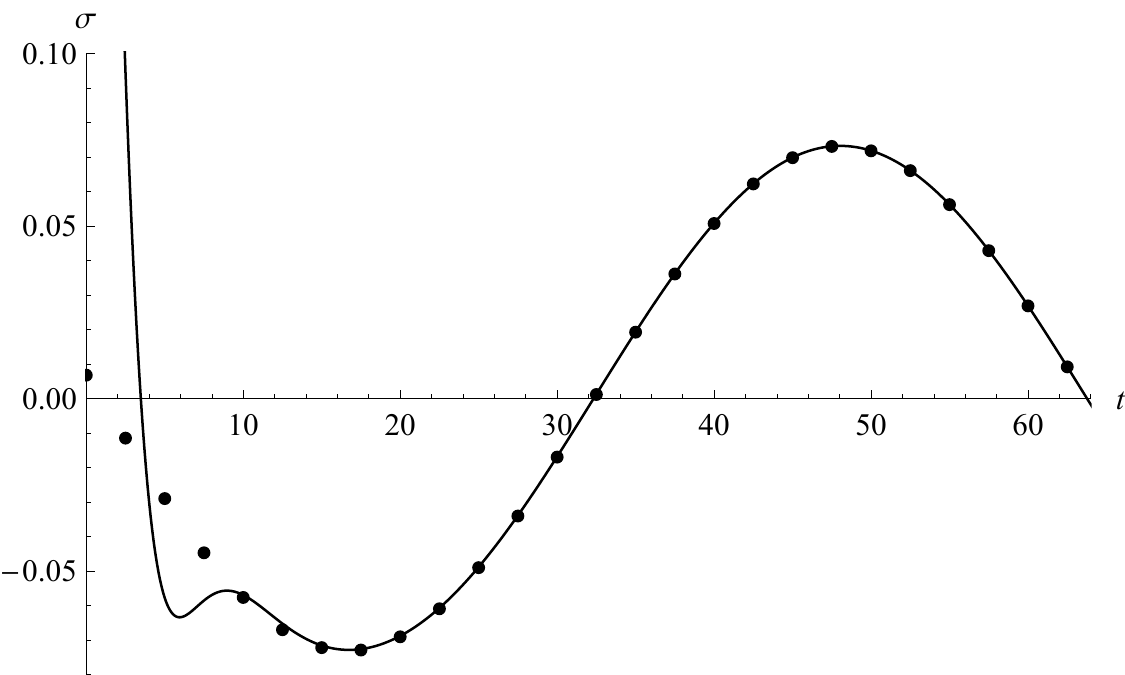}
				\label{sigma-harm-CCP}}
		\end{minipage}
\hfill 
\begin{minipage}{0.46\columnwidth}
				\subfloat[Comparison of transient response and short time asymptotics, depicted by solid and dashed lines respectively.]{
				\includegraphics[width=\columnwidth]{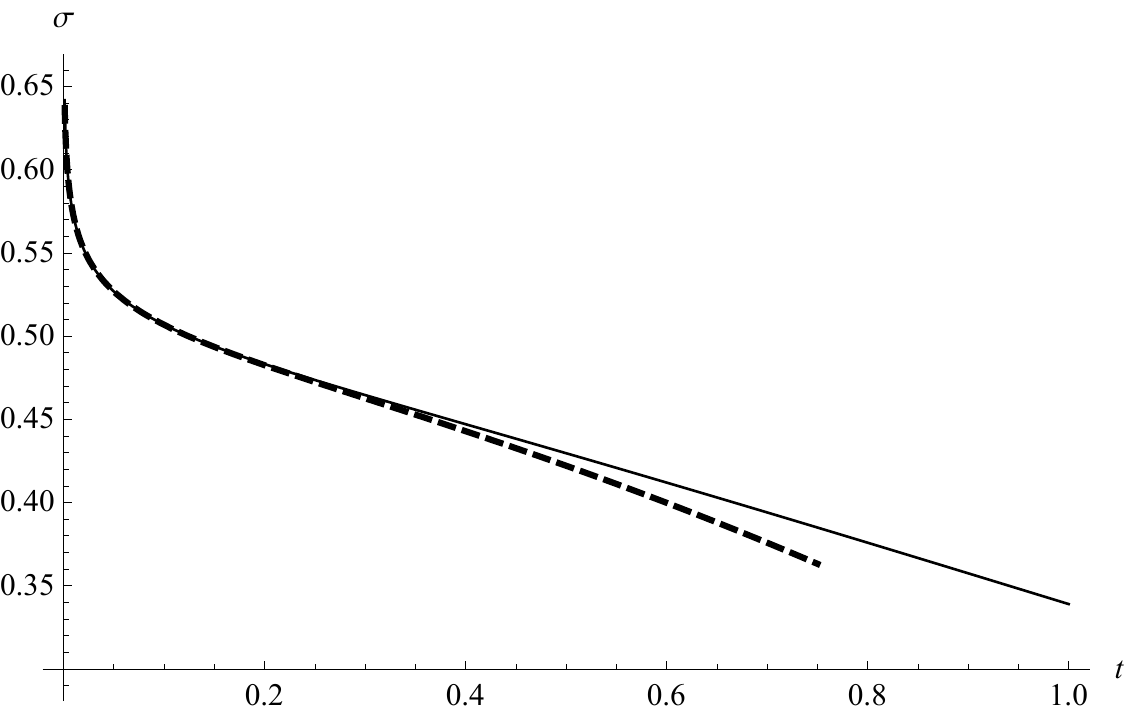}
				\label{sigma-harm-CCP-srsha}}
		\end{minipage}
\end{center}
\caption{Time evolution of stress as a response to strain acting as harmonic
forcing, $\protect\varepsilon \left( t\right) =\protect\varepsilon _{0}\cos
\left( \protect\omega t\right) $, in the case when $\protect\phi _{\protect%
\sigma }$ has a pair of complex conjugated zeros, obtained for model
parameters as in Table \protect\ref{parametri} and angular frequency $%
\protect\omega =0.1$.}
\label{harmonici-CCP}
\end{figure}

Using the model parameters given in Table \ref{parametri}, time profiles of
stress as a transient response to a strain, given as a cosine function, see (%
\ref{Harmeks}), are obtained using (\ref{resp-to-HE}), presented in Figures %
\ref{sigma-harm-NP}, \ref{sigma-harm-RP}, and \ref{sigma-harm-CCP}, and
compared with the stress profiles in the steady state regime, obtained using
(\ref{sigma-isto-kao-sigma-h}). \input{parametri-1.tex}

Figure \ref{sigma-harm-NP} shows the oscillatory character of time profiles
of stress in the case when function $\phi _{\sigma }$ has no zeros, such
that the transient profile tends to the infinity at the initial time
instant, see the short time asymptotics (\ref{asympt-sigma-harm-time-domain}%
) below, while the oscillatory behavior settles into a steady state mode
after about a half of the period, when the contribution of function $\sigma
^{\left( \mathrm{V}\right) }$, see (\ref{sigma-V1}) and (\ref{sigma-V2}), in
expression for stress (\ref{resp-to-HE}) becomes negligible with respect to $%
\sigma ^{\left( \mathrm{H}\right) }$, see (\ref{sigma-H}), so that in order
to describe the oscillatory behavior of stress, it is enough to consider
function $\sigma ^{\left( \mathrm{H}\right) },$ compare (\ref{sigma-H}) and (%
\ref{sigma-isto-kao-sigma-h}). 

Similar oscillatory character of stress, shown in Figure \ref{sigma-harm-RP}%
, is observed in the case when function $\phi _{\sigma }$ has a negative
real zero, while in Figure \ref{sigma-harm-CCP}, which corresponds to the
case when function $\phi _{\sigma }$ has a pair of complex conjugated zeros,
in addition to the oscillations originating from function $\sigma ^{\left( 
\mathrm{H}\right) }$, see (\ref{sigma-H}), there are also superposed
oscillations originating from the damped oscillatory function $\sigma
^{\left( \mathrm{CCP}\right) },$ see (\ref{sigma-ccp}), with a damping so
pronounced that there is only one peak noticeable. Having in addition
function $\sigma ^{\left( \mathrm{V}\right) }$, see (\ref{sigma-V1}) and (%
\ref{sigma-V2}), decreasing to zero, the oscillatory behavior of stress
settles into a steady state regime in less than a quarter of the period.

Figures \ref{sigma-harm-NP-srsha}, \ref{sigma-harm-RP-srsha}, and \ref%
{sigma-harm-CCP-srsha} show good agreement between curves obtained through
analytical expression and through the short time asymptotics in the cases
when function $\phi _{\sigma }$ has no zeros, has a negative real zero $s_{%
\scriptscriptstyle\mathrm{R}{\mathrm{P}}}$, and has a pair of complex
conjugated zeros $s_{\scriptscriptstyle{\mathrm{CCP}}}$ and $\bar{s}_{%
\scriptscriptstyle{\mathrm{CCP}}},$ respectively. The short time
asymptotics, given by the expression%
\begin{align}
\sigma \left( t\right) & =\varepsilon _{0}\frac{b_{2}}{a_{3}}\frac{%
t^{-\left( \nu -\alpha \right) }}{\Gamma \left( 1-\left( \nu -\alpha \right)
\right) }+\varepsilon _{0}\frac{b_{2}}{a_{3}}\left( \frac{b_{1}}{b_{2}}-%
\frac{a_{2}}{a_{3}}\right) \frac{t^{2\alpha +\beta -\nu }}{\Gamma \left(
1+2\alpha +\beta -\nu \right) }  \notag \\
& \qquad +\varepsilon _{0}\frac{b_{2}}{a_{3}}\left( \frac{a_{2}^{2}}{%
a_{3}^{2}}-\frac{a_{1}}{a_{3}}-\frac{a_{2}}{a_{3}}\frac{b_{1}}{b_{2}}\right) 
\frac{t^{3\alpha +2\beta -\nu }}{\Gamma \left( 1+3\alpha +2\beta -\nu
\right) }-\varepsilon _{0}\omega ^{2}\frac{b_{2}}{a_{3}}\frac{t^{2-\left(
\nu -\alpha \right) }}{\Gamma \left( 3-\left( \nu -\alpha \right) \right) }%
+O\left( t^{4\alpha +3\beta -\nu }\right) ,
\label{asympt-sigma-harm-time-domain}
\end{align}%
when $t\rightarrow 0$, is obtained by the use of the theorem that if $\tilde{%
f}\left( s\right) \sim \tilde{g}\left( s\right) $ as $s\rightarrow \infty ,$
then $f\left( t\right) \sim g\left( t\right) $ as $t\rightarrow 0$, from the
stress in the Laplace domain, given by (\ref{sigma-harm-ld}), that for the
model (\ref{model}) transforms as%
\begin{align}
\tilde{\sigma}\left( s\right) & =\varepsilon _{0}s^{\beta +\nu }\frac{%
b_{1}+b_{2}s^{\alpha +\beta }}{a_{1}+a_{2}s^{\alpha +\beta }+a_{3}s^{2\left(
\alpha +\beta \right) }}\frac{s}{s^{2}+\omega ^{2}}  \notag \\
& =\varepsilon _{0}\frac{b_{2}}{a_{3}}\frac{1}{s^{1-\left( \nu -\alpha
\right) }}\frac{1+\frac{b_{1}}{b_{2}}\frac{1}{s^{\alpha +\beta }}}{1+\frac{%
a_{2}}{a_{3}}\frac{1}{s^{\alpha +\beta }}+\frac{a_{1}}{a_{3}}\frac{1}{%
s^{2\left( \alpha +\beta \right) }}}\frac{1}{1+\frac{\omega ^{2}}{s^{2}}} 
\notag \\
& =\varepsilon _{0}\frac{b_{2}}{a_{3}}\frac{1}{s^{1-\left( \nu -\alpha
\right) }}\left( 1+\frac{b_{1}}{b_{2}}\frac{1}{s^{\alpha +\beta }}\right)
\left( 1+\frac{a_{2}}{a_{3}}\frac{1}{s^{\alpha +\beta }}+\frac{a_{1}}{a_{3}}%
\frac{1}{s^{2\left( \alpha +\beta \right) }}\right) ^{-1}\left( 1+\frac{%
\omega ^{2}}{s^{2}}\right) ^{-1}  \notag \\
& =\varepsilon _{0}\frac{b_{2}}{a_{3}}\frac{1}{s^{1-\left( \nu -\alpha
\right) }}\left( 1+\frac{b_{1}}{b_{2}}\frac{1}{s^{\alpha +\beta }}\right) 
\notag \\
& \qquad \times \left( 1-\frac{a_{2}}{a_{3}}\frac{1}{s^{\alpha +\beta }}-%
\frac{a_{1}}{a_{3}}\frac{1}{s^{2\left( \alpha +\beta \right) }}+\frac{%
a_{2}^{2}}{a_{3}^{2}}\frac{1}{s^{2\left( \alpha +\beta \right) }}+O\left(
s^{-3\left( \alpha +\beta \right) }\right) \right) \left( 1-\frac{\omega ^{2}%
}{s^{2}}+O\left( s^{-4}\right) \right)  \notag \\
& =\varepsilon _{0}\frac{b_{2}}{a_{3}}\frac{1}{s^{1-\left( \nu -\alpha
\right) }}\left( 1+\left( \frac{b_{1}}{b_{2}}-\frac{a_{2}}{a_{3}}\right) 
\frac{1}{s^{\alpha +\beta }}+\left( \frac{a_{2}^{2}}{a_{3}^{2}}-\frac{a_{1}}{%
a_{3}}-\frac{a_{2}}{a_{3}}\frac{b_{1}}{b_{2}}\right) \frac{1}{s^{2\left(
\alpha +\beta \right) }}-\frac{\omega ^{2}}{s^{2}}+O\left( s^{-3\left(
\alpha +\beta \right) }\right) \right)  \notag \\
& =\varepsilon _{0}\frac{b_{2}}{a_{3}}\frac{1}{s^{1-\left( \nu -\alpha
\right) }}+\varepsilon _{0}\frac{b_{2}}{a_{3}}\left( \frac{b_{1}}{b_{2}}-%
\frac{a_{2}}{a_{3}}\right) \frac{1}{s^{1+2\alpha +\beta -\nu }}  \notag \\
& \qquad +\varepsilon _{0}\frac{b_{2}}{a_{3}}\left( \frac{a_{2}^{2}}{%
a_{3}^{2}}-\frac{a_{1}}{a_{3}}-\frac{a_{2}}{a_{3}}\frac{b_{1}}{b_{2}}\right) 
\frac{1}{s^{1+3\alpha +2\beta -\nu }}-\varepsilon _{0}\omega ^{2}\frac{b_{2}%
}{a_{3}}\frac{1}{s^{3-\left( \nu -\alpha \right) }}+O\left( s^{-\left(
1+4\alpha +3\beta -\nu \right) }\right) ,\quad \text{when}\quad s\rightarrow
\infty .  \label{asympt-sigma-harm-ld}
\end{align}%
Note, the terms $-\varepsilon _{0}\omega ^{2}\frac{b_{2}}{a_{3}}\frac{1}{%
s^{3-\left( \nu -\alpha \right) }}$ and $-\varepsilon _{0}\omega ^{2}\frac{%
b_{2}}{a_{3}}\frac{t^{2-\left( \nu -\alpha \right) }}{\Gamma \left( 3-\left(
\nu -\alpha \right) \right) },$ respectively appearing in (\ref%
{asympt-sigma-harm-ld}) and (\ref{asympt-sigma-harm-time-domain}), do not
exist if $3\left( \alpha +\beta \right) \leqslant 2$, i.e., if $\alpha
+\beta \leqslant \frac{2}{3}$. 

\section{Time evolution of power per unit volume\label{WiP}}

Energy balance properties of one-dimensional viscoelastic body are
investigated in \cite{SD-2} and power per unit volume, defined by%
\begin{equation}
P\left( t\right) =\sigma \left( t\right) \dot{\varepsilon}\left( t\right) ,
\label{snaga}
\end{equation}%
is obtained in terms of energy $W$ and dissipated power $\mathcal{P}$ per
unit volume as%
\begin{equation}
P\left( t\right) =\frac{\mathrm{d}}{\mathrm{d}t}W\left( t\right) +\mathcal{P}%
\left( t\right) ,  \label{dW+P}
\end{equation}%
with energy and dissipated power per unit volume either expressed through
the relaxation modulus $\sigma _{sr}$ as%
\begin{align}
W\left( t\right) & =\frac{1}{2}\sigma _{sr}\left( t\right) \varepsilon
^{2}\left( t\right) +\frac{1}{2}\int_{0}^{t}\left( -\dot{\sigma}_{sr}\left(
t-t^{\prime }\right) \right) \left( \varepsilon \left( t\right) -\varepsilon
\left( t^{\prime }\right) \right) ^{2}\mathrm{d}t^{\prime }\quad \text{and}
\label{pot-en-epsilon} \\
\mathcal{P}\left( t\right) & =\frac{1}{2}\left( -\dot{\sigma}_{sr}\left(
t\right) \right) \varepsilon ^{2}\left( t\right) +\frac{1}{2}\int_{0}^{t}%
\ddot{\sigma}_{sr}\left( t-t^{\prime }\right) \left( \varepsilon \left(
t\right) -\varepsilon \left( t^{\prime }\right) \right) ^{2}\mathrm{d}%
t^{\prime },  \label{P-epsilon}
\end{align}%
or expressed through the creep compliance $\varepsilon _{cr}$ as%
\begin{align}
W\left( t\right) & =\frac{1}{2}\int_{0}^{t}\dot{\varepsilon}_{cr}\left(
t-t^{\prime }\right) \sigma ^{2}\left( t^{\prime }\right) \mathrm{d}%
t^{\prime }\quad \text{and}  \label{pot-en-sigma} \\
\mathcal{P}\left( t\right) & =\frac{1}{2}\dot{\varepsilon}_{cr}\left(
t\right) \sigma ^{2}\left( t\right) +\frac{1}{2}\int_{0}^{t}\left( -\ddot{%
\varepsilon}_{cr}\left( t-t^{\prime }\right) \right) \left( \sigma \left(
t\right) -\sigma \left( t^{\prime }\right) \right) ^{2}\mathrm{d}t^{\prime }.
\label{P-sigma}
\end{align}

The energy per unit volume, expressed through the relaxation modulus in the
form (\ref{pot-en-epsilon}), consists of the term resembling to the
potential energy of the elastic body with Young's modulus replaced by the
relaxation modulus, as well as of the term which takes into account the
memory of strain, weighted by the derivative of relaxation modulus.
Similarly, the energy per unit volume, expressed through the creep
compliance in the form (\ref{pot-en-sigma}), takes into account the memory
of stress, weighted by the derivative of creep compliance. The dissipated
power per unit volume, expressed in both of the forms (\ref{P-epsilon}) and (%
\ref{P-sigma}), has the instantaneous contribution depending on functions $%
\dot{\sigma}_{sr}$ and $\dot{\varepsilon}_{cr}$ respectively, as well as the
hereditary contribution, with the kernel depending on functions $\ddot{\sigma%
}_{sr}$ and $\ddot{\varepsilon}_{cr}$ respectively.

The positivity of energy, see (\ref{pot-en-epsilon}) and (\ref{pot-en-sigma}%
), and dissipativity of power, i.e., its positivity, see (\ref{P-epsilon})
and (\ref{P-sigma}), is guaranteed by requesting the relaxation modulus to
be a completely monotonic function, thus satisfying%
\begin{equation*}
\sigma _{sr}\left( t\right) \geqslant 0\quad \text{and}\quad \left(
-1\right) ^{k}\frac{\mathrm{d}^{k}}{\mathrm{d}t^{k}}\dot{\sigma}_{sr}\left(
t\right) \leqslant 0,\quad \text{for}\quad t>0,\quad k\in 
\mathbb{N}
_{0},
\end{equation*}%
as well as by requesting the creep compliance to be a Bernstein's function,
i.e., a positive function with completely monotonic first derivative.
Relaxation modulus and creep compliance admit mentioned properties in the
case of fractional models of viscoelastic body having the derivative orders
not exceeding the first order if the thermodynamical restrictions are met,
as proved in \cite{ZO}, while in the case of thermodynamically consistent
fractional Burgers models and fractional anti-Zener and Zener models, where
the fractional derivative orders may be in interval $\left( 1,2\right) ,$
mentioned properties of relaxation modulus and creep compliance are
established assuming that narrowed thermodynamical restrictions are met, see 
\cite{OZ-2} and \cite{SD-2}. Assuming the steady state regime, the
thermodynamical restrictions emerge from the request of non-negativity of
storage and loss modulus for all frequencies of harmonic forcing.

The energies per unit volume, so as the dissipated powers, expressed through
the relaxation modulus and through the creep compliance, see (\ref%
{pot-en-epsilon}) and (\ref{pot-en-sigma}), as well as (\ref{P-epsilon}) and
(\ref{P-sigma}), prove not to be equivalent, as obvious from Figures \ref%
{W-NP}, \ref{P-NP}, \ref{W-RNP}, \ref{P-RNP}, \ref{W-CCP}, and \ref{P-CCP}
below, contrary to the power per unit volume, obtained according to (\ref%
{dW+P}), that yields the same time evolution curves regardless of the form
chosen for energy and dissipated power, see Figures \ref{VelikoP-NP}, \ref%
{VelikoP-RNP}, and \ref{VelikoP-CCP} below. Therefore, a criterion is needed
to differ the stored (potential) energy from the energy-like expression,
when dealing with (\ref{pot-en-epsilon}) and (\ref{pot-en-sigma}), and
similarly one needs to differ the dissipated power from the dissipated
power-like expression, when dealing with (\ref{P-epsilon}) and (\ref{P-sigma}%
).

In order to be able to provide the physical interpretation of energies (\ref%
{pot-en-epsilon}) and (\ref{pot-en-sigma}), as well as of dissipated powers (%
\ref{P-epsilon}) and (\ref{P-sigma}), the equation of motion of
one-dimensional deformable body occupying points $x\in 
\mathbb{R}
,$ namely%
\begin{equation*}
\partial _{x}\sigma \left( x,t\right) =\rho \,\partial _{tt}u\left(
x,t\right) ,
\end{equation*}%
where $u$ is displacement, is multiplied by $\partial _{t}u$ and
subsequently integrated with respect to spatial coordinate along the whole
domain $%
\mathbb{R}
,$ yielding%
\begin{equation*}
\int_{%
\mathbb{R}
}\partial _{x}\sigma \left( x,t\right) \,\partial _{t}u\left( x,t\right) \,%
\mathrm{d}x=\partial _{t}\int_{%
\mathbb{R}
}\frac{\rho \left( \partial _{t}u\left( x,t\right) \right) ^{2}}{2}\mathrm{d}%
x,
\end{equation*}%
which transforms into 
\begin{equation}
\big[\partial _{t}u\left( x,t\right) \,\sigma \left( x,t\right) \big]%
_{x\rightarrow -\infty }^{x\rightarrow +\infty }-\int_{%
\mathbb{R}
}\sigma \left( x,t\right) \,\partial _{t}\varepsilon \left( x,t\right) \,%
\mathrm{d}x=\partial _{t}\int_{%
\mathbb{R}
}\frac{\rho \left( \partial _{t}u\left( x,t\right) \right) ^{2}}{2}\mathrm{d}%
x  \label{eq-mot-int}
\end{equation}%
after integration by parts, where the connection between strain and
displacement, namely 
\begin{equation}
\varepsilon \left( x,t\right) =\partial _{x}u\left( x,t\right) ,
\label{strain}
\end{equation}%
is used, so that the expression (\ref{eq-mot-int}), by the use of boundary
conditions%
\begin{equation*}
\lim_{x\rightarrow \pm \infty }u(x,t)=0\quad \text{and}\quad
\lim_{x\rightarrow \pm \infty }\sigma (x,t)=0,
\end{equation*}%
becomes of the form%
\begin{align}
&\partial _{t}\int_{%
\mathbb{R}
}\frac{\rho \left( \partial _{t}u\left( x,t\right) \right) ^{2}}{2}\mathrm{d}%
x =-\int_{%
\mathbb{R}
}\sigma \left( x,t\right) \,\partial _{t}\varepsilon \left( x,t\right) \,%
\mathrm{d}x=-\int_{%
\mathbb{R}
}P\left( x,t\right) \,\mathrm{d}x=-\int_{%
\mathbb{R}
}\left( \partial _{t}W\left( x,t\right) +\mathcal{P}\left( x,t\right)
\right) \,\mathrm{d}x,\quad \text{i.e.,}  \notag \\
&\partial _{t}\int_{%
\mathbb{R}
}\left( \mathcal{T}\left( x,t\right) +W\left( x,t\right) \right) \,\mathrm{d}%
x =-\int_{%
\mathbb{R}
}\mathcal{P}\left( x,t\right) \,\mathrm{d}x,  \label{ZPE}
\end{align}%
according to (\ref{snaga}) and (\ref{dW+P}), where%
\begin{equation}
\mathcal{T}\left( x,t\right) =\frac{\rho \left( \partial _{t}u\left(
x,t\right) \right) ^{2}}{2}  \label{T}
\end{equation}%
is the kinetic energy per unit volume, implying that the expression (\ref%
{ZPE}) can be interpreted as the law of change of total mechanical energy
per unit area. Note, the power per unit volume $P$, describing both elastic
and viscous properties of material, is decomposed according to (\ref{dW+P})
into the sum of term $\partial _{t}W,$ describing elastic properties of
material and thus representing the time change of potential energy and term $%
\mathcal{P}$, describing viscous properties of the material and thus
representing the dissipation power, see (\ref{ZPE}). Also, the law of change
of total mechanical energy per unit area (\ref{ZPE}), rewritten as 
\begin{equation}
\int_{%
\mathbb{R}
}\big(\partial _{t}\left( \mathcal{T}\left( x,t\right) +W\left( x,t\right)
\right) +\mathcal{P}\left( x,t\right) \big)\mathrm{d}x=0
\label{P+d(W+T)-int}
\end{equation}%
implies the law of change of total mechanical energy per unit volume%
\begin{equation}
\partial _{t}\big(\mathcal{T}\left( x,t\right) +W\left( x,t\right) \big)=-%
\mathcal{P}\left( x,t\right) .  \label{P+d(W+T)}
\end{equation}

The use of expressions for energy (\ref{pot-en-epsilon}) and dissipated
power (\ref{P-epsilon}) along with the kinetic energy per unit volume (\ref%
{T}), transforms the law of change of total mechanical energy (\ref{P+d(W+T)}%
) into 
\begin{align}
& \partial _{t}\left( \frac{\rho \left( \partial _{t}u\left( x,t\right)
\right) ^{2}}{2}+\frac{1}{2}\sigma _{sr}\left( t\right) \left( \partial
_{x}u\left( x,t\right) \right) ^{2}+\frac{1}{2}\int_{0}^{t}\left( -\dot{%
\sigma}_{sr}\left( t-t^{\prime }\right) \right) \left( \partial _{x}u\left(
x,t\right) -\partial _{x}u\left( x,t^{\prime }\right) \right) ^{2}\mathrm{d}%
t^{\prime }\right)   \notag \\
& \qquad \qquad \qquad =-\frac{1}{2}\left( -\dot{\sigma}_{sr}\left( t\right)
\right) \left( \partial _{x}u\left( x,t\right) \right) ^{2}-\frac{1}{2}%
\int_{0}^{t}\ddot{\sigma}_{sr}\left( t-t^{\prime }\right) \left( \partial
_{x}u\left( x,t\right) -\partial _{x}u\left( x,t^{\prime }\right) \right)
^{2}\mathrm{d}t^{\prime },  \label{LCTME-preko-u}
\end{align}%
while the law of change of total mechanical energy (\ref{P+d(W+T)}), using
the expressions for energy (\ref{pot-en-sigma}), dissipated power (\ref%
{P-sigma}), and kinetic energy per unit volume (\ref{T}), becomes%
\begin{align}
& \partial _{t}\left( \frac{\rho \left( \partial _{t}u\left( x,t\right)
\right) ^{2}}{2}+\frac{1}{2}\int_{0}^{t}\dot{\varepsilon}_{cr}\left(
t-t^{\prime }\right) \sigma ^{2}\left( x,t^{\prime }\right) \mathrm{d}%
t^{\prime }\right)   \notag \\
& \qquad \qquad \qquad =-\frac{1}{2}\dot{\varepsilon}_{cr}\left( t\right)
\sigma ^{2}\left( x,t\right) -\frac{1}{2}\int_{0}^{t}\left( -\ddot{%
\varepsilon}_{cr}\left( t-t^{\prime }\right) \right) \left( \sigma \left(
x,t\right) -\sigma \left( x,t^{\prime }\right) \right) ^{2}\mathrm{d}%
t^{\prime }.  \label{LCTME-preko-mix}
\end{align}%
Obviously, the law of change of total mechanical energy is expressed in
terms of displacement in (\ref{LCTME-preko-u}), while in (\ref%
{LCTME-preko-mix}) the stress can be found alongside the displacement and
therefore one needs constitutive equation (\ref{sigma-epsilon})$_{1}$ and
strain (\ref{strain}) in order to eliminate stress from the expression (\ref%
{LCTME-preko-mix}). Therefore, the form of law of change of total mechanical
energy (\ref{LCTME-preko-u}) indicates that the potential energy and
dissipated power in viscoelastic body are given by (\ref{pot-en-epsilon})
and (\ref{P-epsilon}), rather than by (\ref{pot-en-sigma}) and (\ref{P-sigma}%
) representing energy-like and dissipated power-like expressions.

Further, when dissipativity of the hereditary fractional wave equations is
investigated in \cite{ZO}, the expressions 
\begin{equation}
\frac{1}{2}\rho \left\vert \left\vert \partial _{t}u\left( \cdot ,t\right)
\right\vert \right\vert _{L^{2}\left( 
\mathbb{R}
\right) }^{2}+\frac{1}{2}\sigma _{sr}\left( t\right) \left\vert \left\vert
\partial _{x}u\left( \cdot ,t\right) \right\vert \right\vert _{L^{2}\left( 
\mathbb{R}
\right) }^{2}+\frac{1}{2}\int_{0}^{t}\left( -\dot{\sigma}_{sr}\left(
t^{\prime }\right) \right) \left\vert \left\vert \partial _{x}u\left( \cdot
,t\right) \right\vert \right\vert _{L^{2}\left( 
\mathbb{R}
\right) }^{2}\mathrm{d}t^{\prime }\leqslant \frac{1}{2}\rho \left\vert
\left\vert v_{0}\left( \cdot \right) \right\vert \right\vert _{L^{2}\left( 
\mathbb{R}
\right) }^{2}  \label{OZ-preko-epsilon-konacno}
\end{equation}%
and 
\begin{equation}
\frac{1}{2}\rho \left\vert \left\vert \partial _{t}u\left( \cdot ,t\right)
\right\vert \right\vert _{L^{2}\left( 
\mathbb{R}
\right) }^{2}+\int_{0}^{t}\int_{%
\mathbb{R}
}\left( \sigma _{sr}\left( t^{\prime }\right) \ast _{t^{\prime }}\partial
_{t^{\prime }~x}u\left( x,t^{\prime }\right) \right) \partial _{t^{\prime
}~x}u\left( x,t^{\prime }\right) \mathrm{d}x\mathrm{d}t^{\prime }=\frac{1}{2}%
\rho \left\vert \left\vert v_{0}\left( \cdot \right) \right\vert \right\vert
_{L^{2}\left( 
\mathbb{R}
\right) }^{2},  \label{OZ-preko-epsilon-beskonacno}
\end{equation}%
where the following notation is used%
\begin{equation*}
\left\vert \left\vert u\left( \cdot ,t\right) \right\vert \right\vert
_{L^{2}\left( 
\mathbb{R}
\right) }^{2}=\int_{%
\mathbb{R}
}\left( u\left( x,t\right) \right) ^{2}\mathrm{d}x,
\end{equation*}%
are obtained in the case of finite and infinite wave propagation speed
respectively, stating that the kinetic energy per unit area of viscoelastic
body at arbitrary time instant is less than the kinetic energy at the
initial time instant, since $v_{0}=v_{0}\left( x\right) $ is the initial
velocity field, thus implying the dissipativity of the material. In order to
obtain (\ref{OZ-preko-epsilon-konacno}) and (\ref%
{OZ-preko-epsilon-beskonacno}), the fractional wave equation is expressed in
terms of relaxation modulus as 
\begin{gather*}
\rho \,\partial _{tt}u\left( x,t\right) =\sigma _{sr}^{\left( g\right)
}\,\partial _{xx}u\left( x,t\right) +\dot{\sigma}_{sr}\left( t\right) \ast
_{t}\partial _{xx}u\left( x,t\right) \quad \text{and} \\
\rho \,\partial _{tt}u\left( x,t\right) =\sigma _{sr}\left( t\right) \ast
_{t}\partial _{txx}u\left( x,t\right) ,
\end{gather*}%
respectively, while by expressing the fractional wave equation is in terms
of creep compliance as%
\begin{equation}
\rho \varepsilon _{cr}^{\left( g\right) }\,\partial _{tt}u\left( x,t\right)
+\rho \,\dot{\varepsilon}_{cr}\left( t\right) \ast _{t}\partial _{tt}u\left(
x,t\right) =\partial _{xx}u\left( x,t\right) ,  \notag
\end{equation}%
the energy estimate is obtained in the form 
\begin{equation}
\frac{1}{2}\rho \frac{1}{\varepsilon _{cr}\left( t\right) }\partial
_{t}\left( \varepsilon _{cr}\left( t\right) \ast _{t}\left\vert \left\vert
\partial _{t}u\left( \cdot ,t\right) \right\vert \right\vert _{L^{2}\left( 
\mathbb{R}
\right) }^{2}\right) +\frac{1}{2\varepsilon _{cr}\left( t\right) }\left\vert
\left\vert \partial _{x}u\left( \cdot ,t\right) \right\vert \right\vert
_{L^{2}\left( 
\mathbb{R}
\right) }^{2}\leqslant \frac{1}{2}\rho \left\vert \left\vert v_{0}\left(
\cdot \right) \right\vert \right\vert _{L^{2}\left( 
\mathbb{R}
\right) }^{2},  \label{OZ-preko-sigme}
\end{equation}%
that does not contain the kinetic energy at arbitrary time instant and
therefore the conclusion on dissipativity cannot be drawn. Therefore, one
concludes that the energy estimates (\ref{OZ-preko-epsilon-konacno}) and (%
\ref{OZ-preko-epsilon-beskonacno}), expressed in terms of relaxation
modulus, can be interpreted as law of change of total mechanical energy (\ref%
{P+d(W+T)-int}), while the energy estimate expressed in terms of creep
compliance (\ref{OZ-preko-sigme}) cannot. A priori energy estimates (\ref%
{OZ-preko-epsilon-konacno}) and (\ref{OZ-preko-epsilon-beskonacno}), when
compared to (\ref{OZ-preko-sigme}), also indicate that the expressions (\ref%
{pot-en-epsilon}) and (\ref{P-epsilon}), containing relaxation modulus,
correspond to the potential energy and dissipated power, while the
expressions (\ref{pot-en-sigma}) and (\ref{P-sigma}), containing creep
compliance, correspond to the energy-like and dissipated power-like
quantities.

Aiming to obtain curves illustrating the time evolution of energy, given by (%
\ref{pot-en-epsilon}) and (\ref{pot-en-sigma}), as well as of the dissipated
power, given by (\ref{P-epsilon}) and (\ref{P-sigma}), alongside the power
per unit volume, given by (\ref{dW+P}), the deformation is assumed as a sine
function%
\begin{equation}
\varepsilon \left( t\right) =\varepsilon _{0}\sin \left( \omega t\right) ,
\label{HE-sin}
\end{equation}%
which is sufficient to calculate energy, dissipated power, and power per
unit volume expressed in terms of relaxation modulus, according to (\ref%
{pot-en-epsilon}), (\ref{P-epsilon}), and (\ref{dW+P}), since the explicit
forms of relaxation modulus are given by (\ref{sr-opste}), see Appendix \ref%
{RMCC}. On the other hand, the calculation of energy, dissipated power, and
power per unit volume expressed in terms of creep compliance, according to (%
\ref{pot-en-sigma}), (\ref{P-sigma}), and (\ref{dW+P}), apart from the creep
compliance whose explicit forms are given by (\ref{cr-opste}), see Appendix %
\ref{RMCC}, requires stress to be calculated as a response to the
deformation in the form (\ref{HE-sin}), which is performed analogously as in
Section \ref{TRandSSR}.

Namely, the stress as a response to the deformation assumed as a cosine
function, see (\ref{Harmeks}), is derived and given by expression (\ref%
{resp-to-HE}) in Section \ref{TRandSSR}, while in the case of deformation
prescribed as a sine function, see (\ref{HE-sin}), one obtains the response
by using the constitutive equation in the Laplace domain (\ref%
{konst-jednacina}) along with the Laplace transform of the sine function,
yielding%
\begin{equation}
\tilde{\sigma}\left( s\right) =\varepsilon _{0}s^{\xi }\frac{\phi
_{\varepsilon }\left( s\right) }{\phi _{\sigma }\left( s\right) }\frac{%
\omega }{s^{2}+\omega ^{2}}  \label{sigma-sr-ld-sin}
\end{equation}%
and becoming%
\begin{equation}
\sigma (t)=\sigma ^{\left( \mathrm{v}\right) }\left( t\right) +\sigma
^{\left( \mathrm{h}\right) }\left( t\right) +\left\{ \!\!\!%
\begin{tabular}{ll}
$0$, & if $\phi _{\sigma }$ has no zeros, \smallskip \\ 
$\sigma ^{\left( \mathrm{rp}\right) }\left( t\right) $, & if $\phi _{\sigma
} $ has a negative real zero, \smallskip \\ 
$\sigma ^{\left( \mathrm{ccp}\right) }\left( t\right) $, & if $\phi _{\sigma
}$ has a pair of complex conjugated zeros,%
\end{tabular}%
\right.  \label{resp-to-HE-but-sin}
\end{equation}%
after the inverse Laplace transform is applied to (\ref{sigma-sr-ld-sin}),
with the functions%
\begin{align}
\sigma ^{\left( \mathrm{v}\right) }\left( t\right) & =-\frac{\omega
~\varepsilon _{0}}{\pi }\int_{0}^{\infty }\frac{K\left( \rho \right) }{%
\left\vert \phi _{\sigma }\left( \rho \mathrm{e}^{\mathrm{i}\pi }\right)
\right\vert ^{2}}\frac{\rho ^{\xi }}{\rho ^{2}+\omega ^{2}}\mathrm{e}^{-\rho
t}\mathrm{d}\rho ,\quad \text{or}  \label{sigma-V1-but-sin} \\
\sigma ^{\left( \mathrm{v}\right) }\left( t\right) & =-\frac{\omega
~\varepsilon _{0}}{\pi }\int_{0}^{\infty }\frac{\left\vert \phi
_{\varepsilon }\left( \rho \mathrm{e}^{\mathrm{i}\pi }\right) \right\vert }{%
\left\vert \phi _{\sigma }\left( \rho \mathrm{e}^{\mathrm{i}\pi }\right)
\right\vert }\sin \left( \arg \phi _{\varepsilon }\left( \rho \mathrm{e}^{%
\mathrm{i}\pi }\right) -\arg \phi _{\sigma }\left( \rho \mathrm{e}^{\mathrm{i%
}\pi }\right) +\xi \pi \right) \frac{\rho ^{\xi }}{\rho ^{2}+\omega ^{2}}%
\mathrm{e}^{-\rho t}\mathrm{d}\rho ,  \label{sigma-V2-but-sin} \\
\sigma ^{\left( \mathrm{h}\right) }\left( t\right) & =\varepsilon
_{0}\left\vert \hat{E}\left( \omega \right) \right\vert \sin \left( \omega
t+\arg \phi _{\varepsilon }\left( \mathrm{i}\omega \right) -\arg \phi
_{\sigma }\left( \mathrm{i}\omega \right) +\frac{\xi \pi }{2}\right) ,\ 
\text{with}\ \left\vert \hat{E}\left( \omega \right) \right\vert =\omega
^{\xi }\frac{\left\vert \phi _{\varepsilon }\left( \mathrm{i}\omega \right)
\right\vert }{\left\vert \phi _{\sigma }\left( \mathrm{i}\omega \right)
\right\vert },  \label{sigma-H-but-sin} \\
\sigma ^{\left( \mathrm{rp}\right) }\left( t\right) & =\omega ~\varepsilon
_{0}\frac{\left\vert \phi _{\varepsilon }\left( s_{\scriptscriptstyle\mathrm{%
R}{\mathrm{P}}}\right) \right\vert }{\left\vert \phi _{\sigma }^{\prime
}\left( s_{\scriptscriptstyle\mathrm{R}{\mathrm{P}}}\right) \right\vert }%
\mathrm{\cos }\left( \arg \phi _{\varepsilon }\left( s_{\scriptscriptstyle%
\mathrm{R}{\mathrm{P}}}\right) -\arg \phi _{\sigma }^{\prime }\left( s_{%
\scriptscriptstyle\mathrm{R}{\mathrm{P}}}\right) +\xi \pi \right) \frac{\rho
_{\scriptscriptstyle\mathrm{R}{\mathrm{P}}}^{\xi }}{\rho _{\scriptscriptstyle%
\mathrm{R}{\mathrm{P}}}^{2}+\omega ^{2}}\mathrm{e}^{-\rho _{%
\scriptscriptstyle\mathrm{R}{\mathrm{P}}}t},  \label{sigma-rp-but-sin} \\
\sigma ^{\left( \mathrm{ccp}\right) }\left( t\right) & =2\omega ~\varepsilon
_{0}\frac{\left\vert \phi _{\varepsilon }\left( s_{\scriptscriptstyle{%
\mathrm{CCP}}}\right) \right\vert }{\left\vert \phi _{\sigma }^{\prime
}\left( s_{\scriptscriptstyle{\mathrm{CCP}}}\right) \right\vert }\frac{\rho
_{\scriptscriptstyle{\mathrm{CCP}}}^{\xi }}{\left\vert s_{\scriptscriptstyle{%
\mathrm{CCP}}}^{2}+\omega ^{2}\right\vert }\mathrm{e}^{-\left\vert \func{Re}%
s_{\scriptscriptstyle{\mathrm{CCP}}}\right\vert t}  \label{sigma-ccp-but-sin}
\\
& \qquad \times \cos \left( \func{Im}s_{\scriptscriptstyle{\mathrm{CCP}}%
}t+\arg \phi _{\varepsilon }\left( s_{\scriptscriptstyle{\mathrm{CCP}}%
}\right) -\arg \phi _{\sigma }^{\prime }\left( s_{\scriptscriptstyle{\mathrm{%
CCP}}}\right) +\xi \varphi _{\scriptscriptstyle{\mathrm{CCP}}}-\phi \left(
\omega \right) \right) ,  \notag
\end{align}%
where function $K$ is given by (\ref{K}) and $\phi $ is given by (\ref{faza}%
), with $s_{\scriptscriptstyle\mathrm{R}{\mathrm{P}}}=\rho _{%
\scriptscriptstyle\mathrm{R}{\mathrm{P}}}\,\mathrm{e}^{\mathrm{i}\pi }$
being a negative real zero of function $\phi _{\sigma },$ while $s_{%
\scriptscriptstyle{\mathrm{CCP}}}=\rho _{\scriptscriptstyle{\mathrm{CCP}}}\,%
\mathrm{e}^{\mathrm{i}\varphi _{\scriptscriptstyle{\mathrm{CCP}}}}$ and its
complex conjugate $\bar{s}_{\scriptscriptstyle{\mathrm{CCP}}}$ are complex
zeros of function $\phi _{\sigma }$ having negative real part. The
calculation is analogous to the one presented in Section \ref{TRandSSR} and
therefore omitted.

\begin{figure}[h]
\begin{center}
\begin{minipage}{0.5\columnwidth}
			\subfloat[Stored energy expressed in terms of: strain - depicted by solid line and stress - depicted by dashed line.]{
			\includegraphics[width=\columnwidth]{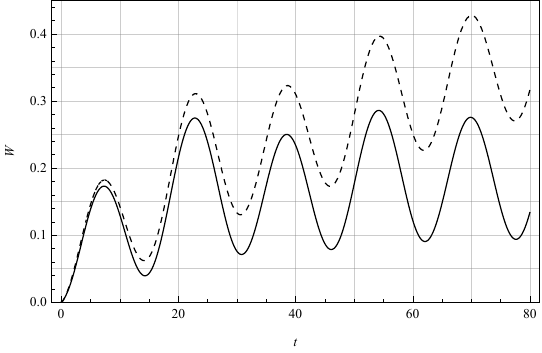}
			\label{W-NP}}
		\end{minipage}
\medskip \vfil
\begin{minipage}{0.46\columnwidth}
				\subfloat[Short time asymptotics of stored energy expressed through strain, depicted by dashed line.]{
				\includegraphics[width=\columnwidth]{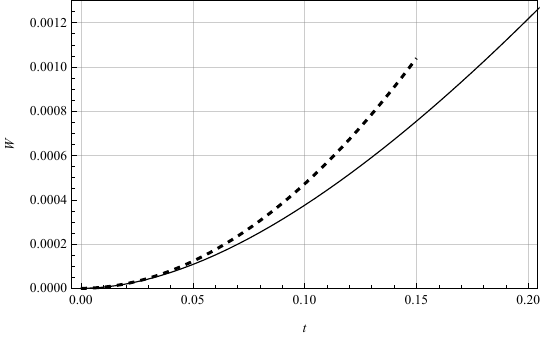}
				\label{W-eps-NP-Srsha}}
		\end{minipage}
\hfill 
\begin{minipage}{0.46\columnwidth}
				\subfloat[Short time asymptotics of stored energy expressed through stress, depicted by dashed line.]{
				\includegraphics[width=\columnwidth]{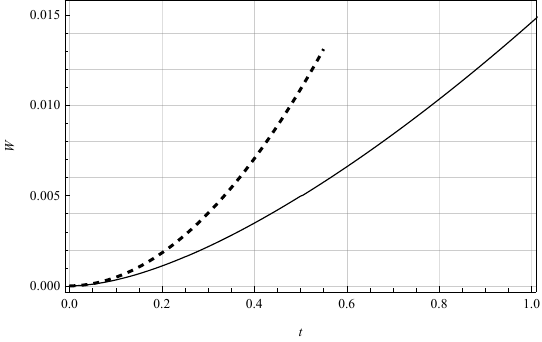}
				\label{W-sig-NP-Srsha}}
		\end{minipage}
\end{center}
\caption{Time evolution of stored energy per unit volume as a response to
strain acting as harmonic forcing, $\protect\varepsilon \left( t\right) =%
\protect\varepsilon _{0}\sin \left( \protect\omega t\right) $, in the case
when $\protect\phi _{\protect\sigma }$ has no zeros, obtained for model
parameters as in Table \protect\ref{parametri}.}
\label{energija-NP}
\end{figure}

In order to illustrate behavior of energy, dissipated power, and power per
unit volume in time, constitutive model I$^{+}$ID.ID, given by (\ref{model})
and having the model parameters as in Table \ref{parametri}, is selected.
Figures \ref{W-NP} and \ref{P-NP} present time profiles of energy and
dissipated power respectively, both of them being positive and oscillatory
functions in the case when $\phi _{\sigma }$ has no zeros. The positivity is
due to the fact that model parameters satisfy narrowed thermodynamical
restrictions, see (\ref{suzena-1}) and (\ref{suzena-2}), guaranteeing that
the relaxation modulus is completely monotonic and that creep compliance is
a Bernstein function, implying positivity of energy, expressed in both forms
(\ref{pot-en-epsilon}) and (\ref{pot-en-sigma}), as well as positivity of
the dissipated power, again in both forms (\ref{P-epsilon}) and (\ref%
{P-sigma}). Discrepancies between the energies, as well as between
dissipated powers, both given in terms of relaxation modulus and creep
compliance are quite small, with the increasing tendency in the case of
energies, however yielding the same time profiles of power, as illustrated
in Figure \ref{VelikoP-NP}. The asymptotic behavior of energies, which is,
according to the asymptotic expressions (\ref{W-srsha-epsilon}) and (\ref%
{W-srsha-sigma}) derived below, proportional to $t^{2-\left( \nu -\alpha
\right) },$ shows good agreement with the time evolution curves, see Figures %
\ref{W-eps-NP-Srsha} and \ref{W-sig-NP-Srsha}, as well as the asymptotic
behavior of dissipated powers, that is proportional to $t^{1-\left( \nu
-\alpha \right) }$ according to the asymptotic expressions (\ref%
{P-srsha-epsilon}) and (\ref{P-srsha-sigma}) below and illustrated in
Figures \ref{P-eps-NP-Srsha} and \ref{P-sig-NP-Srsha}. 
\begin{figure}[p]
\begin{center}
\begin{minipage}{0.5\columnwidth}
			\subfloat[Dissipated power expressed in terms of: strain - depicted by solid line and stress - depicted by dashed line.]{
			\includegraphics[width=\columnwidth]{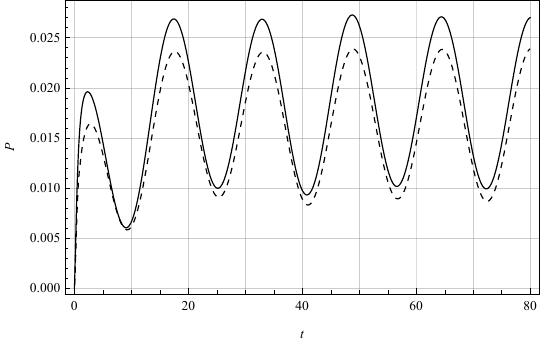}
			\label{P-NP}}
		\end{minipage}
\medskip \vfil
\begin{minipage}{0.46\columnwidth}
				\subfloat[Short time asymptotics of dissipated power expressed through strain, depicted by dashed line.]{
				\includegraphics[width=\columnwidth]{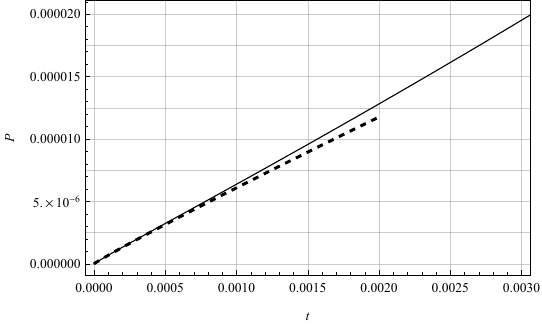}
				\label{P-eps-NP-Srsha}}
		\end{minipage}
\hfill 
\begin{minipage}{0.46\columnwidth}
				\subfloat[Short time asymptotics of dissipated power expressed through stress, depicted by dashed line.]{
				\includegraphics[width=\columnwidth]{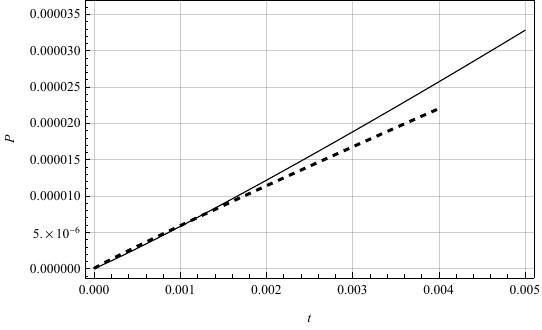}
				\label{P-sig-NP-Srsha}}
		\end{minipage}
\end{center}
\caption{Time evolution of dissipated power per unit volume as a response to
strain acting as harmonic forcing, $\protect\varepsilon \left( t\right) =%
\protect\varepsilon _{0}\sin \left( \protect\omega t\right) $, in the case
when $\protect\phi _{\protect\sigma }$ has no zeros, obtained for model
parameters as in Table \protect\ref{parametri}.}
\label{disipacija-NP}
\end{figure}
\begin{figure}[p]
\begin{center}
\includegraphics[width=0.6\columnwidth]{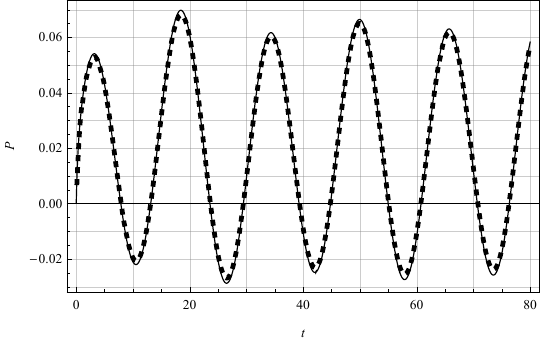}
\end{center}
\caption{Comparison of time evolutions of powers per unit volume expressed
in terms of strain - depicted by solid line and stress - depicted by dashed
line, that are responses to strain acting as harmonic forcing, $\protect%
\varepsilon \left( t\right) =\protect\varepsilon _{0}\sin \left( \protect%
\omega t\right) $, in the case when $\protect\phi _{\protect\sigma }$ has no
zeros, obtained for model parameters as in Table \protect\ref{parametri}.}
\label{VelikoP-NP}
\end{figure}

\begin{figure}[p]
\begin{center}
\includegraphics[width=0.6\columnwidth]{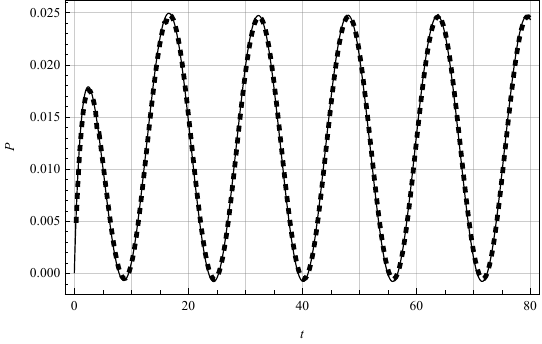}
\end{center}
\caption{Comparison of time evolutions of powers per unit volume expressed
in terms of strain - depicted by solid line and stress - depicted by dashed
line, that are responses to strain acting as harmonic forcing, $\protect%
\varepsilon \left( t\right) =\protect\varepsilon _{0}\sin \left( \protect%
\omega t\right)$, in the case when $\protect\phi _{\protect\sigma }$ has a
negative real zero, obtained for model parameters as in Table \protect\ref%
{parametri}.}
\label{VelikoP-RNP}
\end{figure}

Although model parameters, in the case when $\phi _{\sigma }$ has a negative
real zero, do not satisfy narrowed thermodynamical restrictions, see (\ref%
{suzena-1}) and (\ref{suzena-2}), both energies and dissipated powers are
positive and oscillatory functions, see Figure \ref{W-P-RNP}, illustrating
the fact that narrowed thermodynamical restrictions are sufficient, but not
necessary conditions for relaxation modulus to be completely monotonic and
for creep compliance to be a Bernstein function. Discrepancies in time
profiles of both energies and dissipated powers are more prominent than in
the previous case, with quite large discrepancy between energies, which
increases in time, compare Figures \ref{W-RNP} and \ref{P-RNP}. Time
profiles of power are the same, regardless whether energy and dissipated
power are expressed in terms of relaxation modulus or in terms of creep
compliance, see Figure \ref{VelikoP-RNP}. 
\begin{figure}[p]
\begin{center}
\begin{minipage}{0.46\columnwidth}
				\subfloat[Stored energy expressed in terms of: strain - depicted by solid line and stress - depicted by dashed line.]{
				\includegraphics[width=\columnwidth]{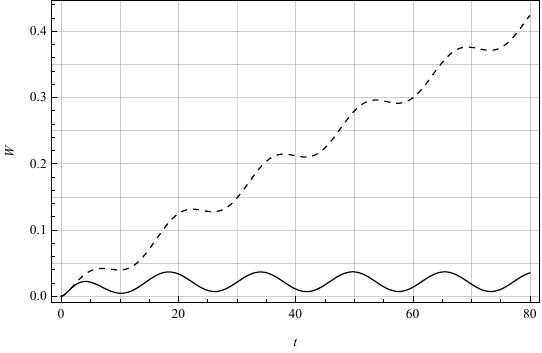}
				\label{W-RNP}}
		\end{minipage}
\hfill 
\begin{minipage}{0.46\columnwidth}
				\subfloat[Stored energy expressed in terms of strain.]{
				\includegraphics[width=\columnwidth]{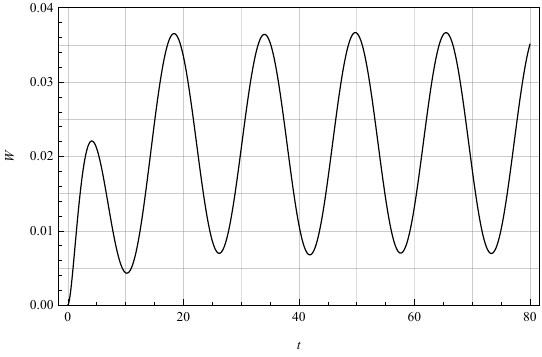}
				\label{W-eps-RNP}}
		\end{minipage}
\medskip \vfil
\begin{minipage}{0.5\columnwidth}
			\subfloat[Dissipated power expressed in terms of: strain - depicted by solid line and stress - depicted by dashed line.]{
			\includegraphics[width=\columnwidth]{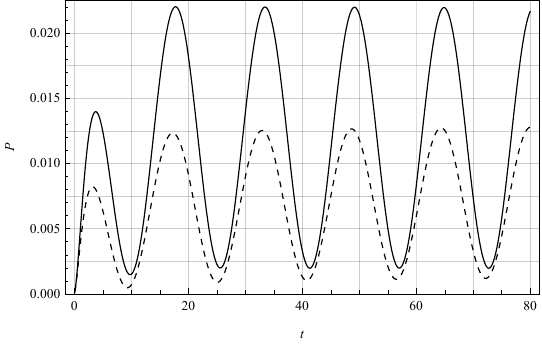}
			\label{P-RNP}}
		\end{minipage}
\end{center}
\caption{Time evolution of stored energy and dissipated power per unit
volume as a response to strain acting as harmonic forcing, $\protect%
\varepsilon \left( t\right) =\protect\varepsilon _{0}\sin \left( \protect%
\omega t\right)$, in the case when $\protect\phi _{\protect\sigma }$ has a
negative real zero, obtained for model parameters as in Table \protect\ref%
{parametri}.}
\label{W-P-RNP}
\end{figure}

Contrary to the two previous cases, if model parameters are such that
function $\phi _{\sigma }$ has a pair of complex conjugated zeros, the
energy expressed through the relaxation modulus is negative on certain time
intervals, so as the dissipated powers, see Figures \ref{W-eps-CCP} and \ref%
{P-CCP}, which is due to the fact that narrowed thermodynamical restrictions
are not satisfied, while the thermodynamical restrictions still apply.
Again, the energy expressed through the creep compliance is positive,
oscillatory and increasing in time, as seen from Figure \ref{W-CCP}.
Although potential energy is strictly non-negative for elastic bodies, the
negativity of potential energy is not uncommon in the mechanics of
particles, while in spite of the fact that the dissipated power is negative
on certain intervals, its integral from the initial time instant,
representing the total dissipated energy, is still positive. Note, the
intervals when power is negative, as well as its absolute values are the
smallest in this case when compared with previous two cases, with the
largest intervals and absolute values obtained in the case when function $%
\phi _{\sigma }$ has no zeros, compare Figure \ref{VelikoP-CCP} with Figures %
\ref{VelikoP-NP} and \ref{VelikoP-RNP}. 
\begin{figure}[p]
\begin{center}
\begin{minipage}{0.46\columnwidth}
				\subfloat[Stored energy expressed in terms of: strain - depicted by solid line and stress - depicted by dashed line.]{
				\includegraphics[width=\columnwidth]{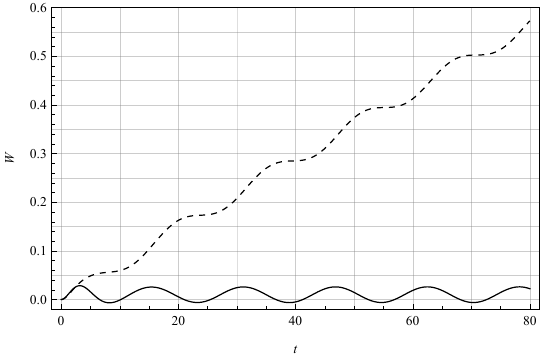}
				\label{W-CCP}}
		\end{minipage}
\hfill 
\begin{minipage}{0.46\columnwidth}
				\subfloat[Stored energy expressed in terms of strain.]{
				\includegraphics[width=\columnwidth]{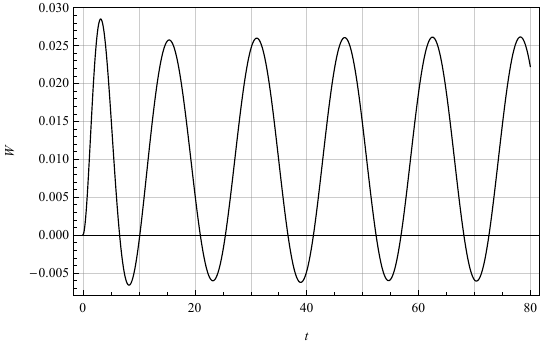}
				\label{W-eps-CCP}}
		\end{minipage}
\medskip \vfil
\begin{minipage}{0.5\columnwidth}
			\subfloat[Dissipated power expressed in terms of: strain - depicted by solid line and stress - depicted by dashed line.]{
			\includegraphics[width=\columnwidth]{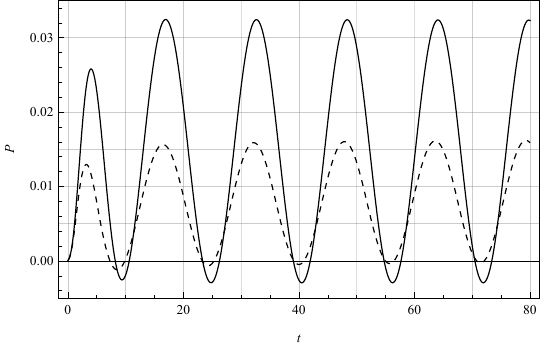}
			\label{P-CCP}}
		\end{minipage}
\end{center}
\caption{Time evolution of stored energy and dissipated power per unit
volume as a response to strain acting as harmonic forcing, $\protect%
\varepsilon \left( t\right) =\protect\varepsilon _{0}\sin \left( \protect%
\omega t\right)$, in the case when $\protect\phi _{\protect\sigma }$ has a
pair of complex conjugated zeros, obtained for model parameters as in Table 
\protect\ref{parametri}.}
\label{W-P-CCP}
\end{figure}
\begin{figure}[p]
\begin{center}
\includegraphics[width=0.6\columnwidth]{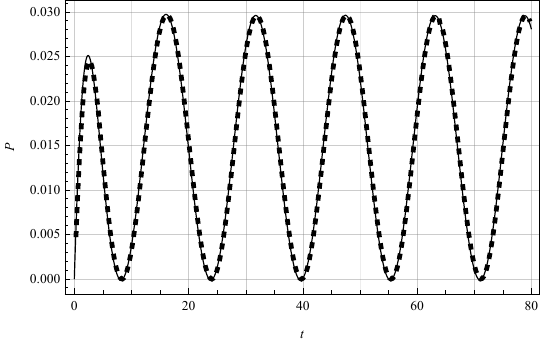}
\end{center}
\caption{Comparison of time evolutions of powers per unit volume expressed
in terms of strain - depicted by solid line and stress - depicted by dashed
line, that are responses to strain acting as harmonic forcing, $\protect%
\varepsilon \left( t\right) =\protect\varepsilon _{0}\sin \left( \protect%
\omega t\right) $, in the case when $\protect\phi _{\protect\sigma }$ has a
pair of complex conjugated zeros, obtained for model parameters as in Table 
\protect\ref{parametri}.}
\label{VelikoP-CCP}
\end{figure}

In order to derive the asymptotic expressions for the energy and dissipated
power per unit volume expressed in terms of relaxation modulus, according to
(\ref{pot-en-epsilon}) and (\ref{P-epsilon}) one needs the asymptotics of
relaxation modulus, given by (\ref{sr-opste}), that is derived in \cite{SD-2}
in the form%
\begin{align}
\sigma _{sr}\left( t\right) & =\frac{b_{2}}{a_{3}}\frac{t^{-\left( \nu
-\alpha \right) }}{\Gamma \left( 1-\left( \nu -\alpha \right) \right) }+%
\frac{b_{2}}{a_{3}}\left( \frac{b_{1}}{b_{2}}-\frac{a_{2}}{a_{3}}\right) 
\frac{t^{2\alpha +\beta -\nu }}{\Gamma \left( 1+2\alpha +\beta -\nu \right) }
\notag \\
& \quad +\frac{b_{2}}{a_{3}}\left( \left( \frac{a_{2}}{a_{3}}\right) ^{2}-%
\frac{a_{1}}{a_{3}}-\frac{a_{2}}{a_{3}}\frac{b_{1}}{b_{2}}\right) \frac{%
t^{3\alpha +2\beta -\nu }}{\Gamma \left( 1+3\alpha +2\beta -\nu \right) }%
+O\left( t^{-\left( 4\alpha +3\beta -\nu \right) }\right) ,\quad \text{when}%
\quad t\rightarrow 0,  \label{sr-sha}
\end{align}%
so that for the asymptotics of energy per unit volume, according to (\ref%
{pot-en-epsilon}) and leading term in (\ref{sr-sha}), one has%
\begin{align}
W\left( t\right) &=\frac{1}{2}\sigma _{sr}\left( t\right) \varepsilon
^{2}\left( t\right) +\frac{1}{2}\int_{0}^{t}\left( -\dot{\sigma}_{sr}\left(
t-t^{\prime }\right) \right) \left( \varepsilon \left( t\right) -\varepsilon
\left( t^{\prime }\right) \right) ^{2}\mathrm{d}t^{\prime }  \notag \\
&\sim \frac{1}{2}\frac{b_{2}}{a_{3}}\frac{t^{-\left( \nu -\alpha \right) }}{%
\Gamma \left( 1-\left( \nu -\alpha \right) \right) }\varepsilon
_{0}^{2}\omega ^{2}t^{2}-\frac{1}{2}\int_{0}^{t}\frac{b_{2}}{a_{3}}\frac{%
\left( t-t^{\prime }\right) ^{-1-\left( \nu -\alpha \right) }}{\Gamma \left(
-\left( \nu -\alpha \right) \right) }\varepsilon _{0}^{2}\omega ^{2}\left(
t-t^{\prime }\right) ^{2}\mathrm{d}t^{\prime }  \notag \\
&\sim \frac{1}{2}\varepsilon _{0}^{2}\omega ^{2}\frac{b_{2}}{a_{3}}\frac{1}{%
\left\vert \Gamma \left( -\left( \nu -\alpha \right) \right) \right\vert }%
\left( \frac{t^{2-\left( \nu -\alpha \right) }}{\nu -\alpha }%
+\int_{0}^{t}\left( t-t^{\prime }\right) ^{1-\left( \nu -\alpha \right) }%
\mathrm{d}t^{\prime }\right)  \notag \\
&\sim \frac{1}{2}\varepsilon _{0}^{2}\omega ^{2}\frac{b_{2}}{a_{3}}\frac{%
t^{2-\left( \nu -\alpha \right) }}{\left\vert \Gamma \left( -\left( \nu
-\alpha \right) \right) \right\vert }\left( \frac{1}{\nu -\alpha }+\frac{1}{%
2-\left( \nu -\alpha \right) }\right)  \notag \\
&\sim \left( 1-\left( \nu -\alpha \right) \right) \frac{b_{2}}{a_{3}}%
\varepsilon _{0}^{2}\omega ^{2}\frac{t^{2-\left( \nu -\alpha \right) }}{%
\Gamma \left( 3-\left( \nu -\alpha \right) \right) },\quad \text{when}\quad
t\rightarrow 0,  \label{W-srsha-epsilon}
\end{align}%
where $\sin x\sim x$, when $x\rightarrow 0$ is used, while for the
asymptotics of dissipated power per unit volume, according to (\ref%
{P-epsilon}) and leading term in (\ref{sr-sha}), one has%
\begin{align}
\mathcal{P}\left( t\right) &=\frac{1}{2}\left( -\dot{\sigma}_{sr}\left(
t\right) \right) \varepsilon ^{2}\left( t\right) +\frac{1}{2}\int_{0}^{t}%
\ddot{\sigma}_{sr}\left( t-t^{\prime }\right) \left( \varepsilon \left(
t\right) -\varepsilon \left( t^{\prime }\right) \right) ^{2}\mathrm{d}%
t^{\prime }  \notag \\
&\sim \frac{1}{2}\frac{b_{2}}{a_{3}}\frac{t^{-1-\left( \nu -\alpha \right) }%
}{\left\vert \Gamma \left( -\left( \nu -\alpha \right) \right) \right\vert }%
\varepsilon _{0}^{2}\omega ^{2}t^{2}+\frac{1}{2}\int_{0}^{t}\frac{b_{2}}{%
a_{3}}\frac{\left( t-t^{\prime }\right) ^{-2-\left( \nu -\alpha \right) }}{%
\Gamma \left( -1-\left( \nu -\alpha \right) \right) }\varepsilon
_{0}^{2}\omega ^{2}\left( t-t^{\prime }\right) ^{2}\mathrm{d}t^{\prime } 
\notag \\
&\sim \frac{1}{2}\varepsilon _{0}^{2}\omega ^{2}\frac{b_{2}}{a_{3}}\frac{1}{%
\Gamma \left( -1-\left( \nu -\alpha \right) \right) }\left( -\frac{%
t^{1-\left( \nu -\alpha \right) }}{-1-\left( \nu -\alpha \right) }%
+\int_{0}^{t}\left( t-t^{\prime }\right) ^{-\left( \nu -\alpha \right) }%
\mathrm{d}t^{\prime }\right)  \notag \\
&\sim \frac{1}{2}\varepsilon _{0}^{2}\omega ^{2}\frac{b_{2}}{a_{3}}\frac{%
t^{1-\left( \nu -\alpha \right) }}{\Gamma \left( -1-\left( \nu -\alpha
\right) \right) }\left( -\frac{1}{-1-\left( \nu -\alpha \right) }+\frac{1}{%
1-\left( \nu -\alpha \right) }\right)  \notag \\
&\sim \left( \nu -\alpha \right) \varepsilon _{0}^{2}\omega ^{2}\frac{b_{2}}{%
a_{3}}\frac{t^{1-\left( \nu -\alpha \right) }}{\Gamma \left( 2-\left( \nu
-\alpha \right) \right) },\quad \text{when}\quad t\rightarrow 0.
\label{P-srsha-epsilon}
\end{align}

On the other hand, the derivation of asymptotics of energy and dissipated
power per unit volume expressed in terms of creep compliance, according to (%
\ref{pot-en-sigma}) and (\ref{P-sigma}), requires the asymptotics of creep
compliance, given by (\ref{cr-opste}), that is derived in \cite{SD-2} in the
form%
\begin{align}
\varepsilon _{cr}\left( t\right) & =\frac{a_{3}}{b_{2}}\frac{t^{\nu -\alpha }%
}{\Gamma \left( 1+\nu -\alpha \right) }+\frac{a_{3}}{b_{2}}\left( \frac{a_{2}%
}{a_{3}}-\frac{b_{1}}{b_{2}}\right) \frac{t^{\beta +\nu }}{\Gamma \left(
1+\beta +\nu \right) }  \notag \\
& \quad +\frac{a_{3}}{b_{2}}\left( \frac{a_{1}}{a_{3}}-\frac{a_{2}}{a_{3}}%
\frac{b_{1}}{b_{2}}+\left( \frac{b_{1}}{b_{2}}\right) ^{2}\right) \frac{%
t^{\alpha +2\beta +\nu }}{\Gamma \left( 1+\alpha +2\beta +\nu \right) }%
+O\left( t^{2\alpha +3\beta +\nu }\right) ,\quad \text{when}\quad
t\rightarrow 0,  \label{cr-sha}
\end{align}%
as well as the asymptotics of stress, taking the form%
\begin{align}
\sigma \left( t\right) &=\varepsilon _{0}\omega \frac{b_{2}}{a_{3}}\frac{%
t^{1-\left( \nu -\alpha \right) }}{\Gamma \left( 2-\left( \nu -\alpha
\right) \right) }+\varepsilon _{0}\omega \frac{b_{2}}{a_{3}}\left( \frac{%
b_{1}}{b_{2}}-\frac{a_{2}}{a_{3}}\right) \frac{t^{1+2\alpha +\beta -\nu }}{%
\Gamma \left( 2+2\alpha +\beta -\nu \right) }  \notag \\
& \quad+\varepsilon _{0}\omega \frac{b_{2}}{a_{3}}\left( \frac{a_{2}^{2}}{%
a_{3}^{2}}-\frac{a_{2}}{a_{3}}\frac{b_{1}}{b_{2}}-\frac{a_{1}}{a_{3}}\right) 
\frac{t^{1+3\alpha +2\beta -\nu }}{\Gamma \left( 2+3\alpha +2\beta -\nu
\right) } -\varepsilon _{0}\omega ^{3}\frac{b_{2}}{a_{3}}\frac{t^{3-\left(
\nu -\alpha \right) }}{\Gamma \left( 4-\left( \nu -\alpha \right) \right) }%
+O\left( t^{1+4\alpha +3\beta -\nu }\right),
\label{asympt-sigma-time-domain}
\end{align}%
when $t\rightarrow 0$, where the stress represents a response to harmonic
strain (\ref{HE-sin}) and it is given by (\ref{resp-to-HE-but-sin}). Hence,
the asymptotics of energy per unit volume, according to (\ref{pot-en-sigma})
and leading terms in (\ref{cr-sha}) and (\ref{asympt-sigma-time-domain}), is
obtained as%
\begin{align}
W\left( t\right) &=\frac{1}{2}\int_{0}^{t}\dot{\varepsilon}_{cr}\left(
t-t^{\prime }\right) \sigma ^{2}\left( t^{\prime }\right) \mathrm{d}%
t^{\prime }  \notag \\
&\sim \frac{1}{2}\int_{0}^{t}\frac{a_{3}}{b_{2}}\frac{\left( t-t^{\prime
}\right) ^{-1+\left( \nu -\alpha \right) }}{\Gamma \left( \nu -\alpha
\right) }\varepsilon _{0}^{2}\omega ^{2}\frac{b_{2}^{2}}{a_{3}^{2}}\frac{%
t^{\prime 2-2\left( \nu -\alpha \right) }}{\Gamma ^{2}\left( 2-\left( \nu
-\alpha \right) \right) }\mathrm{d}t^{\prime }  \notag \\
&\sim\frac{1}{2}\varepsilon _{0}^{2}\omega ^{2}\frac{b_{2}}{a_{3}}\frac{1}{%
\Gamma \left( \nu -\alpha \right) \Gamma ^{2}\left( 2-\left( \nu -\alpha
\right) \right) }\int_{0}^{t}\frac{t^{\prime 2-2\left( \nu -\alpha \right) }%
}{\left( t-t^{\prime }\right) ^{1-\left( \nu -\alpha \right) }}\mathrm{d}%
t^{\prime }  \notag \\
&\sim \frac{1}{2}\varepsilon _{0}^{2}\omega ^{2}\frac{b_{2}}{a_{3}}\frac{%
\Gamma \left( 3-2\left( \nu -\alpha \right) \right) }{\Gamma ^{2}\left(
2-\left( \nu -\alpha \right) \right) \Gamma \left( 3-\left( \nu -\alpha
\right) \right) }t^{2-\left( \nu -\alpha \right) },\quad \text{when}\quad
t\rightarrow 0,  \label{W-srsha-sigma}
\end{align}%
while the asymptotics of dissipated power per unit volume, according to (\ref%
{P-sigma}) and leading terms in (\ref{cr-sha}) and (\ref%
{asympt-sigma-time-domain}), is obtained as%
\begin{align}
\mathcal{P}\left( t\right) &=\frac{1}{2}\dot{\varepsilon}_{cr}\left(
t\right) \sigma ^{2}\left( t\right) +\frac{1}{2}\int_{0}^{t}\left( -\ddot{%
\varepsilon}_{cr}\left( t-t^{\prime }\right) \right) \left( \sigma \left(
t\right) -\sigma \left( t^{\prime }\right) \right) ^{2}\mathrm{d}t^{\prime }
\notag \\
&\sim \frac{1}{2}\frac{a_{3}}{b_{2}}\frac{t^{-1+\left( \nu -\alpha \right) }%
}{\Gamma \left( \nu -\alpha \right) }\varepsilon _{0}^{2}\omega ^{2}\frac{%
b_{2}^{2}}{a_{3}^{2}}\frac{t^{2-2\left( \nu -\alpha \right) }}{\Gamma
^{2}\left( 2-\left( \nu -\alpha \right) \right) }-\frac{1}{2}\int_{0}^{t}%
\frac{a_{3}}{b_{2}}\frac{\left( t-t^{\prime }\right) ^{-2+\left( \nu -\alpha
\right) }}{\Gamma \left( -1+\nu -\alpha \right) }\varepsilon _{0}^{2}\omega
^{2}\frac{b_{2}^{2}}{a_{3}^{2}}\frac{\left( t^{1-\left( \nu -\alpha \right)
}-t^{\prime 1-\left( \nu -\alpha \right) }\right) ^{2}}{\Gamma ^{2}\left(
2-\left( \nu -\alpha \right) \right) }\mathrm{d}t^{\prime }  \notag \\
&\sim \frac{1}{2}\varepsilon _{0}^{2}\omega ^{2}\frac{b_{2}}{a_{3}}\frac{1}{%
\Gamma \left( -1+\nu -\alpha \right) \Gamma ^{2}\left( 2-\left( \nu -\alpha
\right) \right) }  \notag \\
&\quad \quad \times \left( \frac{t^{1-\left( \nu -\alpha \right) }}{-1+\nu
-\alpha }-\int_{0}^{t}\left( t-t^{\prime }\right) ^{-2+\left( \nu -\alpha
\right) }\left( t^{1-\left( \nu -\alpha \right) }-t^{\prime 1-\left( \nu
-\alpha \right) }\right) ^{2}\mathrm{d}t^{\prime }\right)  \notag \\
&\sim \frac{1}{2}\varepsilon _{0}^{2}\omega ^{2}\frac{b_{2}}{a_{3}}\left( 
\frac{1}{-1+\nu -\alpha }-\frac{\pi }{\sin \left( \nu -\alpha \right) \pi }%
\left( 2-\frac{\Gamma \left( 3-2\left( \nu -\alpha \right) \right) }{\Gamma
^{2}\left( 2-\left( \nu -\alpha \right) \right) }-\frac{1}{\Gamma \left( \nu
-\alpha \right) \Gamma \left( 2-\left( \nu -\alpha \right) \right) }\right)
\right)  \notag \\
&\quad \quad \times \frac{t^{1-\left( \nu -\alpha \right) }}{\Gamma \left(
-1+\left( \nu -\alpha \right) \right)\Gamma ^{2}\left( 2-\left( \nu -\alpha
\right) \right) },\quad \text{when}\quad t\rightarrow 0.
\label{P-srsha-sigma}
\end{align}

Note, the asymptotics of stress in the form (\ref{asympt-sigma-time-domain})
is obtained by inverting the asymptotics of stress in the Laplace domain,
see (\ref{sigma-sr-ld-sin}), that for the model I$^{+}$ID.ID, given by (\ref%
{model}), takes the form%
\begin{align}
\tilde{\sigma}\left( s\right) &=\varepsilon _{0}s^{\beta +\nu }\frac{%
b_{1}+b_{2}s^{\alpha +\beta }}{a_{1}+a_{2}s^{\alpha +\beta }+a_{3}s^{2\left(
\alpha +\beta \right) }}\frac{\omega }{s^{2}+\omega ^{2}}  \notag \\
&=\varepsilon _{0}\omega \frac{b_{2}}{a_{3}}\frac{1}{s^{2-\left( \nu -\alpha
\right) }}\left( 1+\frac{b_{1}}{b_{2}}\frac{1}{s^{\alpha +\beta }}\right)
\left( 1+\frac{a_{2}}{a_{3}}\frac{1}{s^{\alpha +\beta }}+\frac{a_{1}}{a_{3}}%
\frac{1}{s^{2\left( \alpha +\beta \right) }}\right) ^{-1}\left( 1+\frac{%
\omega ^{2}}{s^{2}}\right) ^{-1}  \notag \\
&=\varepsilon _{0}\omega \frac{b_{2}}{a_{3}}\frac{1}{s^{2-\left( \nu -\alpha
\right) }}\left( 1+\frac{b_{1}}{b_{2}}\frac{1}{s^{\alpha +\beta }}\right) 
\notag \\
&\quad \quad \times \left( 1-\frac{a_{2}}{a_{3}}\frac{1}{s^{\alpha +\beta }}%
+\left( \frac{a_{2}^{2}}{a_{3}^{2}}-\frac{a_{1}}{a_{3}}\right) \frac{1}{%
s^{2\left( \alpha +\beta \right) }}+O\left( s^{-3\left( \alpha +\beta
\right) }\right) \right) \left( 1-\frac{\omega ^{2}}{s^{2}}+O\left(
s^{-4}\right) \right)  \notag \\
&=\varepsilon _{0}\omega \frac{b_{2}}{a_{3}}\frac{1}{s^{2-\left( \nu -\alpha
\right) }}\left( 1+\left( \frac{b_{1}}{b_{2}}-\frac{a_{2}}{a_{3}}\right) 
\frac{1}{s^{\alpha +\beta }}+\left( \frac{a_{2}^{2}}{a_{3}^{2}}-\frac{a_{1}}{%
a_{3}}-\frac{a_{2}}{a_{3}}\frac{b_{1}}{b_{2}}\right) \frac{1}{s^{2\left(
\alpha +\beta \right) }}-\frac{\omega ^{2}}{s^{2}}+O\left( s^{-3\left(
\alpha +\beta \right) }\right) \right)  \notag \\
&=\varepsilon _{0}\omega \frac{b_{2}}{a_{3}}\frac{1}{s^{2-\left( \nu -\alpha
\right) }}+\varepsilon _{0}\omega \frac{b_{2}}{a_{3}}\left( \frac{b_{1}}{%
b_{2}}-\frac{a_{2}}{a_{3}}\right) \frac{1}{s^{2+2\alpha +\beta -\nu }} 
\notag \\
&\quad \quad +\varepsilon _{0}\omega \frac{b_{2}}{a_{3}}\left( \frac{%
a_{2}^{2}}{a_{3}^{2}}-\frac{a_{2}}{a_{3}}\frac{b_{1}}{b_{2}}-\frac{a_{1}}{%
a_{3}}\right) \frac{1}{s^{2+3\alpha +2\beta -\nu }}-\varepsilon _{0}\omega
^{3}\frac{b_{2}}{a_{3}}\frac{1}{s^{4-\left( \nu -\alpha \right) }}+O\left(
s^{-2-4\alpha -3\beta +\nu }\right) ,\quad \text{when}\quad s\rightarrow
\infty .  \label{asympt-sigma-ld}
\end{align}%
Terms $-\varepsilon _{0}\omega ^{3}\frac{b_{2}}{a_{3}}\frac{1}{s^{4-\left(
\beta +\nu \right) }}$ and $-\varepsilon _{0}\omega ^{3}\frac{b_{2}}{a_{3}}%
\frac{t^{3-\left( \beta +\nu \right) }}{\Gamma \left( 4-\left( \beta +\nu
\right) \right) },$ respectively appearing in (\ref{asympt-sigma-ld}) and (%
\ref{asympt-sigma-time-domain}), do not exist if $3\left( \alpha +\beta
\right) \leqslant 2$, i.e., if $\alpha +\beta \leqslant \frac{2}{3}$.

\section{Conclusion}

Considering the arbitrary fractional anti-Zener and/or Zener model listed in
Appendix \ref{FAZ-ZM}, the time evolution of stress, representing transient
response to a strain prescribed as a harmonic excitation through the cosine
function (\ref{Harmeks}), is derived in Section \ref{TRandSSR} in the form (%
\ref{resp-to-HE}) constituted either by functions $\sigma ^{\left( \mathrm{V}%
\right) }$ and $\sigma ^{\left( \mathrm{H}\right) },$ or by either $\sigma
^{\left( \mathrm{RP}\right) }$ or $\sigma ^{\left( \mathrm{CCP}\right) }$ in
addition to the previously mentioned functions. Functions $\sigma ^{\left( 
\mathrm{V}\right) }$, $\sigma ^{\left( \mathrm{H}\right) }$, $\sigma
^{\left( \mathrm{RP}\right) }$, and $\sigma ^{\left( \mathrm{CCP}\right) },$
given by (\ref{sigma-V1}) - (\ref{sigma-ccp}), display different qualitative
properties: the first one is non-oscillatory decreasing function, possibly
also non-monotonic, the second one is an oscillatory function having the
angular frequency of the excitation, the third one is either a positive
exponentially decreasing function or a negative function exponentially
increasing to zero, while the fourth one is a damped oscillatory function,
with angular frequency determined by the imaginary part of zero of function $%
\phi _{\sigma }$ and damping constant determined by its real part. Also, in
Section \ref{TRandSSR} the stress, as a harmonic function (\ref%
{sigma-isto-kao-sigma-h}), is obtained in the steady state regime as a
consequence of a strain assumed in the form (\ref{fazori})$_{1}$ and it is
noted that the stress as a response in the steady state regime in the form (%
\ref{sigma-isto-kao-sigma-h}) coincides with the function $\sigma ^{\left( 
\mathrm{H}\right) },$ obtained in the transient regime in the form (\ref%
{sigma-H}), being the only term not tending to zero in the expression for
stress (\ref{resp-to-HE}).

Time evolution profiles of transient responses to a cosine-type excitation
are presented in Figures \ref{harmonici-NP}, \ref{harmonici-RP}, and \ref%
{harmonici-CCP}, along with the steady state profiles and corresponding
short time asymptotics, for the thermodynamically consistent fractional
anti-Zener and Zener model I$^{+}$ID.ID, see (\ref{model}), in the cases
when function $\phi _{\sigma }$ has no zeros, has a negative real zero, and
has a pair of complex conjugated zeros, respectively, illustrating the
establishment of the steady state regime due to the dissipation properties
of material described by the model I$^{+}$ID.ID. Note, the form of function $%
\phi _{\sigma }$ depends on the chosen model and forms of function $\phi
_{\sigma },$ along with the forms of function $\phi _{\varepsilon },$ are
listed in Table \ref{skupina} for all anti-Zener and Zener models.

In Section \ref{WiP}, the time evolutions of powers per unit volume are
analyzed in Figures \ref{VelikoP-NP}, \ref{VelikoP-RNP}, and \ref%
{VelikoP-CCP} and the equivalence between expression (\ref{dW+P}), with
energy and dissipated power expressed by (\ref{pot-en-epsilon}) and (\ref%
{P-epsilon}) through the relaxation modulus, and expression (\ref{dW+P}),
with energy and dissipated power expressed by (\ref{pot-en-sigma}) and (\ref%
{P-sigma}) through the creep compliance, is examined and confirmed, where
the relaxation modulus, given by (\ref{sr-opste}) is constituted either of
function $\sigma _{sr}^{\left( \mathrm{NP}\right) },$ or having an
additional term either $\sigma _{sr}^{\left( \mathrm{RP}\right) },$ or $%
\sigma _{sr}^{\left( \mathrm{CCP}\right) },$ see (\ref{sigma-NP}) - (\ref%
{sigma-CCP}), while the creep compliance, given by (\ref{cr-opste}) is
constituted either of function $\varepsilon _{cr}^{\left( \mathrm{NP}\right)
},$ or having an additional term either $\varepsilon _{cr}^{\left( \mathrm{RP%
}\right) },$ or $\varepsilon _{cr}^{\left( \mathrm{CCP}\right) },$ see (\ref%
{epsilon-NP}) - (\ref{epsilon-CCP}). Since the stored energy and dissipated
power in expressions (\ref{pot-en-epsilon}) and (\ref{P-epsilon}) are given
in terms of arbitrary strain, the previously mentioned figures are produced
for strain assumed as a sine function, see (\ref{HE-sin}), while the curves
for comparison are obtained according to the expressions (\ref{pot-en-sigma}%
) and (\ref{P-sigma}), where the energy and dissipated power are given in
terms of stress, which is a transient response to strain assumed as a sine
function (\ref{HE-sin}) and obtained in the form (\ref{resp-to-HE-but-sin}),
constituted either by functions $\sigma ^{\left( \mathrm{v}\right) }$ and $%
\sigma ^{\left( \mathrm{h}\right) },$ or by either $\sigma ^{\left( \mathrm{%
rp}\right) }$ or $\sigma ^{\left( \mathrm{ccp}\right) }$ in addition to the
previously mentioned functions, see (\ref{sigma-V1-but-sin}) - (\ref%
{sigma-ccp-but-sin}).

Assuming the strain as a sine function (\ref{HE-sin}) and stress in the form
(\ref{resp-to-HE-but-sin}) as a response to it, the equivalence between two
forms of energy, obtained by (\ref{pot-en-epsilon}) and (\ref{pot-en-sigma}%
), as well as of the dissipated power, obtained by (\ref{P-epsilon}) and (%
\ref{P-sigma}), is examined by comparing time evolution profiles in Figures %
\ref{energija-NP} and \ref{disipacija-NP} in the case when $\phi _{\sigma }$
has no zeros, as well as in Figures \ref{W-P-RNP} and \ref{W-P-CCP} in the
case when $\phi _{\sigma }$ has a negative real zero and when $\phi _{\sigma
}$ has a pair of complex conjugated zeros, and the discrepancies between
time profiles are found. The law of change of total mechanical energy (\ref%
{P+d(W+T)}) is used in order to differ the stored energy from the
energy-like expression, as well as the dissipated power from the dissipated
power-like expression and it is concluded, according to (\ref{LCTME-preko-u}%
), that the expressions (\ref{pot-en-epsilon}) and (\ref{P-epsilon})
correspond to the stored energy and dissipated power, while, according to (%
\ref{LCTME-preko-mix}), the expressions (\ref{pot-en-sigma}) and (\ref%
{P-sigma}) are energy-like and dissipated power-like expressions.
Nevertheless, it is clear that the power per unit volume of viscoelastic
body is the same whether expressed through the relaxation modulus or through
the creep compliance, since the agreement between time evolution curves is
obtained in Figures \ref{VelikoP-NP}, \ref{VelikoP-RNP}, and \ref%
{VelikoP-CCP}.

\section*{Acknowledgements}

\noindent The work is supported by the Ministry of Science, Technological
Development and Innovation of the Republic of Serbia under grant
451-03-47/2023-01/200125 (SJ and DZ).

\section*{ORCID}

\noindent Sla\dj an Jeli\'{c} https://orcid.org/0000-0002-0877-6781

\noindent Du\v{s}an Zorica https://orcid.org/0000-0002-9117-8589


\appendix

\section{Fractional anti-Zener and Zener models\label{FAZ-ZM}}

Fractional anti-Zener and Zener models, quoted in the sequel, are derived
and checked for thermodynamical consistency in \cite{SD-1} while the
narrowed thermodynamical restrictions on model parameters are posed in \cite%
{SD-2}.

\subsection{Symmetric models}

\textbf{Model ID.ID} takes the form%
\begin{equation*}
\left( a_{1}\,_{0}\mathrm{I}_{t}^{\alpha }+a_{2}\,_{0}\mathrm{D}_{t}^{\beta
}\right) \sigma \left( t\right) =\left( b_{1}\,_{0}\mathrm{I}_{t}^{\mu
}+a_{2}\,_{0}\mathrm{D}_{t}^{\alpha +\beta -\mu }\right) \varepsilon \left(
t\right) ,
\end{equation*}%
with narrowed thermodynamical restrictions%
\begin{gather*}
0\leqslant \alpha +\beta -\mu \leqslant 1,\quad \mu \leqslant \alpha ,\quad
\beta +\mu \leqslant 1, \\
-\frac{a_{1}}{a_{2}}\frac{\cos \frac{\left( 2\alpha +\beta -\mu \right) \pi 
}{2}}{\cos \frac{\left( \beta +\mu \right) \pi }{2}}\leqslant \frac{b_{1}}{%
b_{2}}\leqslant \frac{a_{1}}{a_{2}}\frac{\sin \frac{\left( 2\alpha +\beta
-\mu \right) \pi }{2}}{\sin \frac{\left( \beta +\mu \right) \pi }{2}}\frac{%
\cos \frac{\left( 2\alpha +\beta -\mu \right) \pi }{2}}{\cos \frac{\left(
\beta +\mu \right) \pi }{2}}\leqslant \frac{a_{1}}{a_{2}}\frac{\sin \frac{%
\left( 2\alpha +\beta -\mu \right) \pi }{2}}{\sin \frac{\left( \beta +\mu
\right) \pi }{2}}.
\end{gather*}

\noindent \textbf{Model ID.DD}$^{{}^{+}}$\noindent \textbf{\ }takes the form%
\begin{equation*}
\left( a_{1}\,_{0}\mathrm{I}_{t}^{\alpha }+a_{2}\,_{0}\mathrm{D}_{t}^{\beta
}\right) \sigma \left( t\right) =\left( b_{1}\,_{0}\mathrm{D}_{t}^{\mu
}+b_{2}\,_{0}\mathrm{D}_{t}^{\alpha +\beta +\mu }\right) \varepsilon \left(
t\right) ,
\end{equation*}%
with narrowed thermodynamical restrictions%
\begin{gather*}
1\leqslant \alpha +\beta +\mu \leqslant 2,\quad \beta \leqslant \mu
\leqslant 1-\alpha , \\
\frac{a_{1}}{a_{2}}\frac{\left\vert \cos \frac{\left( 2\alpha +\beta +\mu
\right) \pi }{2}\right\vert }{\cos \frac{\left( \mu -\beta \right) \pi }{2}}%
\leqslant \frac{a_{1}}{a_{2}}\frac{\left\vert \cos \frac{\left( 2\alpha
+\beta +\mu \right) \pi }{2}\right\vert }{\cos \frac{\left( \mu -\beta
\right) \pi }{2}}\frac{\sin \frac{\left( 2\alpha +\beta +\mu \right) \pi }{2}%
}{\sin \frac{\left( \mu -\beta \right) \pi }{2}}\leqslant \frac{b_{1}}{b_{2}}%
.
\end{gather*}

\noindent \textbf{Model IID.IID} takes the form%
\begin{equation*}
\left( a_{1}\,_{0}\mathrm{I}_{t}^{\alpha }+a_{2}\,_{0}\mathrm{I}_{t}^{\beta
}+a_{3}\,_{0}\mathrm{D}_{t}^{\gamma }\right) \sigma \left( t\right) =\left(
b_{1}\,_{0}\mathrm{I}_{t}^{\alpha +\gamma -\eta }+b_{2}\,_{0}\mathrm{I}%
_{t}^{\beta +\gamma -\eta }+b_{3}\,_{0}\mathrm{D}_{t}^{\eta }\right)
\varepsilon \left( t\right) ,
\end{equation*}%
with narrowed thermodynamical restrictions%
\begin{gather*}
\beta <\alpha ,\quad \gamma \leqslant \eta ,\quad 0\leqslant \beta +\gamma
-\eta \leqslant \alpha +2\gamma -\eta \leqslant 1,\quad \alpha +\gamma
\leqslant \beta +\eta , \\
\frac{a_{3}}{a_{1}}\leqslant \frac{b_{3}}{b_{1}}\frac{\sin \frac{\left(
\alpha +\eta \right) \pi }{2}}{\sin \frac{\left( \alpha +2\gamma -\eta
\right) \pi }{2}}\frac{\cos \frac{\left( \alpha +\eta \right) \pi }{2}}{\cos 
\frac{\left( \alpha +2\gamma -\eta \right) \pi }{2}}\leqslant \frac{b_{3}}{%
b_{1}}\frac{\sin \frac{\left( \alpha +\eta \right) \pi }{2}}{\sin \frac{%
\left( \alpha +2\gamma -\eta \right) \pi }{2}}, \\
\frac{a_{3}}{a_{2}}\leqslant \frac{b_{3}}{b_{2}}\frac{\sin \frac{\left(
\beta +\eta \right) \pi }{2}}{\sin \frac{\left( \beta +2\gamma -\eta \right)
\pi }{2}}\frac{\cos \frac{\left( \beta +\eta \right) \pi }{2}}{\cos \frac{%
\left( \beta +2\gamma -\eta \right) \pi }{2}}\leqslant \frac{b_{3}}{b_{2}}%
\frac{\sin \frac{\left( \beta +\eta \right) \pi }{2}}{\sin \frac{\left(
\beta +2\gamma -\eta \right) \pi }{2}},
\end{gather*}%
valid if $\beta +\eta <\alpha +\eta <1.$ Otherwise, the narrowed
thermodynamical restrictions cannot be guaranteed and constraints on model
parameters are given by the thermodynamical restrictions%
\begin{gather*}
\beta <\alpha ,\quad \gamma \leqslant \eta ,\quad 0\leqslant \beta +\gamma
-\eta \leqslant \alpha +2\gamma -\eta \leqslant 1,\quad \alpha +\gamma
\leqslant \beta +\eta , \\
\frac{b_{3}}{b_{1}}\frac{\left\vert \cos \frac{\left( \alpha +\eta \right)
\pi }{2}\right\vert }{\cos \frac{\left( \alpha +2\gamma -\eta \right) \pi }{2%
}}\leqslant \frac{a_{3}}{a_{1}}\leqslant \frac{b_{3}}{b_{1}}\frac{\sin \frac{%
\left( \alpha +\eta \right) \pi }{2}}{\sin \frac{\left( \alpha +2\gamma
-\eta \right) \pi }{2}}, \\
\frac{b_{3}}{b_{2}}\frac{\left\vert \cos \frac{\left( \beta +\eta \right)
\pi }{2}\right\vert }{\cos \frac{\left( \beta +2\gamma -\eta \right) \pi }{2}%
}\leqslant \frac{a_{3}}{a_{2}}\leqslant \frac{b_{3}}{b_{2}}\frac{\sin \frac{%
\left( \beta +\eta \right) \pi }{2}}{\sin \frac{\left( \beta +2\gamma -\eta
\right) \pi }{2}}.
\end{gather*}

\noindent \textbf{Model IDD.IDD} takes the form%
\begin{equation*}
\left( a_{1}\,_{0}\mathrm{I}_{t}^{\alpha }+a_{2}\,_{0}\mathrm{D}_{t}^{\beta
}+a_{3}\,_{0}\mathrm{D}_{t}^{\gamma }\right) \sigma \left( t\right) =\left(
b_{1}\,_{0}\mathrm{I}_{t}^{\mu }+b_{2}\,_{0}\mathrm{D}_{t}^{\alpha +\beta
-\mu }+b_{3}\,_{0}\mathrm{D}_{t}^{\alpha +\gamma -\mu }\right) \varepsilon
\left( t\right) ,
\end{equation*}%
with narrowed thermodynamical restrictions%
\begin{gather*}
0\leqslant \alpha +\gamma -\mu \leqslant 1,\quad \beta <\gamma ,\quad \mu
\leqslant \alpha ,\quad \gamma +\mu \leqslant \alpha +\beta ,\quad \gamma
+\mu \leqslant 1, \\
\frac{a_{2}}{a_{1}}\leqslant \frac{b_{2}}{b_{1}}\frac{\sin \frac{\left(
2\alpha +\beta -\mu \right) \pi }{2}}{\sin \frac{\left( \beta +\mu \right)
\pi }{2}}\frac{\cos \frac{\left( 2\alpha +\beta -\mu \right) \pi }{2}}{\cos 
\frac{\left( \beta +\mu \right) \pi }{2}}\leqslant \frac{b_{2}}{b_{1}}\frac{%
\sin \frac{\left( 2\alpha +\beta -\mu \right) \pi }{2}}{\sin \frac{\left(
\beta +\mu \right) \pi }{2}}, \\
\frac{a_{3}}{a_{1}}\leqslant \frac{b_{3}}{b_{1}}\frac{\sin \frac{\left(
2\alpha +\gamma -\mu \right) \pi }{2}}{\sin \frac{\left( \gamma +\mu \right)
\pi }{2}}\frac{\cos \frac{\left( 2\alpha +\gamma -\mu \right) \pi }{2}}{\cos 
\frac{\left( \gamma +\mu \right) \pi }{2}}\leqslant \frac{b_{3}}{b_{1}}\frac{%
\sin \frac{\left( 2\alpha +\gamma -\mu \right) \pi }{2}}{\sin \frac{\left(
\gamma +\mu \right) \pi }{2}},
\end{gather*}%
valid if $2\alpha +\beta -\mu <2\alpha +\gamma -\mu <1.$ Otherwise, the
narrowed thermodynamical restrictions cannot be guaranteed and constraints
on model parameters are given by the thermodynamical restrictions%
\begin{gather*}
0\leqslant \alpha +\gamma -\mu \leqslant 1,\quad \beta <\gamma ,\quad \mu
\leqslant \alpha ,\quad \gamma +\mu \leqslant \alpha +\beta ,\quad \gamma
+\mu \leqslant 1, \\
\frac{b_{2}}{b_{1}}\frac{\left\vert \cos \frac{\left( 2\alpha +\beta -\mu
\right) \pi }{2}\right\vert }{\cos \frac{\left( \beta +\mu \right) \pi }{2}}%
\leqslant \frac{a_{2}}{a_{1}}\leqslant \frac{b_{2}}{b_{1}}\frac{\sin \frac{%
\left( 2\alpha +\beta -\mu \right) \pi }{2}}{\sin \frac{\left( \beta +\mu
\right) \pi }{2}}, \\
\frac{b_{3}}{b_{1}}\frac{\left\vert \cos \frac{\left( 2\alpha +\gamma -\mu
\right) \pi }{2}\right\vert }{\cos \frac{\left( \gamma +\mu \right) \pi }{2}}%
\leqslant \frac{a_{3}}{a_{1}}\leqslant \frac{b_{3}}{b_{1}}\frac{\sin \frac{%
\left( 2\alpha +\gamma -\mu \right) \pi }{2}}{\sin \frac{\left( \gamma +\mu
\right) \pi }{2}}.
\end{gather*}

\noindent \textbf{Model IID.IDD} takes the form%
\begin{equation*}
\left( a_{1}\,_{0}\mathrm{I}_{t}^{\alpha }+a_{2}\,_{0}\mathrm{I}_{t}^{\beta
}+a_{3}\,_{0}\mathrm{D}_{t}^{\gamma }\right) \sigma \left( t\right) =\left(
b_{1}\,_{0}\mathrm{I}_{t}^{\mu }+b_{2}\,_{0}\mathrm{D}_{t}^{\nu }+b_{3}\,_{0}%
\mathrm{D}_{t}^{\alpha +\gamma -\mu }\right) \varepsilon \left( t\right) ,
\end{equation*}%
with narrowed thermodynamical restrictions%
\begin{gather*}
\mu \leqslant \beta <\alpha ,\quad \gamma \leqslant \nu ,\quad \alpha +\beta
+\gamma \leqslant 1+\mu ,\quad \mu +\nu -\gamma <\alpha \leqslant 1-\nu , \\
0\leqslant \left\{ 
\begin{tabular}{l}
$\alpha -\beta -\gamma -\mu $ \smallskip  \\ 
$\alpha -2\mu -\nu $ \smallskip 
\end{tabular}%
\right\} \leqslant 2\alpha -\beta -2\mu -\nu \leqslant \left\{ 
\begin{tabular}{l}
$2\alpha -\beta -\mu $ \smallskip  \\ 
$2\alpha +\gamma -2\mu -\nu $ \smallskip 
\end{tabular}%
\right\} \leqslant 2\alpha +\gamma -\mu <1, \\
-\frac{b_{3}}{b_{1}}\frac{\cos \frac{\left( 2\alpha +\gamma -\mu \right) \pi 
}{2}}{\cos \frac{\left( \gamma +\mu \right) \pi }{2}}\leqslant \frac{a_{3}}{%
a_{1}}\leqslant \frac{b_{3}}{b_{1}}\frac{\sin \frac{\left( 2\alpha +\gamma
-\mu \right) \pi }{2}}{\sin \frac{\left( \gamma +\mu \right) \pi }{2}}\frac{%
\cos \frac{\left( 2\alpha +\gamma -\mu \right) \pi }{2}}{\cos \frac{\left(
\gamma +\mu \right) \pi }{2}}\leqslant \frac{b_{3}}{b_{1}}\frac{\sin \frac{%
\left( 2\alpha +\gamma -\mu \right) \pi }{2}}{\sin \frac{\left( \gamma +\mu
\right) \pi }{2}}.
\end{gather*}

\noindent \textbf{Model I}$^{{}^{+}}$\textbf{ID.I}$^{{}^{+}}$\textbf{ID}
takes the form%
\begin{equation*}
\left( a_{1}\,_{0}\mathrm{I}_{t}^{1+\alpha }+a_{2}\,_{0}\mathrm{I}_{t}^{%
\frac{1+\alpha -\gamma }{2}}+a_{3}\,_{0}\mathrm{D}_{t}^{\gamma }\right)
\sigma \left( t\right) =\left( b_{1}\,_{0}\mathrm{I}_{t}^{1+\mu }+b_{2}\,_{0}%
\mathrm{I}_{t}^{\frac{1+\mu -\left( \alpha +\gamma -\mu \right) }{2}%
}+b_{3}\,_{0}\mathrm{D}_{t}^{\alpha +\gamma -\mu }\right) \varepsilon \left(
t\right) ,
\end{equation*}%
with narrowed thermodynamical restrictions%
\begin{gather*}
\mu \leqslant \alpha ,\quad \alpha +\gamma +2\left( \alpha -\mu \right)
=3\alpha +\gamma -2\mu \leqslant 1, \\
\frac{a_{2}}{a_{1}}\leqslant \frac{b_{2}}{b_{1}}\frac{\cos \frac{\left(
1-3\alpha -\gamma +2\mu \right) \pi }{4}}{\cos \frac{\left( 1+\alpha -\gamma
-2\mu \right) \pi }{4}}\frac{\sin \frac{\left( 1-3\alpha -\gamma +2\mu
\right) \pi }{4}}{\sin \frac{\left( 1+\alpha -\gamma -2\mu \right) \pi }{4}}%
\leqslant \frac{b_{2}}{b_{1}}\frac{\cos \frac{\left( 1-3\alpha -\gamma +2\mu
\right) \pi }{4}}{\cos \frac{\left( 1+\alpha -\gamma -2\mu \right) \pi }{4}},
\\
\frac{a_{3}}{a_{2}}\leqslant \frac{b_{3}}{b_{2}}\frac{\sin \frac{\left(
1+3\alpha +\gamma -2\mu \right) \pi }{4}}{\sin \frac{\left( 1-\alpha +\gamma
+2\mu \right) \pi }{4}}\frac{\cos \frac{\left( 1+3\alpha +\gamma -2\mu
\right) \pi }{4}}{\cos \frac{\left( 1-\alpha +\gamma +2\mu \right) \pi }{4}}%
\leqslant \frac{b_{3}}{b_{2}}\frac{\sin \frac{\left( 1+3\alpha +\gamma -2\mu
\right) \pi }{4}}{\sin \frac{\left( 1-\alpha +\gamma +2\mu \right) \pi }{4}},
\\
a_{3}b_{1}\cos \frac{\left( \gamma +\mu \right) \pi }{2}\leqslant
a_{2}b_{2}\sin \frac{\left( \alpha -\mu \right) \pi }{2}+a_{1}b_{3}\cos 
\frac{\left( 2\alpha +\gamma -\mu \right) \pi }{2}, \\
a_{1}b_{3}\sin \frac{\left( 2\alpha +\gamma -\mu \right) \pi }{2}\leqslant
a_{2}b_{2}\cos \frac{\left( \alpha -\mu \right) \pi }{2}-a_{3}b_{1}\sin 
\frac{\left( \gamma +\mu \right) \pi }{2}, \\
a_{1}b_{3}\sin \frac{\left( 2\alpha +\gamma -\mu \right) \pi }{2}\leqslant
a_{2}b_{2}\cos \frac{\left( \alpha -\mu \right) \pi }{2}\frac{\sin \frac{%
\left( \alpha -\mu \right) \pi }{2}}{\cos \frac{\left( 2\alpha +\gamma -\mu
\right) \pi }{2}}+a_{3}b_{1}\sin \frac{\left( \gamma +\mu \right) \pi }{2}%
\frac{\cos \frac{\left( \gamma +\mu \right) \pi }{2}}{\cos \frac{\left(
2\alpha +\gamma -\mu \right) \pi }{2}}.
\end{gather*}

\noindent \textbf{Model IDD}$^{{}^{+}}$\textbf{.IDD}$^{{}^{+}}$ takes the
form%
\begin{equation*}
\left( a_{1}\,_{0}\mathrm{I}_{t}^{\alpha }+a_{2}\,_{0}\mathrm{D}_{t}^{\frac{%
1+\gamma -\alpha }{2}}+a_{3}\,_{0}\mathrm{D}_{t}^{1+\gamma }\right) \sigma
\left( t\right) =\left( b_{1}\,_{0}\mathrm{I}_{t}^{\alpha +\gamma -\eta
}+b_{2}\,_{0}\mathrm{D}_{t}^{\frac{1+\eta -\left( \alpha +\gamma -\eta
\right) }{2}}+b_{3}\,_{0}\mathrm{D}_{t}^{1+\eta }\right) \varepsilon \left(
t\right) ,
\end{equation*}%
with narrowed thermodynamical restrictions%
\begin{gather*}
0\leqslant \alpha +\gamma -\eta \leqslant 1,\quad \gamma \leqslant \eta
,\quad \alpha +\eta +\left( \eta -\gamma \right) =\alpha -\gamma +2\eta
\leqslant 1, \\
\frac{a_{2}}{a_{1}}\leqslant \frac{b_{2}}{b_{1}}\frac{\sin \frac{\left(
1+\alpha -\gamma +2\eta \right) \pi }{4}}{\sin \frac{\left( 1+\alpha
+3\gamma -2\eta \right) \pi }{4}}\frac{\cos \frac{\left( 1+\alpha -\gamma
+2\eta \right) \pi }{4}}{\cos \frac{\left( 1+\alpha +3\gamma -2\eta \right)
\pi }{4}}\leqslant \frac{b_{2}}{b_{1}}\frac{\sin \frac{\left( 1+\alpha
-\gamma +2\eta \right) \pi }{4}}{\sin \frac{\left( 1+\alpha +3\gamma -2\eta
\right) \pi }{4}}, \\
\frac{a_{3}}{a_{2}}\leqslant \frac{b_{3}}{b_{2}}\frac{\cos \frac{\left(
1-\alpha +\gamma -2\eta \right) \pi }{4}}{\cos \frac{\left( 1-\alpha
-3\gamma +2\eta \right) \pi }{4}}\frac{\sin \frac{\left( 1-\alpha +\gamma
-2\eta \right) \pi }{4}}{\sin \frac{\left( 1-\alpha -3\gamma +2\eta \right)
\pi }{4}}\leqslant \frac{b_{3}}{b_{2}}\frac{\cos \frac{\left( 1-\alpha
+\gamma -2\eta \right) \pi }{4}}{\cos \frac{\left( 1-\alpha -3\gamma +2\eta
\right) \pi }{4}}, \\
a_{3}b_{1}\cos \frac{\left( \alpha +2\gamma -\eta \right) \pi }{2}%
-a_{2}b_{2}\sin \frac{\left( \eta -\gamma \right) \pi }{2}\leqslant
a_{1}b_{3}\cos \frac{\left( \alpha +\eta \right) \pi }{2}, \\
a_{1}b_{3}\sin \frac{\left( \alpha +\eta \right) \pi }{2}\leqslant
a_{2}b_{2}\cos \frac{\left( \eta -\gamma \right) \pi }{2}-a_{3}b_{1}\sin 
\frac{\left( \alpha +2\gamma -\eta \right) \pi }{2}, \\
a_{1}b_{3}\sin \frac{\left( \alpha +\eta \right) \pi }{2}\leqslant
a_{2}b_{2}\cos \frac{\left( \eta -\gamma \right) \pi }{2}\frac{\sin \frac{%
\left( \eta -\gamma \right) \pi }{2}}{\cos \frac{\left( \alpha +\eta \right)
\pi }{2}}+a_{3}b_{1}\sin \frac{\left( \alpha +2\gamma -\eta \right) \pi }{2}%
\frac{\cos \frac{\left( \alpha +2\gamma -\eta \right) \pi }{2}}{\cos \frac{%
\left( \alpha +\eta \right) \pi }{2}}.
\end{gather*}

\noindent \textbf{Model I}$^{{}^{+}}$\textbf{ID.IDD}$^{{}^{+}}$ takes the
form%
\begin{equation*}
\left( a_{1}\,_{0}\mathrm{I}_{t}^{1+\alpha }+a_{2}\,_{0}\mathrm{I}_{t}^{%
\frac{1+\alpha -\gamma }{2}}+a_{3}\,_{0}\mathrm{D}_{t}^{\gamma }\right)
\sigma \left( t\right) =\left( b_{1}\,_{0}\mathrm{I}_{t}^{\alpha +\gamma
-\eta }+b_{2}\,_{0}\mathrm{D}_{t}^{\frac{1+\eta -\left( \alpha +\gamma -\eta
\right) }{2}}+b_{3}\,_{0}\mathrm{D}_{t}^{1+\eta }\right) \varepsilon \left(
t\right) ,
\end{equation*}%
with narrowed thermodynamical restrictions%
\begin{gather*}
\eta \leqslant \gamma ,\quad \alpha +\gamma +2\left( \gamma -\eta \right)
=\alpha +3\gamma -2\eta \leqslant 1, \\
\frac{a_{1}}{b_{1}}\frac{\sin \frac{\left( 1+\alpha -\gamma +2\eta \right)
\pi }{4}}{\cos \frac{\left( 1-\alpha -3\gamma +2\eta \right) \pi }{4}}%
\leqslant \frac{a_{1}}{b_{1}}\frac{\sin \frac{\left( 1+\alpha -\gamma +2\eta
\right) \pi }{4}}{\cos \frac{\left( 1-\alpha -3\gamma +2\eta \right) \pi }{4}%
}\frac{\cos \frac{\left( 1+\alpha -\gamma +2\eta \right) \pi }{4}}{\sin 
\frac{\left( 1-\alpha -3\gamma +2\eta \right) \pi }{4}}\leqslant \frac{a_{2}%
}{b_{2}}, \\
\frac{a_{2}}{b_{2}}\leqslant \frac{a_{3}}{b_{3}}\frac{\cos \frac{\left(
1-\alpha -3\gamma +2\eta \right) \pi }{4}}{\sin \frac{\left( 1+\alpha
-\gamma +2\eta \right) \pi }{4}}\frac{\sin \frac{\left( 1-\alpha -3\gamma
+2\eta \right) \pi }{4}}{\cos \frac{\left( 1+\alpha -\gamma +2\eta \right)
\pi }{4}}\leqslant \frac{a_{3}}{b_{3}}\frac{\cos \frac{\left( 1-\alpha
-3\gamma +2\eta \right) \pi }{4}}{\sin \frac{\left( 1+\alpha -\gamma +2\eta
\right) \pi }{4}}, \\
a_{1}b_{3}\cos \frac{\left( \alpha +\eta \right) \pi }{2}-a_{2}b_{2}\sin 
\frac{\left( \gamma -\eta \right) \pi }{2}\leqslant a_{3}b_{1}\cos \frac{%
\left( \alpha +2\gamma -\eta \right) \pi }{2}, \\
a_{3}b_{1}\sin \frac{\left( \alpha +2\gamma -\eta \right) \pi }{2}\leqslant
a_{2}b_{2}\cos \frac{\left( \gamma -\eta \right) \pi }{2}-a_{1}b_{3}\sin 
\frac{\left( \alpha +\eta \right) \pi }{2}, \\
a_{3}b_{1}\sin \frac{\left( \alpha +2\gamma -\eta \right) \pi }{2}\leqslant
a_{2}b_{2}\cos \frac{\left( \gamma -\eta \right) \pi }{2}\frac{\sin \frac{%
\left( \gamma -\eta \right) \pi }{2}}{\cos \frac{\left( \alpha +2\gamma
-\eta \right) \pi }{2}}+a_{1}b_{3}\sin \frac{\left( \alpha +\eta \right) \pi 
}{2}\frac{\cos \frac{\left( \alpha +\eta \right) \pi }{2}}{\cos \frac{\left(
\alpha +2\gamma -\eta \right) \pi }{2}}.
\end{gather*}

\subsection{Asymmetric models}

\noindent \textbf{Model IID.ID} takes the form%
\begin{equation*}
\left( a_{1}\,_{0}\mathrm{I}_{t}^{\alpha +\beta -\gamma }+a_{2}\,_{0}\mathrm{%
I}_{t}^{\nu }+a_{3}\,_{0}\mathrm{D}_{t}^{\gamma }\right) \sigma \left(
t\right) =\left( b_{1}\,_{0}\mathrm{I}_{t}^{\alpha }+b_{2}\,_{0}\mathrm{D}%
_{t}^{\beta }\right) \varepsilon \left( t\right) ,
\end{equation*}%
with narrowed thermodynamical restrictions%
\begin{gather*}
0\leqslant \alpha \leqslant \nu <\alpha +\beta -\gamma \leqslant 1,\quad
\beta +\nu \leqslant 1, \\
\frac{b_{1}}{b_{2}}\leqslant \frac{a_{1}}{a_{3}}\frac{\sin \frac{\left(
\alpha +2\beta -\gamma \right) \pi }{2}}{\sin \frac{\left( \alpha +\gamma
\right) \pi }{2}}\frac{\cos \frac{\left( \alpha +2\beta -\gamma \right) \pi 
}{2}}{\cos \frac{\left( \alpha +\gamma \right) \pi }{2}}\leqslant \frac{a_{1}%
}{a_{3}}\frac{\sin \frac{\left( \alpha +2\beta -\gamma \right) \pi }{2}}{%
\sin \frac{\left( \alpha +\gamma \right) \pi }{2}},
\end{gather*}%
valid if $\alpha +2\beta -\gamma <1,$ while if $\alpha +2\beta -\gamma
\geqslant 1,$ then the narrowed thermodynamical restrictions cannot be
guaranteed and constraints on model parameters are given by the
thermodynamical restrictions%
\begin{gather*}
0\leqslant \alpha \leqslant \nu <\alpha +\beta -\gamma \leqslant 1,\quad
\beta +\nu \leqslant 1, \\
\frac{a_{1}}{a_{3}}\frac{\left\vert \cos \frac{\left( \alpha +2\beta -\gamma
\right) \pi }{2}\right\vert }{\cos \frac{\left( \alpha +\gamma \right) \pi }{%
2}}\leqslant \frac{b_{1}}{b_{2}}\leqslant \frac{a_{1}}{a_{3}}\frac{\sin 
\frac{\left( \alpha +2\beta -\gamma \right) \pi }{2}}{\sin \frac{\left(
\alpha +\gamma \right) \pi }{2}}.
\end{gather*}

\noindent \textbf{Model IDD.DD}$^{{}^{+}}$ takes the form%
\begin{equation*}
\left( a_{1}\,_{0}\mathrm{I}_{t}^{\alpha }+a_{2}\,_{0}\mathrm{D}_{t}^{\beta
}+a_{3}\,_{0}\mathrm{D}_{t}^{\gamma }\right) \sigma \left( t\right) =\left(
b_{1}\,_{0}\mathrm{D}_{t}^{\mu }+b_{2}\,_{0}\mathrm{D}_{t}^{\alpha +\beta
+\mu }\right) \varepsilon \left( t\right) ,
\end{equation*}%
with narrowed thermodynamical restrictions%
\begin{gather*}
1\leqslant \alpha +\beta +\mu \leqslant 2,\quad \beta <\gamma \leqslant \mu
\leqslant 1-\alpha , \\
\frac{a_{1}}{a_{2}}\frac{\left\vert \cos \frac{\left( 2\alpha +\beta +\mu
\right) \pi }{2}\right\vert }{\cos \frac{\left( \mu -\beta \right) \pi }{2}}%
\leqslant \frac{a_{1}}{a_{2}}\frac{\left\vert \cos \frac{\left( 2\alpha
+\beta +\mu \right) \pi }{2}\right\vert }{\cos \frac{\left( \mu -\beta
\right) \pi }{2}}\frac{\sin \frac{\left( 2\alpha +\beta +\mu \right) \pi }{2}%
}{\sin \frac{\left( \mu -\beta \right) \pi }{2}}\leqslant \frac{b_{1}}{b_{2}}%
.
\end{gather*}

\noindent \textbf{Model I}$^{{}^{+}}$\textbf{ID.ID} takes the form%
\begin{equation*}
\left( a_{1}\,_{0}\mathrm{I}_{t}^{\alpha +\beta +\nu }+a_{2}\,_{0}\mathrm{I}%
_{t}^{\nu }+a_{3}\,_{0}\mathrm{D}_{t}^{\alpha +\beta -\nu }\right) \sigma
\left( t\right) =\left( b_{1}\,_{0}\mathrm{I}_{t}^{\alpha }+b_{2}\,_{0}%
\mathrm{D}_{t}^{\beta }\right) \varepsilon \left( t\right) ,
\end{equation*}%
with narrowed thermodynamical restrictions%
\begin{gather*}
0\leqslant \alpha +\beta -\nu \leqslant 1,\quad 1\leqslant \alpha +\beta
+\nu \leqslant 2,\quad \alpha \leqslant \nu \leqslant 1-\beta , \\
\frac{a_{1}}{a_{2}}\frac{\left\vert \cos \frac{\left( \alpha +2\beta +\nu
\right) \pi }{2}\right\vert }{\cos \frac{\left( \nu -\alpha \right) \pi }{2}}%
\leqslant \frac{a_{1}}{a_{2}}\frac{\left\vert \cos \frac{\left( \alpha
+2\beta +\nu \right) \pi }{2}\right\vert }{\cos \frac{\left( \nu -\alpha
\right) \pi }{2}}\frac{\sin \frac{\left( \alpha +2\beta +\nu \right) \pi }{2}%
}{\sin \frac{\left( \nu -\alpha \right) \pi }{2}}\leqslant \frac{b_{1}}{b_{2}%
}, \\
\frac{b_{1}}{b_{2}}\leqslant \frac{a_{2}}{a_{3}}\frac{\sin \frac{\left(
\beta +\nu \right) \pi }{2}}{\sin \frac{\left( 2\alpha +\beta -\nu \right)
\pi }{2}}\frac{\cos \frac{\left( \beta +\nu \right) \pi }{2}}{\cos \frac{%
\left( 2\alpha +\beta -\nu \right) \pi }{2}}\leqslant \frac{a_{2}}{a_{3}}%
\frac{\sin \frac{\left( \beta +\nu \right) \pi }{2}}{\sin \frac{\left(
2\alpha +\beta -\nu \right) \pi }{2}}.
\end{gather*}

\noindent \textbf{Model IDD}$^{{}^{+}}$\textbf{.DD}$^{{}^{+}}$ takes the form%
\begin{equation*}
\left( a_{1}\,_{0}\mathrm{I}_{t}^{\alpha }+a_{2}\,_{0}\mathrm{D}_{t}^{\beta
}+a_{3}\,_{0}\mathrm{D}_{t}^{\alpha +2\beta }\right) \sigma \left( t\right)
=\left( b_{1}\,_{0}\mathrm{D}_{t}^{\mu }+b_{2}\,_{0}\mathrm{D}_{t}^{\alpha
+\beta +\mu }\right) \varepsilon \left( t\right) ,
\end{equation*}%
with narrowed thermodynamical restrictions%
\begin{gather*}
1\leqslant \alpha +2\beta \leqslant 2,\quad 1\leqslant \alpha +\beta +\mu
\leqslant 2,\quad \beta \leqslant \mu \leqslant 1-\alpha , \\
\frac{a_{1}}{a_{2}}\frac{\left\vert \cos \frac{\left( 2\alpha +\beta +\mu
\right) \pi }{2}\right\vert }{\cos \frac{\left( \mu -\beta \right) \pi }{2}}%
\leqslant \frac{a_{1}}{a_{2}}\frac{\left\vert \cos \frac{\left( 2\alpha
+\beta +\mu \right) \pi }{2}\right\vert }{\cos \frac{\left( \mu -\beta
\right) \pi }{2}}\frac{\sin \frac{\left( 2\alpha +\beta +\mu \right) \pi }{2}%
}{\sin \frac{\left( \mu -\beta \right) \pi }{2}}\leqslant \frac{b_{1}}{b_{2}}%
, \\
\frac{b_{1}}{b_{2}}\leqslant \frac{a_{2}}{a_{3}}\frac{\sin \frac{\left(
\alpha +\mu \right) \pi }{2}}{\sin \frac{\left( \alpha +2\beta -\mu \right)
\pi }{2}}\frac{\cos \frac{\left( \alpha +\mu \right) \pi }{2}}{\cos \frac{%
\left( \alpha +2\beta -\mu \right) \pi }{2}}\leqslant \frac{a_{2}}{a_{3}}%
\frac{\sin \frac{\left( \alpha +\mu \right) \pi }{2}}{\sin \frac{\left(
\alpha +2\beta -\mu \right) \pi }{2}},
\end{gather*}%
valid if $2\alpha +\beta -\mu <1,$ while if $2\alpha +\beta -\mu \geqslant 1,
$then the narrowed thermodynamical restrictions cannot be guaranteed and
constraints on model parameters are given by the thermodynamical restrictions%
\begin{gather*}
1\leqslant \alpha +2\beta \leqslant 2,\quad 1\leqslant \alpha +\beta +\mu
\leqslant 2,\quad \beta \leqslant \mu \leqslant 1-\alpha , \\
\frac{a_{1}}{a_{2}}\frac{\left\vert \cos \frac{\left( 2\alpha +\beta +\mu
\right) \pi }{2}\right\vert }{\cos \frac{\left( \mu -\beta \right) \pi }{2}}%
\leqslant \frac{b_{1}}{b_{2}}\leqslant \frac{a_{2}}{a_{3}}\frac{\sin \frac{%
\left( \alpha +\mu \right) \pi }{2}}{\sin \frac{\left( \alpha +2\beta -\mu
\right) \pi }{2}}.
\end{gather*}

\noindent \textbf{Model ID.IDD} takes the form%
\begin{equation*}
\left( a_{1}\,_{0}\mathrm{I}_{t}^{\alpha }+a_{2}\,_{0}\mathrm{D}_{t}^{\beta
}\right) \sigma \left( t\right) =\left( b_{1}\,_{0}\mathrm{I}_{t}^{\mu
}+b_{2}\,_{0}\mathrm{D}_{t}^{\nu }+b_{3}\,_{0}\mathrm{D}_{t}^{\alpha +\beta
-\mu }\right) \varepsilon \left( t\right) ,
\end{equation*}%
with narrowed thermodynamical restrictions%
\begin{gather*}
0\leqslant \beta \leqslant \nu <\alpha +\beta -\mu \leqslant 1,\quad \mu
\leqslant \alpha \leqslant 1-\nu , \\
-\frac{a_{1}}{a_{2}}\frac{\cos \frac{\left( 2\alpha +\beta -\mu \right) \pi 
}{2}}{\cos \frac{\left( \beta +\mu \right) \pi }{2}}\leqslant \frac{b_{1}}{%
b_{3}}\leqslant \frac{a_{1}}{a_{2}}\frac{\sin \frac{\left( 2\alpha +\beta
-\mu \right) \pi }{2}}{\sin \frac{\left( \beta +\mu \right) \pi }{2}}\frac{%
\cos \frac{\left( 2\alpha +\beta -\mu \right) \pi }{2}}{\cos \frac{\left(
\beta +\mu \right) \pi }{2}}\leqslant \frac{a_{1}}{a_{2}}\frac{\sin \frac{%
\left( 2\alpha +\beta -\mu \right) \pi }{2}}{\sin \frac{\left( \beta +\mu
\right) \pi }{2}},
\end{gather*}%
valid if $2\alpha +\beta -\mu <1,$ while if $2\alpha +\beta -\mu \geqslant 1,
$then the narrowed thermodynamical restrictions cannot be guaranteed and
constraints on model parameters are given by the thermodynamical restrictions%
\begin{gather*}
0\leqslant \beta \leqslant \nu <\alpha +\beta -\mu \leqslant 1,\quad \mu
\leqslant \alpha \leqslant 1-\nu , \\
\frac{a_{1}}{a_{2}}\frac{\left\vert \cos \frac{\left( 2\alpha +\beta -\mu
\right) \pi }{2}\right\vert }{\cos \frac{\left( \beta +\mu \right) \pi }{2}}%
\leqslant \frac{b_{1}}{b_{3}}\leqslant \frac{a_{1}}{a_{2}}\frac{\sin \frac{%
\left( 2\alpha +\beta -\mu \right) \pi }{2}}{\sin \frac{\left( \beta +\mu
\right) \pi }{2}}.
\end{gather*}

\noindent \textbf{Model ID.DDD}$^{{}^{+}}$ takes the form%
\begin{equation*}
\left( a_{1}\,_{0}\mathrm{I}_{t}^{\alpha }+a_{2}\,_{0}\mathrm{D}_{t}^{\beta
}\right) \sigma \left( t\right) =\left( b_{1}\,_{0}\mathrm{D}_{t}^{\mu
}+b_{2}\,_{0}\mathrm{D}_{t}^{\nu }+b_{3}\,_{0}\mathrm{D}_{t}^{\alpha +\beta
+\nu }\right) \varepsilon \left( t\right) ,
\end{equation*}%
with narrowed thermodynamical restrictions%
\begin{gather*}
1\leqslant \alpha +\beta +\nu \leqslant 2,\quad \beta \leqslant \mu <\nu
\leqslant 1-\alpha , \\
\frac{a_{1}}{a_{2}}\frac{\left\vert \cos \frac{\left( 2\alpha +\beta +\nu
\right) \pi }{2}\right\vert }{\cos \frac{\left( \nu -\beta \right) \pi }{2}}%
\leqslant \frac{a_{1}}{a_{2}}\frac{\left\vert \cos \frac{\left( 2\alpha
+\beta +\nu \right) \pi }{2}\right\vert }{\cos \frac{\left( \nu -\beta
\right) \pi }{2}}\frac{\sin \frac{\left( 2\alpha +\beta +\nu \right) \pi }{2}%
}{\sin \frac{\left( \nu -\beta \right) \pi }{2}}\leqslant \frac{b_{2}}{b_{3}}%
.
\end{gather*}

\noindent \textbf{Model ID.IDD}$^{{}^{+}}$ takes the form%
\begin{equation*}
\left( a_{1}\,_{0}\mathrm{I}_{t}^{\alpha }+a_{2}\,_{0}\mathrm{D}_{t}^{\beta
}\right) \sigma \left( t\right) =\left( b_{1}\,_{0}\mathrm{I}_{t}^{\alpha
+\beta -\nu }+b_{2}\,_{0}\mathrm{D}_{t}^{\nu }+b_{3}\,_{0}\mathrm{D}%
_{t}^{\alpha +\beta +\nu }\right) \varepsilon \left( t\right) ,
\end{equation*}%
with narrowed thermodynamical restrictions%
\begin{gather*}
0\leqslant \alpha +\beta -\nu \leqslant 1,\quad 1\leqslant \alpha +\beta
+\nu \leqslant 2,\quad \beta \leqslant \nu \leqslant 1-\alpha , \\
\frac{b_{1}}{b_{2}}\frac{\sin \frac{\left( \alpha +2\beta -\nu \right) \pi }{%
2}}{\sin \frac{\left( \alpha +\nu \right) \pi }{2}}\leqslant \frac{b_{1}}{%
b_{2}}\frac{\sin \frac{\left( \alpha +2\beta -\nu \right) \pi }{2}}{\sin 
\frac{\left( \alpha +\nu \right) \pi }{2}}\frac{\cos \frac{\left( \alpha
+2\beta -\nu \right) \pi }{2}}{\cos \frac{\left( \alpha +\nu \right) \pi }{2}%
}\leqslant \frac{a_{1}}{a_{2}}, \\
\frac{a_{1}}{a_{2}}\leqslant \frac{b_{2}}{b_{3}}\frac{\cos \frac{\left( \nu
-\beta \right) \pi }{2}}{\left\vert \cos \frac{\left( 2\alpha +\beta +\nu
\right) \pi }{2}\right\vert }\frac{\sin \frac{\left( \nu -\beta \right) \pi 
}{2}}{\sin \frac{\left( 2\alpha +\beta +\nu \right) \pi }{2}}\leqslant \frac{%
b_{2}}{b_{3}}\frac{\cos \frac{\left( \nu -\beta \right) \pi }{2}}{\left\vert
\cos \frac{\left( 2\alpha +\beta +\nu \right) \pi }{2}\right\vert }.
\end{gather*}

\section{Relaxation modulus and creep compliance \label{RMCC}}

Considering fractional anti-Zener and Zener models, listed in Appendix \ref%
{FAZ-ZM}, the explicit forms of relaxation modulus and creep compliance are
derived in \cite{SD-2} and quoted in the sequel. The relaxation modulus
takes the form%
\begin{equation}
\sigma _{sr}(t)=\sigma _{sr}^{\left( \mathrm{NP}\right) }\left( t\right)
+\left\{ \!\!\!%
\begin{tabular}{ll}
$0$, & if $\phi _{\sigma }$ has no zeros, \smallskip \\ 
$\sigma _{sr}^{\left( \mathrm{RP}\right) }\left( t\right) $, & if $\phi
_{\sigma }$ has a negative real zero, \smallskip \\ 
$\sigma _{sr}^{\left( \mathrm{CCP}\right) }\left( t\right) $, & if $\phi
_{\sigma }$ has a pair of complex conjugated zeros,%
\end{tabular}%
\ \right.  \label{sr-opste}
\end{equation}%
with the corresponding functions $\sigma _{sr}^{\left( \mathrm{NP}\right) },$
$\sigma _{sr}^{\left( \mathrm{RP}\right) },$ and $\sigma _{sr}^{\left( 
\mathrm{CCP}\right) }$ given by%
\begin{align}
\sigma _{sr}^{\left( \mathrm{NP}\right) }\left( t\right) & =\frac{1}{\pi }%
\int_{0}^{\infty }\frac{1}{\rho ^{1-\xi }}\frac{\left\vert \phi
_{\varepsilon }\left( \rho \mathrm{e}^{\mathrm{i}\pi }\right) \right\vert }{%
\left\vert \phi _{\sigma }\left( \rho \mathrm{e}^{\mathrm{i}\pi }\right)
\right\vert }\sin \left( \arg \phi _{\varepsilon }\left( \rho \mathrm{e}^{%
\mathrm{i}\pi }\right) -\arg \phi _{\sigma }\left( \rho \mathrm{e}^{\mathrm{i%
}\pi }\right) +\xi \pi \right) \mathrm{e}^{-\rho t}\mathrm{d}\rho ,
\label{sigma-NP} \\
\sigma _{sr}^{\left( \mathrm{NP}\right) }\left( t\right) & =\frac{1}{\pi }%
\int_{0}^{\infty }\frac{1}{\rho ^{1-\xi }}\frac{K\left( \rho \right) }{%
\left\vert \phi _{\sigma }\left( \rho \mathrm{e}^{\mathrm{i}\pi }\right)
\right\vert ^{2}}\mathrm{e}^{-\rho t}\mathrm{d}\rho ,  \label{sigma-NP-1} \\
\sigma _{sr}^{\left( \mathrm{RP}\right) }\left( t\right) & =-\frac{1}{\rho _{%
\scriptscriptstyle\mathrm{R}{\mathrm{P}}}^{1-\xi }}\frac{\left\vert \phi
_{\varepsilon }\left( s_{\scriptscriptstyle\mathrm{R}{\mathrm{P}}}\right)
\right\vert }{\left\vert \phi _{\sigma }^{\prime }\left( s_{%
\scriptscriptstyle\mathrm{R}{\mathrm{P}}}\right) \right\vert }\cos \left(
\arg \phi _{\varepsilon }\left( s_{\scriptscriptstyle\mathrm{R}{\mathrm{P}}%
}\right) -\arg \phi _{\sigma }^{\prime }\left( s_{\scriptscriptstyle\mathrm{R%
}{\mathrm{P}}}\right) +\xi \pi \right) \mathrm{e}^{-\rho _{\scriptscriptstyle%
\mathrm{R}{\mathrm{P}}}t},  \label{sigma-RP} \\
\sigma _{sr}^{\left( \mathrm{CCP}\right) }\left( t\right) & =2\frac{1}{\rho
_{\scriptscriptstyle{\mathrm{CCP}}}^{1-\xi }}\frac{\left\vert \phi
_{\varepsilon }\left( s_{\scriptscriptstyle{\mathrm{CCP}}}\right)
\right\vert }{\left\vert \phi _{\sigma }^{\prime }\left( s_{%
\scriptscriptstyle{\mathrm{CCP}}}\right) \right\vert }\mathrm{e}%
^{-\left\vert \func{Re}s_{\scriptscriptstyle{\mathrm{CCP}}}\right\vert
t}\cos \left( \func{Im}s_{\scriptscriptstyle{\mathrm{CCP}}}t+\arg \phi
_{\varepsilon }\left( s_{\scriptscriptstyle{\mathrm{CCP}}}\right) -\arg \phi
_{\sigma }^{\prime }\left( s_{\scriptscriptstyle{\mathrm{CCP}}}\right)
-\left( 1-\xi \right) \varphi _{\scriptscriptstyle{\mathrm{CCP}}}\right) ,
\label{sigma-CCP}
\end{align}%
where $\phi _{\sigma }^{\prime }\left( s\right) =\frac{\mathrm{d}}{\mathrm{d}%
s}\phi _{\sigma }\left( s\right) $ and where $s_{\scriptscriptstyle\mathrm{R}%
{\mathrm{P}}}=\rho _{\scriptscriptstyle\mathrm{R}{\mathrm{P}}}\,\mathrm{e}^{%
\mathrm{i}\pi }$ is a negative real zero of function $\phi _{\sigma },$
while $s_{\scriptscriptstyle{\mathrm{CCP}}}=\rho _{\scriptscriptstyle{%
\mathrm{CCP}}}\,\mathrm{e}^{\mathrm{i}\varphi _{\scriptscriptstyle{\mathrm{%
CCP}}}}$ is a complex zero of function $\phi _{\sigma }$ having negative
real part, with function $K$ defined by (\ref{K}).

On the other hand, the creep compliance takes the form%
\begin{equation}
\varepsilon _{cr}(t)=\varepsilon _{cr}^{\left( \mathrm{NP}\right) }\left(
t\right) +\left\{ \!\!\!%
\begin{tabular}{ll}
$0$, & if $\phi _{\varepsilon }$ has no zeros, \smallskip \\ 
$\varepsilon _{cr}^{\left( \mathrm{RP}\right) }\left( t\right) $, & if $\phi
_{\varepsilon }$ has a negative real zero, \smallskip \\ 
$\varepsilon _{cr}^{\left( \mathrm{CCP}\right) }\left( t\right) $, & if $%
\phi _{\varepsilon }$ has a pair of complex conjugated zeros,%
\end{tabular}%
\ \right.  \label{cr-opste}
\end{equation}%
with the corresponding functions $\varepsilon _{cr}^{\left( \mathrm{NP}%
\right) },$ $\varepsilon _{cr}^{\left( \mathrm{RP}\right) },$ and $%
\varepsilon _{cr}^{\left( \mathrm{CCP}\right) }$ given by%
\begin{align}
\varepsilon _{cr}^{\left( \mathrm{NP}\right) }\left( t\right) & =\frac{1}{%
\pi }\int_{0}^{\infty }\frac{1}{\rho ^{1+\xi }}\frac{\left\vert \phi
_{\sigma }\left( \rho \mathrm{e}^{\mathrm{i}\pi }\right) \right\vert }{%
\left\vert \phi _{\varepsilon }\left( \rho \mathrm{e}^{\mathrm{i}\pi
}\right) \right\vert }\sin \left( \arg \phi _{\varepsilon }\left( \rho 
\mathrm{e}^{\mathrm{i}\pi }\right) -\arg \phi _{\sigma }\left( \rho \mathrm{e%
}^{\mathrm{i}\pi }\right) +\xi \pi \right) \left( 1-\mathrm{e}^{-\rho
t}\right) \mathrm{d}\rho ,  \label{epsilon-NP} \\
\varepsilon _{cr}^{\left( \mathrm{NP}\right) }\left( t\right) & =\frac{1}{%
\pi }\int_{0}^{\infty }\frac{1}{\rho ^{1+\xi }}\frac{K\left( \rho \right) }{%
\left\vert \phi _{\varepsilon }\left( \rho \mathrm{e}^{\mathrm{i}\pi
}\right) \right\vert ^{2}}\left( 1-\mathrm{e}^{-\rho t}\right) \mathrm{d}%
\rho ,  \label{epsilon-NP-1} \\
\varepsilon _{cr}^{\left( \mathrm{RP}\right) }\left( t\right) & =-\frac{1}{%
\rho _{\scriptscriptstyle\mathrm{R}{\mathrm{P}}}^{1+\xi }}\frac{\left\vert
\phi _{\sigma }\left( s_{\scriptscriptstyle\mathrm{R}{\mathrm{P}}}\right)
\right\vert }{\left\vert \phi _{\varepsilon }^{\prime }\left( s_{%
\scriptscriptstyle\mathrm{R}{\mathrm{P}}}\right) \right\vert }\cos \left(
\arg \phi _{\varepsilon }^{\prime }\left( s_{\scriptscriptstyle\mathrm{R}{%
\mathrm{P}}}\right) -\arg \phi _{\sigma }\left( s_{\scriptscriptstyle\mathrm{%
R}{\mathrm{P}}}\right) +\xi \pi \right) \left( 1-\mathrm{e}^{-\rho _{%
\scriptscriptstyle\mathrm{R}{\mathrm{P}}}t}\right) ,  \label{epsilon-RP} \\
\varepsilon _{cr}^{\left( \mathrm{CCP}\right) }\left( t\right) & =2\frac{1}{%
\rho _{\scriptscriptstyle{\mathrm{CCP}}}^{1+\xi }}\frac{\left\vert \phi
_{\sigma }\left( s_{\scriptscriptstyle{\mathrm{CCP}}}\right) \right\vert }{%
\left\vert \phi _{\varepsilon }^{\prime }\left( s_{\scriptscriptstyle{%
\mathrm{CCP}}}\right) \right\vert }  \notag \\
& \qquad \times \bigg(\mathrm{e}^{-\left\vert \func{Re}s_{\scriptscriptstyle{%
\mathrm{CCP}}}\right\vert t}\cos \left( \func{Im}s_{\scriptscriptstyle{%
\mathrm{CCP}}}t-\arg \phi _{\varepsilon }^{\prime }\left( s_{%
\scriptscriptstyle{\mathrm{CCP}}}\right) +\arg \phi _{\sigma }\left( s_{%
\scriptscriptstyle{\mathrm{CCP}}}\right) -\left( 1+\xi \right) \varphi _{%
\scriptscriptstyle{\mathrm{CCP}}}\right)  \notag \\
& \qquad \qquad -\cos \left( \arg \phi _{\varepsilon }^{\prime }\left( s_{%
\scriptscriptstyle{\mathrm{CCP}}}\right) -\arg \phi _{\sigma }\left( s_{%
\scriptscriptstyle{\mathrm{CCP}}}\right) +\left( 1+\xi \right) \varphi _{%
\scriptscriptstyle{\mathrm{CCP}}}\right) \bigg),  \label{epsilon-CCP}
\end{align}%
where $\phi _{\varepsilon }^{\prime }\left( s\right) =\frac{\mathrm{d}}{%
\mathrm{d}s}\phi _{\varepsilon }\left( s\right) $ and where $s_{%
\scriptscriptstyle\mathrm{R}{\mathrm{P}}}=\rho _{\scriptscriptstyle\mathrm{R}%
{\mathrm{P}}}\,\mathrm{e}^{\mathrm{i}\pi }$ is a negative real zero of
function $\phi _{\varepsilon },$ while $s_{\scriptscriptstyle{\mathrm{CCP}}%
}=\rho _{\scriptscriptstyle{\mathrm{CCP}}}\,\mathrm{e}^{\mathrm{i}\varphi _{%
\scriptscriptstyle{\mathrm{CCP}}}}$ is a complex zero of function $\phi
_{\varepsilon }$ having negative real part.

The property of relaxation modulus to be completely monotonic and creep
compliance to be a Bernstein function is guaranteed by requiring that
function $K,$ defined by (\ref{K}) and appearing in functions $\sigma
_{sr}^{\left( \mathrm{NP}\right) }$ and $\varepsilon _{cr}^{\left( \mathrm{NP%
}\right) },$ see (\ref{sigma-NP-1}) and (\ref{epsilon-NP-1}), is
non-negative, i.e., by requiring%
\begin{equation*}
K\left( \rho \right) \geqslant 0,\quad \text{i.e.,}\quad \sin \left( \arg
\phi _{\varepsilon }\left( \rho \mathrm{e}^{\mathrm{i}\pi }\right) -\arg
\phi _{\sigma }\left( \rho \mathrm{e}^{\mathrm{i}\pi }\right) +\xi \pi
\right) \geqslant 0\quad \text{for}\quad \rho \geqslant 0,
\end{equation*}%
in addition to the request that functions $\phi _{\sigma }$ and $\phi
_{\varepsilon }$ do not have zeros, see (\ref{sr-opste}) and (\ref{cr-opste}%
). The character of relaxation modulus (creep compliance) in time domain
changes if function $\phi _{\sigma }$ ($\phi _{\varepsilon }$) has zeros, so
that in addition to $\sigma _{sr}^{\left( \mathrm{NP}\right) }$ ($%
\varepsilon _{cr}^{\left( \mathrm{NP}\right) }$) there is either a term,
namely $\sigma _{sr}^{\left( \mathrm{RP}\right) }$ ($\varepsilon
_{cr}^{\left( \mathrm{RP}\right) }$), decaying exponentially in time in the
case of negative real zero, or a term, namely $\sigma _{sr}^{\left( \mathrm{%
CCP}\right) }$ ($\varepsilon _{cr}^{\left( \mathrm{CCP}\right) }$),
displaying damped oscillatory character in the case of complex conjugated
zeros with negative real part.

\end{document}

%% file: tabela-svih-modela.tex
 \begin{table}[p]
 \begin{center}
 \begin{tabular}{|c|c|c|c|c|c|c|c|}

 \hline \xrowht{14pt}

 & Function  $\phi _{\sigma }$ & Function $\phi _{\varepsilon }$ & Model &  & Order $\xi $ & Order $\lambda $ & Order $\kappa $ \\ 

 \hhline{|=|=|=|=|=|=|=|=|} \xrowht{14pt}

\multirow{8}{*}{\multirowcell{8}{\rotatebox{90}{Symmetric models}}} & \multirow{2}{*}{\multirowcell{2}{$ a_{1}+a_{2}s^{\alpha +\beta}$ }}  & \multirow{2}{*}{\multirowcell{2}{$ b_{1}+b_{2}s^{\alpha +\beta}$} } 
& ID.ID & $*$ & $\alpha -\mu $ & $-$ & $-$ \\

\cline{4-8} \xrowht{14pt}

 & & & ID.DD$^{+}$ &  & $\alpha +\mu <1$ & $-$ & $-$ \\

\Xcline{2-8}{2\arrayrulewidth} \xrowht{14pt}

& \multirow{3}{*}{\multirowcell{3}{$a_{1}+a_{2}s^{\lambda }+a_{3}s^{\alpha +\gamma }$}} &  \multirow{2}{*}{\multirowcell{2}{$b_{1}+b_{2}s^{\lambda }+b_{3}s^{\alpha +\gamma }$}} 
 & IID.IID & $*$ & $\eta-\gamma $ & $\alpha -\beta $ & $-$ \\

\cline{4-8} \xrowht{14pt}

 & & & IDD.IDD & $*$ & $\alpha -\mu $ & $\alpha +\beta <1$ & $-$ \\

\cline{3-8} \xrowht{14pt}

 & & $b_{1}+b_{2}s^{\kappa }+b_{3}s^{\alpha +\gamma }$ & IID.IDD &  & $\alpha -\mu $ & $\mu +\nu <1$ & $\alpha -\beta $ \\ 

\cline{2-8} \xrowht{14pt}

 & \multirow{3}{*}{\multirowcell{3}{$a_{1}+a_{2}s^{\frac{1+\alpha +\gamma }{2}}+a_{3}s^{1+\alpha +\gamma }$}} 
 &\multirow{3}{*}{\multirowcell{3}{$b_{1}+b_{2}s^{\frac{1+\alpha +\gamma }{2}}+b_{3}s^{1+\alpha +\gamma }$}} & I$^{+}$ID.I$^{+}$ID &  & $\alpha -\mu $ & $-$ & $-$ \\

\cline{4-8} \xrowht{14pt} 

 &  &  & IDD$^{+}$.IDD$^{+}$ &  & $\eta -\gamma $ & $-$ & $-$ \\ 

\cline{4-8} \xrowht{14pt}

& & & I$^{+}$ID.IDD$^{+}$ &  & $1-\left( \gamma -\eta \right) $ & $-$ & $-$ \\ 

\Xhline{4\arrayrulewidth} \xrowht{14pt}

 \multirow{7}{*}{\multirowcell{7}{\rotatebox{90}{Asymmetric models}}}  
 & \multirow{2}{*}{\multirowcell{2}{$a_{1}+a_{2}s^{\lambda }+a_{3}s^{\kappa }$}} 
 & \multirow{4}{*}{\multirowcell{4}{$b_{1}+b_{2}s^{\alpha +\beta }$ }}
 & IID.ID &  & $\beta -\gamma $ & $\left( \alpha +\beta \right)-\left( \nu +\gamma \right) $ & $\alpha +\beta <1$ \\ 

\cline{4-8} \xrowht{14pt}

 & & & IDD.DD$^{+}$ &  & $\alpha +\mu <1$ & $\alpha +\beta <1$ & $\alpha +\mu <1$ \\ 

\cline{2-2} \cline{4-8} \xrowht{14pt}

 & \multirow{2}{*}{\multirowcell{2}{$a_{1}+a_{2}s^{\alpha +\beta }+a_{3}s^{2\left( \alpha +\beta \right) }$}}
 & & I$^{+}$ID.ID &  & $\beta +\nu <1$ & $-$ & $-$ \\

\cline{4-8} \xrowht{14pt}

 & & & IDD$^{+}$.DD$^{+}$ &  & $\alpha +\mu <1$ & $-$ & $-$ \\

\Xcline{2-8}{2\arrayrulewidth} \xrowht{14pt}

 & \multirow{3}{*}{\multirowcell{3}{$a_{1}+a_{2}s^{\alpha +\beta }$}} & \multirow{2}{*}{\multirowcell{2}{$b_{1}+b_{2}s^{\lambda }+b_{3}s^{\kappa }$}} & ID.IDD &  & $\alpha-\mu $ & $\mu +\nu <1$ & $\alpha +\beta <1$ \\ 

\cline{4-8} \xrowht{14pt}

 & & & ID.DDD$^{+}$ &  & $\alpha +\mu <1$ & $\nu -\mu $ & $\alpha +\nu-\left( \mu -\beta \right) $ \\

\cline{3-8} \xrowht{14pt} 

  & & $b_{1}+b_{2}s^{\alpha +\beta}+b_{3}s^{2\left( \alpha +\beta \right) }$ & ID.IDD$^{+}$ &  & $\nu -\beta $ & $-$ & $-$ \\

\hline

 \end{tabular}
 \end{center}
 \caption{Summary of constitutive functions $\phi _{\sigma }$ and $\phi _{\varepsilon }$, along with the order $\xi$, corresponding to the thermodynamically consistent fractional anti-Zener and Zener models. The notation $*$ means that the orders $\alpha +\beta$ and $\alpha +\gamma$ belong either to the interval $(0,1)$ or interval $(1,2)$.}
 \label{skupina}
 \end{table}

%% file: parametri-1.tex
 \begin{table}[h]
 \begin{center}

\begin{tabular}{|c|c|c|c|c|c|c|c|c|}

 \hline \xrowht{14pt}

Case when $\phi _{\sigma }$ has & $\alpha $ & $\beta $ & $\nu $ & $a_{1}$ & $a_{2}$ & $a_{3}$ & $b_{1}$ & $b_{2}$ \\

 \hhline{|=|=|=|=|=|=|=|=|=|} \xrowht{14pt}

no zeros & \multirow{3}{*}{\multirowcell{3}{$0.35$}} & \multirow{3}{*}{\multirowcell{3}{$0.55$}} & \multirow{3}{*}{\multirowcell{3}{$0.4$}} & $0.05$ & $1.5$ & $0.45$ & $0.7$ & $0.95$ \\  \xrowht{14pt}
a negative real zero &  &  &  & $11$ & $28.4029\dots $ & $20.27$ & $7$ & $9.5$ \\  \xrowht{14pt}
\makecell{a pair of complex\\ conjugated zeros} & &  &  & $11$ & $15$ & $20.27$ & $7$ & $9.5$\\ 

\hline

\end{tabular}

 \end{center}
 \caption{Model parameters used for numerical examples.}
 \label{parametri}
 \end{table}